\documentclass[aps,prc,superscriptaddress,nofootinbib,dvipsnames,notitlepage]{revtex4-2}
\usepackage{graphicx}% Include figure files
\usepackage{physics}
\usepackage{amssymb}
\usepackage{amsmath}
\usepackage{amsthm}
\usepackage{color}
\usepackage{tikz}
\usepackage{tikz-3dplot}
\usetikzlibrary{3d,shapes,arrows.meta,decorations.pathreplacing,decorations.markings}
\usetikzlibrary{arrows.meta, decorations.markings}

\usepackage[colorlinks=true,  linkcolor=blue,citecolor=blue,urlcolor=blue]{hyperref}

\newtheorem{remark}{Remark}
\newtheorem{lemma}{Lemma}
\newtheorem{proposition}{Proposition}
\newtheorem{corollary}{Corollary}
\newtheorem{theorem}{Theorem}
\newcommand{\nnb}{\nonumber}
\newcommand{\1}{^\prime}
\newcommand{\2}{^{\prime\prime}}
\newcommand{\ZZ}{\mathbb{Z}}

\newcommand{\wt}{\widetilde}

\tikzstyle{midarrow} = [decoration={markings, mark= at position 0.6 with {\arrow{Stealth}}}, postaction={decorate}]
\begin{document}

%\preprint{APS/123-QED}

\title{
Topological quantum computation assisted by phase transitions
}% Force line breaks with \\

\author{Yuanjie \surname{Ren}}%
\email[]{yuanjie@mit.edu}
\author{Peter \surname{Shor}}%
\email[]{shor@math.mit.edu}
\affiliation{Department of Physics, Massachusetts Institute of Technology, Cambridge, Massachusetts 02139, USA}%Lines break automatically or can be forced with \\
\date{\today}% It is always \today, today,
\begin{abstract}
In this paper, we explore topological quantum computation augmented by subphases and phase transitions. We commence by investigating the anyon tunneling map, denoted as $\varphi$, between subphases of the quantum double model $\mathcal{D}(G)$ for any arbitrary finite group $G$. Subsequently, we delve into the relationship between $\varphi$ and the Floquet code, and extend the Abelian Floquet code to encompass non-abelian cases. We conclude by demonstrating how phase transitions in both the temporal and spatial directions can enhance the diversity of topological gates for general topological orders described by modular tensor categories.
\end{abstract}
%\pacs{25.75.Cj, 25.75.Dw, 25.75.Ld }% PACS, the Physics and Astronomy
\maketitle
%---------Use revtex4---------------------------
\section{Introduction}
Topological quantum computation (TQC) presents a robust paradigm for the design and realization of quantum computers, where logical information is encoded within topological degrees of freedom \cite{kitaev2003, fowler2012}. Essentially, it leverages the global properties of quantum systems, which are inherently resistant to local sources of noise, making them attractive candidates for fault-tolerant quantum computation.

There exist two predominant approaches in this domain. The first harnesses anyons within topological phases of matter \cite{shawn2015,RevModPhys.80.1083}. In this framework, topological qudits are represented by anyon fusion diagrams, and quantum gates are realized through the process of anyon braiding. The second approach involves the introduction of gapped boundaries \cite{PhysRevA.86.032324,PhysRevLett.119.170504}. Here, the braiding process engages gapped boundaries or punctures, with anyons condensed to them.

Recent advancements have demonstrated the power of Floquet code in the context of topological quantum computation \cite{Hastings2021dynamically, kesselring:2022, davydova2023,joseph2023}. This dynamical perspective oppens the possibility of constructing topological quantum computers through dynamical codes. These can be interpreted as phase transitions of isomorphic codes where logical qudits remain preserved at every phase transition. Such codes promise tangible engineering advantages, such as the potential for low-weight syndrome extraction \cite{Hastings2021dynamically} and the enrichment of gate sets via automorphisms \cite{davydova2023quantum}.

Motivated by these groundbreaking findings, our work aims to develop a comprehensive theoretical framework that incorporates phase transitions in the process of quantum computing. This scheme is feasible for general modular tensor categories (MTC), but for the scope of this work, we work mostly with the example of quantum double models $\mathcal{D}(G)$ of an arbitrary finite group $G$.
This work implies that quantum gates can arise  from various phases of quantum matter in a quantum-computing process, introducing the great prospect that certain gates, which are challenging to implement in one phase via elementary operations (like braiding or wrapping), might be efficiently realized in an alternative phase. By mixing these phases, one can access the computational advantages of both simultaneously. This "phase mixing" can manifest both spatially and temporally. In temporal terms, this concept extends the notion of Floquet codes, allowing for general schedules of sub-phases upon requirements met. Spatially, this is an augmentation of the surface code paradigm, where traditional holes (or punctures) of trivial phases (related to Lagrangian algebra) now expand to contain non-trivial subphases of the parent system (pertaining to non-Lagrangian algebra). We will also explore the integration of the two cases, namely, phase transitions in both temporal and spatial dimensions.

The paper is organized by the following.  In Sect.~\ref{sect_the_system}, we review the quantum double system and its derived phases. In Sect.~\ref{condnesation_and_tunneling_map},  we study the condensation processes and the anyon-tunneling map.  In Sect.~\ref{sect_Floquet_code}, we apply this theory to generalize Floquet codes, which has only been studied in the abelian cases. In Sect.~\ref{sect_topological_quantum_computation}, we investigate  topological logical operations enriched by introducing subphases in both temporal and spatial directions. 
%--------
%   Review of the quantum double system
%--------
\section{Review of the quantum double system}\label{sect_the_system}
In Sections~\ref{sect_the_system} to \ref{sect_Floquet_code}, we will focus exclusively on a specific modular tensor category: the representation categories of quantum double models, denoted as $Rep\mathcal{D}(G)$, for any given finite group 
$G$. The associated lattice model, referred to as the \textbf{Quantum Double Model} (QDM), has applications in fault-tolerant quantum computation as outlined by Kitaev in \cite{kitaev2003}. Subsequent mathematical intricacies of the QDM were elaborated upon by H. Bombin and M.A. Martin-Delgado in \cite{bombin:2008}. To ensure the comprehensiveness of this paper, this section provides a concise overview of the key mathematical aspects that underpin the discussions in the subsequent sections. Readers already familiar with the QDM might choose to skip this portion. It's also worth noting that throughout this paper, we use either $\overline{g}$ or $g^{-1}$ to signify the group inverse of a member 
$g$ of $G$.  In the context of the quantum double algebra $\mathcal{D}(G)$, we utilize the pair $(\overline{a},R)$ or $(C,R)$ to indicate a representation of $\mathcal{D}(G)$, where $C\equiv \overline{a}$ stands for the conjugacy class of $a\in G$ and $R$ is a representation of the centralizer group $Z(a)$. 
Lastly, we use $e$ to denote the unity of a group $G$ or an edge in a lattice.

The lattice representation of the QDM can be instantiated on any arbitrary lattice, denoted as 
$\Lambda$, which possesses an orientation. An orientation is also granted  to its dual, $\Lambda^*$, where the faces of $\Lambda$ are treated as the vertices of $\Lambda^*$ and vice versa. The physical degrees of freedom are positioned on the edges of 
$\Lambda$.  For the sake of clarity, we'll adopt the square lattice in our discussions. Nevertheless, the findings articulated in these sections are not contingent upon the specific choice of $\Lambda$. This is because the data we derive for quantum computation is inherently categorical.

The orientation selected for the square lattice is depicted in Fig.~\ref{fig_the_lattice_and_the_tunneling_ribbon}. Here, $\Lambda$ is represented using solid lines, while the dual lattice, $\Lambda^*$, is illustrated with dashed lines. At the bottom of this figure, two distinct triangle types are demonstrated. The one featuring the long edge aligned with lattice 
$\Lambda$ (or $\Lambda^*$) is termed the \textbf{direct triangle} $\tau\1$ (or \textbf{direct triangle} $\tau$). It's vital to note that the orientation of triangles isn't determined by their placement within the lattice. Instead, it's an intrinsic attribute of the triangle itself, symbolized by the double-arrow $\Rightarrow$ within the illustration. For any given triangle $t$, the arrow originates from the shorter edge $\partial_0 t$
 and points toward the other shorter edge $\partial_1 t$. A \textbf{site} is defined as the line segment that begins at a vertex $v$ on the lattice and stretches to one of its proximate plaquette or face centers, represented as $f$, and denoted as $s=(v,p)$. If the edge length of each square within 
$\Lambda$ stands at a unit length, then the length of a site becomes $1/\sqrt{2}$. For example, the shorter edges of a triangle $t$, represented as $\partial_i t$
 for $i=0,1$, are considered sites.

%----------
For each edge $e$ in the lattice $\Lambda$ we allocate an unnormalized qudit represented by $\mathcal{H}_e=\mathbb{C}^{|G|}=\{\ket{g}:g\in G\}$, where $|G|$ is the order of $G$. The Hilbert space for the QDM, $\mathcal{D}(G)$, is defined as $\mathcal{H}_\Lambda:=\bigotimes_v\mathcal{H}_v$, which is the tensor product of $\mathcal{H}_e$ across all edges within $\Lambda$. Next, we proceed to construct the ribbon operator on the lattice. We define two multiplication operators: $L^h_+$ which maps $g$ to $gh$, and $L^h$, which maps $g$ to $gh^{-1}$.
Their lattice realization for qubit on edge $e\in \Lambda$ is
\begin{align}
L_+^h(e)=\sum_{g\in G}\ket{hg}\bra{g},\quad  L_-^h(e)=\sum_{g\in G}\ket{gh^{-1}}\bra{g}
\end{align}
We also define one-dimensional projection operators to a specific  direction on edge $e$
\begin{align}
T^g_+(e)=\ket{g}\bra{g},\quad T^g_-(e)=\ket{g^{-1}}\bra{g^{-1}}.
\end{align}
Let's denote the long edge of a direct triangle $\tau\1$ (dual  triangle $\tau$) by $e_\tau$ ($e_{\tau\1}$).
With this notation in place, we can define operators that carry a triangle subscript: $L^h_\tau:=L^h_\pm(e_\tau)$ and $T^g_{\tau\1}:=T^g_\pm(e_{\tau\1})$. 
The choice of sign, $\pm$, is contingent on the alignment of the triangles's orientation, denoted $t=\tau,\tau\1$ (as represented by the double arrow $\Rightarrow$), with the orientation of edge $e$ (symbolized by $\rightarrow$ or $\dashrightarrow$ on $\Lambda$ or $\Lambda^*$, respectively).  For instance, in Fig.~\ref{fig_the_lattice_and_the_tunneling_ribbon}, we have 
 $L_\tau^h=L_+^h(e_\tau)$ and $T_{\tau\1}^g=T_+^h(e_{\tau\1})$.

Given $h$ and $g$ as eleemnts within $G$, for the  dual triangle $\tau$ and the direct triangle $\tau^\prime$, we establish the following definitions
\begin{align}
F_\tau^{h,g}=\delta_{1,g}L_\tau^h,\quad F_{\tau^\prime}^{h,g}=T_{\tau^\prime}^g.
\end{align}
A general ribbon $\rho$ consists of a sequence of triangles, $t_i$, where each  $t_i$ can be either a dual triangle $\tau$ or a direct triangle $\tau\1$:
\begin{align}
    \rho=t_0t_1\cdots t_n.
\end{align}
The arrows $\Rightarrow$ within these triangles maintain consistent direction throughout the sequence of triangles, and 
\begin{align}
  \partial_0t_i=\partial_1 t_{i-1},\quad 
  \partial_0\rho:=\partial_0 t_0,\quad 
  \partial_1\rho:=\partial_1 t_n.
\end{align}
If $\rho$ consists solely of dual (or direct) triangles, we refer to it as a dual (or direct) ribbon.
The $F$-ribbon operator on ribbon $\rho$ is defined recursively by the following glue-formula
\begin{align}
    F_\rho^{h,g}:=\sum_{k\in G}F_{\rho_1}^{h,k}F_{\rho_2}^{\overline{k} hk,\overline{k},g}.
\end{align}
Specifically, appending a dual triangle to the tail or the head is
\begin{align}    F_{\rho\tau}^{h,g}=F_\rho^{h,g}L_\tau^{\overline{g}h g},\quad F_{\tau \rho} =L_\tau^h F_\rho^{h,g},\label{one_triangle_gluing}
\end{align}
respectively. 
Let's define the ribbon operator with a subset $S$ of group $G$ as a parameter:
\begin{align}
    F^{S,g}:=\frac{1}{|S|}\sum_{s\in S} F^{s,g},\quad F^{g,S}:=\sum_{s\in S}F^{g,s}.
\end{align}
 Let $s=(v,p)$ represent the site associated with vertex $v$ and plaquette $p$. Let 
 $\alpha_s$ (or $\beta_s$) denote the uniquely smallest dual (or direct) ribbon that both starts and terminates at site $s$. 
The loop ribbon $\alpha_s$ (or $\beta_s$), composed of  four triangles forms a  unit square within the lattice $\Lambda^*$ (or $\Lambda$). Accordingly, we define:
\begin{align}
A_s^h:=F_{\alpha_s}^{h,e},\quad B_s^g:=F_{\beta_s}^{e,g^{-1}}.
\end{align}
The operator $A_s^h$  factors into four $L^h_\pm$  operators, which act concurrently on the four associated physical qudits. Among these, two undergo left multiplication while the other two experience right multiplication.   The operator $B_s^g$ verifies that the group multiplication of its four supporting qudits yields $g\in G$. Depending on the sign $\pm$, two of these qudits are inverted. 

Let $M$ be a subgroup of $G$. Let $N$ be a normal subgroup of $G$. We can define for $s=(v,p)$
\begin{align}
    A^M_v:=\frac{1}{|M|}\sum_{m\in M} A^m_s,\quad B^N_p:=\sum_{n\in N} B_s^n,
\end{align}
For $M=G$, $N=\{e\}$, we obtain the Hamiltonian for Kitaev's original Quantum double model as cited in \cite{kitaev2003}:
\begin{align}\label{Hamiltonian_of_the_original_QDM}
H^{G,e}=-\sum_{v\in \Lambda} A_v - \sum_{p\in\Lambda} B_p,
\end{align}
where $A_v=A_v^G$ and $B_p=B_p^e$.
This is a commuting projector model, meaning that each term functions as a projector, and all pairs of terms within the Hamiltonian commute. More will be discussed in the following section, Sect.~\ref{condnesation_and_tunneling_map}).

 Let $\rho$ be a ribbon. Define algebra
\begin{align}
    \mathcal{F}_\rho:=Span\{F^{h,g}_\rho:h,g\in G\}.
\end{align}
This algebraic structure represents the algebra of two-point excitations as described in \cite{kitaev2003,bombin:2008}. Specifically, for any given $F_\rho\in\mathcal{F}_\rho$, we have 
$[F_\rho, A_s]=[F_\rho,B_s]=0$ for every site $s=(v,p)$, unless $s$ is positioned at the terminus of $\rho$. It's noteworthy that the anyonic types in the QDM can be characterized by the pair $(C,R)$. Here, $C$ denotes a conjugacy class within $G$, while 
$R$ symbolizes a representation of the centralizer group $Z(r_C)$ for a representative $r_C\in C$ (the selection of representatives isn't important). Although the basis $F^{h,g}_\rho$ of $\mathcal{F}_\rho$ isn't the eigen-basis corresponding to anyonic types (or representations of 
$\mathcal{D}(G)$), appropriate linear transformations, specifically Eq.(B66) and Eq.(B67) from \cite{bombin:2008}, enable a transition from the $(h,g)$-basis to the $(C,R,u,v)$-basis. In this setup, 
$u$ and $v$ label the local degrees of freedom at $\partial_0\rho$ and $\partial_1\rho$ respectively. Navigating within the 
$(C,R,u,v)$-basis is more intricate than in the $(h,g)$-basis. Thus,  we turn to the character theory of algebras, which reveals information about anyon types before and after the tunneling, regardless of the bases chosen.  This method is detailed further in Sect.~\ref{condnesation_and_tunneling_map}.
%----------------
\section{Condensation and the anyon tunneling map}\label{condnesation_and_tunneling_map}
When introducing multiple phases to the TQC system, additional features emerge. Temporally, we aim to understand how an anyon evolves during phase transitions. Spatially, it's essential to discern the ground state as described by fusion trees. 
This necessitates a thorough investigation into how the anyon transitions from one topological order to another.
%----------
\subsection{The derived phase \texorpdfstring{$\mathcal{D}(M/N)$}{Lg} and its ribbon operators}
We present the subphases of $\mathcal{D}(G)$.  They are specified by the Hamiltonian  \cite{bombin:2008}
\begin{align}\label{H_GMN}
H^{M,N}=-\sum_v A_v^M - \sum_p B_p^N-U\sum_e (T_e^M+L_e^N). 
\end{align}
This is a commuting projector model, since
\begin{align}&
&O_i^2=O_i, \quad [O_i,O_j]=0,\quad \forall \ O_i,O_j\in \{A^M_v, B_p^N, T^M_e, L^N_e:v,p,e\in \Lambda\}.
\end{align}
When we set $M=G$ and $N=\{e\}$, the terms $T^M_e$ and $L_e^N$ become trivial, reducing to Kitaev's original quantum double model. 
Our choice of making $N$ a normal subgroup ensures the single-qudit operator exhibits the relaition $L_+^N=L_-^N$.
%------
The presented Hamiltonian characterizes subphases $\mathcal{D}(M/N)$ of the parent phase $D(G)$. This arises from our decision to consider the types of excitations as solely point-like, and hence no confinement involves. Analogous to the excitations in $\mathcal{D}(G)$, any given ribbon operator $F_\rho$ in the derived phase merely excites only the two endpoints $\partial_0\rho $ and $\partial_1\rho$.
The interior of the ribbon remains commutative with the Hamiltonian. The confinement terms $T^M$ and $L^N$ remain as $+1$ eigenvalues at each edge (including those located at the endpoints of $\rho$). The only difference is that the parameter $(h,g)$ now lies in the quotient group $M/N$.

 This model does not capture the entirety of possible quantum double subphases within $\mathcal{D}(G)$, categorically. Specifically, there might be instances where it overlooks certain subphases that are isomorphic to a given quantum double model but do not fit into the $\mathcal{D}(M/N)$ framework. Such omissions can emerge from unique features of $\mathcal{D}(G)$ for a specific $G$.

To encompass a broader range of scenarios, the terms $A_v$ and $B_p$ in the parent phase can be expressed as sums of closed ribbon operators $F^{R,C}_\sigma$ within the $(C,R)$-basis for a closed ribbon $\sigma$. Here, $F^{R,C}_\sigma:=\sum_{u,v}\delta_{u,v}F^{R,C;u,v}_\sigma$, and $F^{R,C;u,v}$ is detailed in Appendix~B.8 of \cite{bombin:2008}. 
To be more precise,
\begin{align}
H=-\sum_s A_s-\sum_s B_s,\quad 
    A^G_s=\frac{1}{|G|}\sum_{h\in G}F^{h,e}_{\alpha_s}= \frac{1}{|G|} \sum_{C,R}F^{R,C},
    \quad 
    B^e_s=F_{\beta_s}^{e,e}=\sum_{R\in Irr(G)} F^{R,[e]}
\end{align}
where in the equation for $A^G_s$, $R$ is summed over all irreducible representations of the centeralizer $Z(r_C)$ and $C$ is summed over all conjugacy classes of $G$. In the equation for $B^e_s$, we sum  over $R$ for all irreducible representations of $G=Z(e)$, and $[e]$ is the conjugacy classes of $e\in G$. If  a duality exists between two anyons $a_1=(C_1,R_1)$ and $a_2=(C_2,R_C)$,  one may interchange the two ribbon operators $F^{a_1}_\sigma$ and $F^{a_2}_\sigma$ in the Hamiltonian to reflect this duality, and the resulted Hamiltonian might not go back to the form of $-\sum_sA_s-\sum_s B_s$.  We will see a specific example  of this phenomenon in Eq.~\ref{sect_Floquet_code_DZ2Z2}.  
%------
\subsection{Anyon tunneling map}\label{subsect_anyon_tunneling}
\begin{figure*}[htb]
    \centering
    \begin{tikzpicture}
    \def\d{1.4}
    \def\h{4}
    \def\w{8}
    \def\ar{0.3}
    \def\arn{0.2}
    \def\del{0.01}
    %-----
    %-------------------draw subphases-----
    \fill[fill=gray, fill opacity=0.5] (0,0) rectangle (2*\d,\h*\d);
    \fill[fill=gray, fill opacity=0.5] (6*\d,0) rectangle (\w*\d,\h*\d);
    % Draw the horizontal lines
    \foreach \y in {0,...,\h} {
        \draw (0,\d*\y) -- (\w*\d,\d*\y);
    }
    % Draw the vertical lines
    \foreach \x in {0,...,\w} {
        \draw (\d*\x,0) -- (\d*\x,\h*\d);
    }
    %------
    \draw[dashed] (0,0.5*\d) -- (\w*\d,0.5*\d);
    \draw[dashed] (0,1.5*\d) -- (\w*\d,1.5*\d);
    \draw[dashed] (0,2.5*\d) -- (\w*\d,2.5*\d);
    \draw[dashed] (0,3.5*\d) -- (\w*\d,3.5*\d);
    %\draw[dashed] (0,4.5*\d) -- (\w*\d,4.5*\d);
    %-------
    \draw[dashed] (0.5*\d,0) -- (0.5*\d,\h*\d);
    \draw[dashed] (1.5*\d,0) -- (1.5*\d,\h*\d);
    \draw[dashed] (2.5*\d,0) -- (2.5*\d,\h*\d);
    \draw[dashed] (3.5*\d,0) -- (3.5*\d,\h*\d);
    \draw[dashed] (4.5*\d,0) -- (4.5*\d,\h*\d);
    \draw[dashed] (5.5*\d,0) -- (5.5*\d,\h*\d);
    \draw[dashed] (6.5*\d,0) -- (6.5*\d,\h*\d);
    \draw[dashed] (7.5*\d,0) -- (7.5*\d,\h*\d);
    %-------draw orientations-----
     \draw[midarrow](\ar*\d-\del,0)-- (\ar*\d+\del,0);
     \draw[midarrow](0,\arn*\d+\del)-- (0,\arn*\d-\del);
     %----
    \draw[midarrow](\ar*\d-\del,0.5*\d)-- (\ar*\d+\del,0.5*\d);
    \draw[midarrow](0.5*\d,\ar*\d-\del)-- (0.5*\d,\ar*\d+\del);
    %----draw ribbons----
    \filldraw[fill=BlueGreen!50,fill opacity=0.8] (2.5*\d ,1.5*\d) -- (5.5*\d,1.5*\d) -- (6*\d,2*\d)-- (2*\d ,2*\d) -- cycle;
    %\filldraw[fill=blue!30,fill opacity=0.8] (4*\d ,1.5*\d) -- (5.5*\d,1.5*\d) -- (6*\d,2*\d)-- (4.5*\d ,2*\d) -- cycle;
    \filldraw[fill=BurntOrange!30,fill opacity=0.8]  (5.5*\d,1.5*\d) -- (6*\d,2*\d)-- (6.5*\d ,1.5*\d) -- cycle;
    \filldraw[fill=ForestGreen!30,fill opacity=0.8]   (6*\d,2*\d)-- (6.5*\d ,1.5*\d) --(7*\d,2*\d)-- cycle;
    %-------
    %\filldraw[fill=blue!30,fill opacity=0.8] (4*\d ,1.5*\d) -- (2.5*\d,1.5*\d) -- (2*\d,2*\d)-- (4.5*\d ,2*\d) -- cycle;
    \filldraw[fill=BurntOrange!30,fill opacity=0.8]  (2.5*\d,1.5*\d) -- (2*\d,2*\d)-- (1.5*\d ,1.5*\d) -- cycle;
    \filldraw[fill=ForestGreen!30,fill opacity=0.8]   (2*\d,2*\d)-- (1.5*\d ,1.5*\d) --(1*\d,2*\d)-- cycle;
    %---------------
    \node[text=black] at (6.5*\d,1.79*\d) {\large $\tau\1_1$};
    \node[text=black] at (6.*\d,1.7*\d) {\large$\tau_1$};
    %\node[text=black] at (5.*\d,1.75*\d) {\large $\rho_1$};
    \node[text=black] at (4.*\d,1.75*\d) {\large $\rho$};
    \node[text=black] at (1.5*\d,1.79*\d) {\large $\tau\1_2$};
    \node[text=black] at (2.*\d,1.7*\d) {\large$\tau_2$};
    %---------------
    \node[text=black] at (1*\d,3.5*\d) {\Huge $\mathcal{P}_2$};
    \node[text=black] at (4*\d,3.5*\d) {\Huge $\mathcal{P}_0$};
    \node[text=black] at (7.0*\d,3.5*\d) {\Huge $\mathcal{P}_1$};
    %----------
    \filldraw[fill=teal!30,fill opacity=0.8]   (2.5*\d,0.5*\d)-- (3.5*\d ,0.5*\d) --(3.*\d,1*\d)-- cycle;
    \node[text=black] at (2.65*\d,0.8*\d) {\large $\tau$};
    \node[text=black] at (3.*\d,0.7*\d) {\large $\Rightarrow$};
    \filldraw [black] (3.*\d,0.5*\d) circle (0.05*\d);
    \node at (2.8*\d,0.28*\d) {$L_+^h$}; 
    %------------------------------
    \filldraw[fill=teal!30,fill opacity=0.8]   (4*\d,1*\d)-- (5.0*\d ,1*\d) --(4.5*\d,0.5*\d)-- cycle;
    \node[text=black] at (4.85*\d,0.65*\d) {\large$\tau\1$};
    \node[text=black] at (4.5*\d,0.79*\d) {\large$\Rightarrow$};
    \filldraw [black] (4.5*\d,1*\d) circle (0.05*\d);
    \node at (4.72*\d,1.17*\d) {$T_+^g$}; 
    %-----
\end{tikzpicture}
\caption{A patch of the lattice, containing parts of subphases $\mathcal{P}_1=\mathcal{D}(M/N)$, $\mathcal{P}_2=\mathcal{D}(M\1/N\1)$, and their parent phase $\mathcal{P}_0=\mathcal{D}(G)$. In the middle of the figure is a ribbon, denoted as $\gamma:=\tau\1_2\tau_2\rho\tau_1\tau\1_1$, which tunnels through both subphases. On the bottom left, the diagram shows the orientation (or directionality) of both the primary lattice and its dual counterpart. At the bottom center of the illustration, conventions related to the orientation of the dual (and direct) triangles are depicted, with respective labels as $\tau$ and $\tau\1$. The direction of triangles is fixed onto the triangle, regardless of their placement and orientations in the lattice. The two boundary sites/edges of each triangle $\tau$ ($\tau\1$) are directed by the convention $\partial_0\Rightarrow \partial_1$. }\label{fig_the_lattice_and_the_tunneling_ribbon} 
\end{figure*}
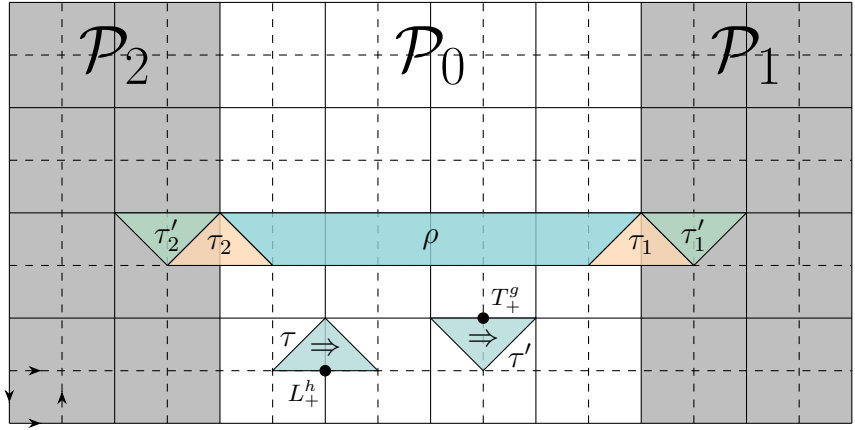

In this section, we delve into the anyon map of tunnelings through domwan walls.  In general, the structure of the phases (MTCs) of a quantum system can be envisioned as a tree diagram.
The most basic representation involves two subphases nested within a parent phase. This can be visualized as a "Y"-shaped tree diagram, where the two branches represent the two subphases and the stem or root signifies the parent phase. However, this "Y" shape, or level-1 tree, is just a rudimentary example. In more complex scenarios, the tree diagram can span multiple levels and exhibit a far more intricate structure.
At  a given level (say level $n$), there is a parent phase $\mathcal{P}_0$, which corresponds to the root of the subtree. Within this parent phase are several embedded subphases, denoted as $\{\mathcal{P}_i\}$. Within each $\mathcal{P}_i$, further sub-subphases can be found, making $\mathcal{P}_i$ the root of level $n+1$ on the phase tree diagram. For simplicity, we will dconsider just two subphases within the parent phase $\mathcal{P}_0$. This setup is illustrated in Fig.~\ref{fig_the_lattice_and_the_tunneling_ribbon}.
%---------
Consider the parent phase $\mathcal{P}_0$ as $\mathcal{D}(G)$. Let's take two subphases: $\mathcal{P}_1$ given by $\mathcal{D}(M/N)$ and $\mathcal{P}_2$ by $\mathcal{D}(M\1/N\1)$. Here, $M$ and $M\1$ are subgroups of $G$, and $N$ and $N\1$ are normal subgroups of $M$ and $M\1$, repsectively. The Hamiltonian for each subphase $\mathcal{P}_i$ is specified by Eq.~\ref{H_GMN}. The  Hamiltonian  of the entire system is 
\begin{align}\label{whole_hamiltonian_on_the_surface}
H=H^{M\1,N\1}(\mathcal{P}_2)+H^{G,e}(\mathcal{P}_0)+H^{M,N}(\mathcal{P}_1)+H(\partial\mathcal{P}_1,M)+H(\partial\mathcal{P}_2,M\1)
\end{align} 
where the  Hamiltonian for a domain wall between the parent phase $\mathcal{P}_0$ and a subphase $\mathcal{P}_i$ is 
\begin{align}
    H(\partial\mathcal{P}_i,M)=- \sum_{v\in \partial \mathcal{P}_i} A_v^M -\sum_{e\in \partial \mathcal{P}_i} T_e^M.
\end{align}
This construction realizes a commuting projector model. To see this, it suffices to examine the commutation relation between $A^M_v$ and $T^M_e$ on the domain wall, in relation to their neighboring terms $A^G_{v\1}$ and $B_p^e$ within $\mathcal{P}_0$.

The term $A^M_v$ commutes with $A^G_{v\1}$ since group left multiplications inherently commute with right multiplications. Additionally, $A^M_v$ commutes with $B_p^e$, which can be perceived as a specific instance of $[A_v^M,B_p^{N={e}}]=0$ when considering the trivial normal subgroup.

Further, the operator $T^M_e$ commutes with $B_p^e$ because both operators are diagonal in the basis characterized by $\{\ket{g}:g\in G\}$. Notably, the term $T^M_e$, for edges $e$ on the boundary of $\mathcal{P}_i$, does not intersect with any $A^G_v$ terms present in $\mathcal{P}_0$. Conclusively, the model presented indeed represents a commuting projector.

The ground state(s) (the degeneracy of the ground state depends on the topology of the surface) for this Hamiltonian is defined as the state as the common eigenvector of all the following operators
\begin{align}\label{gorund_state_defined}
\ket{\Omega}=A_v(\mathcal{P}_0)\ket{\Omega}=B_p(\mathcal{P}_0)\ket{\Omega}=A_v^M(\mathcal{P}_1, \partial \mathcal{P}_1)\ket{\Omega}=B_p^N(\mathcal{P}_1)\ket{\Omega}=T^M_e(\mathcal{P}_1,\partial\mathcal{P}_1)\ket{\Omega}=L^N(\mathcal{P}_1)\ket{\Omega}.
\end{align}
%------------
%------------
For a ribbon operator $F_{\rho\tau_1}$ to commute with $B_{\partial_1\rho}=B_{s_1}$, it is both necessary and sufficient that 
\begin{align}
    F_{\rho\tau_1}^{h,g}=F^{h,g}_\rho L^{g^{-1}hg}_{\tau_1}, \quad F_{\tau_2\rho}^{h,g}=L_{\tau_2}^{h} F^{h,g}_\rho
\end{align}
(as stated in Proposition 7(b) of \cite{bombin:2008}).
We now extend our discussion by appending the ribbon operator with $\tau^\prime$. At this juncture, the single-edge operator $L^N$ becomes pivotal. Within the regime of no confinement, $L^N$ demands an equal superposition across the elements of group $N$, denoted as $\ket{N}:=\sum_{n\in N}\ket{n}/\sqrt{|N|}$.
%-------
\begin{lemma}\label{lemma_quotient_by_Ncap}
    Let $N_\cap :=N\cap N\1$. Then $N/N_\cap$ and $N\1/N_\cap$ are normal in $G/N_\cap$.
%    \begin{proof}    Let $gN_\cap\in G/N_\cap$ and $nN_\cap\in N/N_\cap$, then $(gN_\cap)^{-1}(nN_\cap)(gN_\cap)=g^{-1}ngN_\cap \in N/N_\cap$. Hence $N/N_\cap$ is normal in $G/N_\cap$, and a similar statement holds for $N\1$. \end{proof}
\end{lemma}
In both subphases, as they quotient out the normal subgroup $N_\cap$, the local degrees of freedom that describe the anyons perceive $N_\cap$ as the group unity in the eigenspace of the operators $L^N_e$ and $L^{N\1}_e$. Formally, in these subphases, the ribbon operator 
  adheres to the equivalence relation:
\begin{align}
    F^{h, g}_\rho= F^{hN_\cap, gN_\cap}
\end{align}
Therefore, without loss of generality, we can safely assume
\begin{align}
    N_\cap\equiv N\cap N\1=\{e\}
\end{align}
Should $N_\cap$ turn out to be nontrivial in practice, we can readily adopt $G/N_\cap$, $N/N_\cap$, $N\1/N_\cap$, $M/N_\cap$, and $M\1/N_\cap$ as our input group data.
%---------

Before we continue to construct ribbon operators of the anyon-tunneling process, we present another necessary lemma. This lemma emerges as a direct corollary to Lemma~5 found in \cite{bombin:2008}.
\begin{lemma}\label{lemma_OA_OL}
    Define  algebra for dual triangle $\tau$ and direct triangle $\tau\1$
    \begin{align}
        \mathcal{A}_{\tau}:=Span \{L^h_\tau:h\in G\},\quad 
        \mathcal{A}_{\tau\1}:=Span \{T^g_{\tau\1}:g\in G\},
    \end{align}
and define $\mathcal{A}_\rho:=\otimes_i \mathcal{A}_{t_i}$ with $\rho$ a sequence of direct triangles or dual triangles denoted by $t_i$. 
Let $\rho$ be an open ribbon on the lattice. Let $s=(v,f)$ be a site and $\tau$ a dual triangle. Let $O\in \mathcal{A}_\rho$, define
\begin{align}
    O_k:=A_v^k OA_v^{k^{-1}},\quad O\1_k:= L_\tau^k O_\tau^{\overline{k}}
\end{align}
The lemma goes as follows:
    For subgroup $H\subset G$,
    \begin{align}
    &    [O,A_v^H]=0 \quad \Leftrightarrow \quad O=\frac{1}{|H|} \sum_{k\in H} O_k\\
    & [O,L_\tau^H]=0 \quad \Leftrightarrow \quad   O=\frac{1}{|H|} \sum_{k\in H} O\1_k
    \end{align}
\end{lemma}
Define $\xi_1:=\rho\tau_1\tau\1_1$ and $\xi_2:=\tau\1_2\tau_2\rho$, as depicted in Fig.~\ref{fig_the_lattice_and_the_tunneling_ribbon}. Given that $\mathcal{F}_{\xi_i}\subset \mathcal{A}_{\xi_i}$ for $i=1$ and $2$ and using Lemma~\ref{lemma_OA_OL}, operators in $\mathcal{F}_{\xi_1}$ ($\mathcal{F}_{\xi_2}$) that commute with $A^M$ and $L^N$ (or $A^{M\1}$ and $L^{N\1}$ respectively) can be represented in the following general form:
\begin{align}
  &G_{\xi_1}=\sum_{m\in M} \sum_{n\in N} c_{h_1,g_1,g_2} L^{n}_{\tau\1_1}(A^m_{\partial_1\tau_1}G^{h,g}_{\xi_1}A^{\overline{m}}_{\partial_1\tau_1})L^{\overline{n}}_{\tau\1_1}
  %=\sum_{m\in M}\sum_{n\in N}\sum_{h_1,g_1,g_2} c_{h_1,g_1,g_2}F^{h_1,g_1 m^{-1}}_{\rho\tau_1}T^{mng_2}_{\tau\1}\nnb\\
  =\sum_{m\in M}\sum_{h_1,g_1,g_2}c_{h_1,g_1,g_2} F^{h_1,g_1 m^{-1}}_{\rho\tau_1}T^{mg_2N}_{\tau\1_1}   \label{G_rhotautaup}\\
  &G_{\xi_2}= \sum_{m\in M\1}\sum_{n\1\in N\1} L_{\tau\1_{2}}^n (A^{m\1}_{\partial_0\tau_2} G_{\tau\1_{2}\tau_{2}\rho} A^{\overline{m\1}}_{\partial_0\tau_2}) L^{\overline{n\1}}_{\tau\1_{2}}=\sum_{h_1,g_1,g_2}f_{h_1,g_1,g_2}
  \sum_{m\1\in M\1}  T^{N\1 g_2 \overline{m\1}}_{\tau\1_{2}} F^{m\1 h_1 \overline{m\1}, m\1 g_1}_{\tau_{2}\rho}  \label{G_taupm1taum1rho}
\end{align}
It's not obvious to see if further constraints are needed on the three parameters $h,g_1, g_2\in G$.  
However, if one defines the inner product for the algebra ribbons by 
\begin{align}
\langle O_1 | O_2 \rangle:=\bra{\Omega}O_1^\dagger O_2\ket{\Omega}
\end{align}
for ground state $\ket{\Omega}$. It turns out that 
\begin{align}
    &\langle G^{h,k_1,k_2}_{\xi_1}| G^{h\1,g_1,g_2}_{\xi_1}\rangle 
\propto \delta_{k_1k_2N,g_1g_2N}\\
    &\langle G^{h,k_1,k_2}_{\xi_2}|
G^{h\1,g_1,g_2}_{\xi_2}\rangle
\propto \delta_{N\1k_2k_1,N\1 g_2g_1}
\end{align}
For this reason, we remove one effective variable from $\{g_1,g_2\}$ for each ribbon. There are more than one way to achieve this. As subsequent evaluations will reveal, summing over the cosets $G/M$ ($M\1\backslash G$, repsectively) yields an orthonormal basis for the algebra.
\begin{align}
G^{h,g}_{\xi_1}:=& \sum_{g_1\in G/M} G^{h,g_1,\overline{g_1}g}_{\xi_1}
%= \sum_{g_1\in G/M} \sum_{m\in M}  F^{h, g_1 m^{-1}}_{\rho\tau} T^{m \overline{g_1} g N}=\sum_{k\in G} F^{h, k} T^{\overline{k} g N}
=\sum_{n\in N}F^{h,k n}_{\xi_1}.\\
G^{h,g}_{\xi_2}:=& 
\sum_{g_2\in M\1\backslash G} G^{\overline{g_2}hg_2, \overline{g_2} g, g_2} 
%=\sum_{n\1\in N\1}\sum_k T^{\overline{n\1} k}_{\tau\1_{2}} F^{\overline{k} h k,\overline{k}g_1}_{\tau_{2}\rho}
%=\sum_{n\1\in N\1}\sum_k T^{ k}_{\tau\1_{2}} F^{\overline{ k}( \overline{n\1}  hn\1) k,\overline{k}(\overline{n\1}g)}_{\tau_{2}\rho}
=\sum_{n\1\in N\1} F^{\overline{n\1}h n\1,\overline{n\1}g}_{\xi_2}
\end{align}
We glue the two ribbons in the phase of $\mathcal{D}(G)$. To this end, define $\gamma:=\xi_2 \cup \xi_1=\tau_2\1 \tau_2 \rho\tau_1\tau_1\1$.
\begin{align}
  % G^{h,g}_\gamma=&\sum_{k\in G} G_{\xi_2}^{h,k} G^{\overline{k}h k ,\overline{k} g}_{\xi_1}.\\
    G^{h,g}_\gamma=&\sum_{k_1\in G} \sum_{k_2\in G}T^{N\1 k_1} F^{\overline{k_1}hk_1, \overline{k_1}k_2}  T^{\overline{k}_2 gN}
    =\sum_{n\in N}\sum_{n\1\in N\1} F^{\overline{n\1} hn\1 ,\overline{n\1}gn}_\gamma
\end{align}
Therefore the algebra of anyon-tunneling is the restriction of 
$ \mathcal{F}_\gamma$
to 
\begin{align}    \mathcal{G}_\gamma:=Span\qty{G^{h,g}_\gamma:(h,g)\in S},
\end{align}
where $S:=\{(h,g):h\in M\1\cap gM\overline{g},g\in G\}$.
The restriction of $(h,g)\in S$, or equivalenlty $\overline{g}hg\in M$ and $h\in M\1$, comes from the requirement that  $[T^M_{\tau_1},G_\gamma^{h,g}]=[T^M_{\tau_2},G_\gamma^{h,g}]=0$.

Before delving into the algebraic properties of these operators, it's pertinent to discuss the equivalence relations stemming from the two normal subgroups  $N$ and $N\1$.
The pair $(h,g)$ is subject to the following equivalence defined as
\begin{align}
(h,g)\sim_{N,N\1}(\overline{n\1} h n\1,\overline{n\1}gn),\quad \forall n\in N, \ n\1\in N\1.
\end{align}
When constructing a set of representatives $Q(\sim_{N,N\1})$ for the equivalence relation, we first consider $Q(NN\1)$, which is a representative set of the double coset $N\1\backslash G/N\equiv G/(NN\1)$. With a fixed $g\in Q(NN\1)$, the orbit of $(h,g)$ under the action $(h,g)\mapsto (\overline{n\1} hn\1, \overline{n\1} gn)$ for every $n\in N$ and $n\1\in N$ is uniquely determined by its "starting point" $h$ of the orbit. Alternatively, we can use the term $\overline{g}hg$ as the starting point since $g$ is fixed on this orbit. The distinction between opting for $h$ or $\overline{g}hg$ becomes significant when proving Theorem~\ref{theorem_chi_of_phase_tunneling_map}.) Consequently, the set of representatives for the equivalence relation is:
\begin{align}
Q(\sim_{N,N\1})=\{(h,g):g\in Q(NN\1), h\in G\}.
\end{align}
It's useful to symbolize the representatives of relation $\sim_{N\1,N}$ restricted to the above subset $S$  as $Q_S(\sim_{N\1,N})$. 
We use $[(h,g)]_{N,N\1}$ to
represent the equivalence class (or orbit) of $(h,g)$ under the relation $\sim_{N,N\1}$. 
This relation corresponds to the identity:
\begin{align}
G^{h,g}_\gamma= G_\gamma^{\overline{n\1}hn\1, \overline{n\1}gn},\quad \forall n\in N,\quad n\1\in N\1.
\end{align}
%----
An  alternative basis that works equally well can be established by the following definition:
\begin{align}
    H^{h,g}_\gamma:=G^{gh\overline{g},g}= \sum_{n,n\1} F^{(n\1 g) h (\overline{g}  \overline{n\1}), n\1 gn}_\gamma,
\end{align} 
and the inverse is $G^{h,g}=H^{\overline{g}hg,g}$.  It has the property
\begin{align}
H^{h,g}=&    H^{nh\overline{n},n\1g\overline{n}},\quad \forall n\in N,n\1\in N\1,
\end{align}
corresponding to the equivalence class
$(h,g)\sim_{N,N\1} (nh\overline{n}, n\1 g \overline{n})$. 
%-------
Using (B41) and (B42) of \cite{bombin:2008},
\begin{align}
%calculation is hidden in the comment
&A_{\partial_0\gamma}^{\ell N\1}G^{h,g}_\gamma
=G^{\ell h \overline{\ell }, \ell g}_\gamma A^{\ell N\1}_{\partial_0 \gamma}\quad \quad B_{\partial_0\gamma}^{\ell N\1} G^{h,g}_\gamma
=G^{h,g}_\gamma B_{\partial_0\gamma}^{\ell h N\1}  \label{eq_AGBG_left}\\
&A_{\partial_1\gamma}^{\ell N}G^{h,g}_\gamma
=G^{ h ,  g\overline{\ell}}_\gamma A^{\ell N}_{\partial_1 \gamma}\quad \quad \ \ 
B_{\partial_1\gamma}^{\ell N} G^{h,g}_\gamma 
=G^{h,g}_\gamma B_{\partial_1\gamma}^{\overline{g}\overline{h}g\ell  N } \label{eq_AGBG_right}
\end{align}
Utilizing the $H^{h,g}_\gamma$ basis leads to a similar algebraic structure, but with the data $(M/N, \partial_1\gamma)$ and $(M\1/N\1,\partial_0\gamma)$ swapped. Consequently, the transformation $(h,g)\leftrightarrow (\overline{g}hg,g)$ equates to exchanging the left and right endpoints of ribbon operators, or alternatively, reversing the predefined orientations of direct and dual triangles.
Given that the two sets of bases produce the same algebra, we'll stick to the $G^{h,g}_\gamma$ basis for the rest of this section.

%-----------
The ground state $\ket{\Omega}$ defined by Eq.~\ref{gorund_state_defined} has the following properties:
\begin{align}
&A^{\ell N}_{\partial_1\gamma} \ket{\Omega}=\ket{\Omega},\quad B^{\ell N}_{\partial_1\gamma}\ket{\Omega}=\delta_{\ell N, N} \ket{\Omega}\\
    &A^{\ell N\1 }_{\partial_0\gamma}\ket{\Omega}=\ket{\Omega},\quad B^{\ell N\1  }_{\partial_0\gamma}\ket{\Omega}=\delta_{\ell N\1 ,N\1}\ket{\Omega}.
\end{align}
The proof is  straightforward and is omitted here.
%————
%\begin{lemma}\label{lemma_double_coset_equivalence}    $g_1\sim_N g_2$ and $g_1\sim_{N\1}g_2$ iff  $N\1  g_1 N=N\1 g_2 N$. \end{lemma}
%--------
\begin{proposition}\label{proposition_list_of_AgN_BgN_properties}
Here is the list of properties of the algebra of $A^{hN}$ and $B^{gN}$.
\begin{itemize}
\item[1.] $A^{gN}A^{g\1 N}=A^{gg\1 N}$.
\item[2.]  $B^{gN}B^{g\1 N}=\delta_{gN,g\1 N}B^{gN}$.
\item[3.] $B^{gN}A^{hN}=A^{hN}B^{h^{-1}ghN}$. 
\item[4.] The algebra given by multiplications of terms $A^{hN}B^{gN}$ on the same site is isomorphic to the algebra of $\mathcal{D}(M/N)$. Namely 
\begin{align} 
(A^{h_1 N}B^{g_1N})(A^{h_2 N}B^{g_2N})=\delta_{g_2 N,  \ h_2^{-1} g_1 h_2N}\  A^{h_1 h_2 N} B^{g_2 N}.
\end{align}
Replacing $M\to M\1$ and $N\to N\1$, we get similar algebra for $\mathcal{D}(M/N)$.  Taking $N=\{e\}$ and $M=G$, this reduces back to the usual expression for $\mathcal{D}(G)$.
\end{itemize}
\end{proposition}
The proof is in Remark.~\ref{proof_of_proposition_list_of_AgN_BgN_properties}.
%----------

Define 
$\ket{G^{h,g}}:=G_{\gamma}^{h,g}\ket{\Omega}.$
Employing the commutation relationships from Eq.\ref{eq_AGBG_left} and Eq.\ref{eq_AGBG_right}, the resulting operations from the quantum double operators can be determined as:
%---------
\begin{align}\label{AlN_RHS}
(A^{h_1 N}B^{g_1 N})_{\partial_1\gamma}\ket{G^{h,g}}=&\delta_{g_1 \overline{g} hg N, N}  \ket{G^{h,g\overline{h_1}}}\\
(A^{h_2 N\1} B^{g_2 N\1})_{\partial_0\gamma }\ket{G^{h,g}}=&\delta_{\overline{h}g_2 N\1, N\1} \ket{G^{h_2 h\overline{h_2},h_2 g}}
\end{align}
It's straightforward to see the actions satisfy the closure of the algebra and the compatibility condition.
\begin{proposition}\label{proposition_properties_of_G_gamma}
    The generator $F_{\xi}$ of algebra $\mathcal{D}_\xi$ has the following properties.
    \begin{itemize}
        \item[1.]     (Closure of the algebra $\mathcal{G}_\gamma$). $\mathcal{G}_\gamma$ holds.  For pairs $(h,g),\ (h\1,g\1)\in Q_S(\sim_{N\1,N})$,
        \begin{align}
            G_\gamma^{h,g}G_\gamma^{h\1,g\1N}=\delta_{N\1 g_1N,N\1  g_2N } G^{hh\1, g}_\gamma.
        \end{align}
        \item[2.] (Hermitian conjugate). $(F^{h,gN}_\xi)^\dagger=F^{h^{-1},g N}_\xi$ for any $g\in G$ and $h\in gM\overline{g}$.
    \end{itemize}
\end{proposition}
The proof is given by Remark~\ref{proof_of_proposition_properties_of_G_gamma} in the appendix.
\begin{lemma}\label{lemma_expval_of_G}
%Recall the equivalence class $(h,g)\sim_{N,N\1} (\overline{n\1} hn\1,\overline{n\1} gn)$. Let $\pi:G^*\times G\to (G^*\times G)/\sim_{N,N\1}$ be the canonical projection of the equivlence class. Then
$
\langle \Omega | G^{h, g} |\Omega\rangle =\delta_{h,e} C_{M\1 gM}$
for some constant $C_{M\1 gM}$ that depends on the double coset in $M\1\backslash G/M$.
\end{lemma}
The proof is given in Remark.~\ref{proof_of_lemma_expval_of_G}. 
%---
%---------
\begin{proposition}
The set of vectors given by
\begin{align}
   \{ \ket{\psi^{h,g}}:=(C_{M\1 gM})^{-1/2}G^{h,g}_\gamma\ket{\Omega} :(h,g)\in Q(\sim_{N,N\1})\}
\end{align}
constitutes an orthonormal basis for the algebra associated with two-point excitations at the endpoints of $\gamma$. Specifically, these vectors are the eigenstates with eigenvalue 
$+1$ for all terms in the Hamiltonian given by Eq.~\ref{whole_hamiltonian_on_the_surface}, with the exception of  $A^{M\1}_{\partial_0\gamma}$, $B^{N\1}_{\partial_0\gamma}$, 
$A^{M}_{\partial_1\gamma}$ and $B^{N}_{\partial_1\gamma}$. 
\begin{proof} 
Using Proposition~\ref{proposition_properties_of_G_gamma} and Lemma~\ref{lemma_expval_of_G}
%    \begin{align}    \langle \psi^{h_j,g_j}|\psi^{h_i,g_i}\rangle    =&(C_{M\1 g_j M}C_{M\1 g_i M})^{-1/2}\bra{\Omega}(G^{h_j,g_j}_\gamma)^\dagger F^{h_2,g_2}_\gamma\ket{\Omega}    =(C_{M\1 g_j M}C_{M\1 g_i M})^{-1/2} \bra{\Omega}G^{h^{-1}_j,g_j}_\gamma G^{h_i,g_i}_\gamma\ket{\Omega} \nnb\\    =&\delta_{N\1 g_iN ,N\1 g_jN } (C_{M\1 g_j M}C_{M\1 g_i M})^{-1/2}\bra{\Omega}G^{(\overline{n\1}_j h_j^{-1}n\1_j) h_i,g_j}_\gamma\ket{\Omega}    =\delta_{h_j,h_i } \delta_{g_j NN\1,g_i NN\1}\end{align}
%-----------------
\begin{align}
    \langle \psi^{h\1,g\1}|\psi^{h,g}\rangle
    =&(C_{M\1 g\1 M}C_{M\1 g M})^{-1/2}\bra{\Omega}(G^{h\1,g\1}_\gamma)^\dagger G^{h,g}_\gamma\ket{\Omega}
    =(C_{M\1 g\1 M}C_{M\1 g M})^{-1/2} \bra{\Omega}\psi^{(h\1)^{-1},g\1}_\gamma \psi^{h,g}_\gamma\ket{\Omega} \nnb\\
    =&\delta_{g,g\1 } (C_{M\1 g\1 M}C_{M\1 g M})^{-1/2}\bra{\Omega}G^{(h\1)^{-1} h,g}_\gamma\ket{\Omega}
    =\delta_{h,h\1 } \delta_{g,g\1}
\end{align}
%Note that $\delta_{hN,N}\delta_{hN\1,N}=\delta_{N\1 hN, N\1 eN}$ by Lemma~\ref{lemma_double_coset_equivalence}.
\end{proof}
\end{proposition}
\begin{theorem}\label{theorem_chi_of_phase_tunneling_map}
Define $S:=\{(h,g):\overline{g}hg\in M, h\in M\1:g\in G\}$.
Given input $(h_2g_2^*,h_1g_1^*)\in \mathcal{D}(G)\otimes \mathcal{D}(M/N)$, the character $\chi_{\mathcal{G}}$ is
%-----
\begin{align}
    \chi_{\mathcal{G}}(h_2g_2^*,h_1g_1^*)
    =&(|N|\ |N\1|)^{-1}
    \delta_{h_2g_2N\1,g_2h_2N\1 } \delta_{g_1h_1N,h_1g_1N}
\sum_{(h,g)\in S} \delta_{  g_1N,\overline{g}hgN}\delta_{g_2hN\1,N\1}
       \delta_{NgN\1 , N h_2gh_1^{-1} N\1}.
\end{align}
Refer to Remark~\ref{calculation_of_chi_of_phase_tunneling_map} for detailed proof.  For the benefit of the reader, it's worth noting that this formula remains valid even when $N_\cap\equiv N\cap N\1$ is nontrivial, irrespective of whether one applies the quotient of groups by $N_\cap$ as delineated following Lemma~\ref{lemma_quotient_by_Ncap}. This is because, during the proof, we've correspondingly relaxed the condition $\delta_{h,h\1}$
  to be $\delta_{hN_\cap, h\1 N_\cap}$ within the inner product 
$\langle \psi^{h\1,g\1} | \psi^{h,g}\rangle$. Further discussion on this can be found in Remark~\ref{calculation_of_chi_of_phase_tunneling_map}.
\end{theorem}
%---
\begin{corollary}
Let $\sim_{N}$ denote the equivalence relation of two group elements residing in the same coset of $G/N$.     For $M\1=G$ and $N\1=\{e\}$,
\begin{align}
    \chi_{\mathcal{G}}(h_2g_2^*,h_1g_1^*)=|N|^{-1}
    \delta_{h_2g_2,g_2h_2}  \delta_{g_1h_1 N, h_1g_1N} 
    \sum_{g\in G}\delta_{gg_1\overline{g} \sim_N \overline{g_2}  }  \delta_{gh_1\overline{g}\sim_N h_2 } .
\end{align}
Furthermore, if $G$ is Abelian,
\begin{align}
    \chi_{\mathcal{G}}(h_2g_2^*,h_1g_1^*)=\frac{|G|}{|N|}  \delta_{g_1 \sim_N \overline{g_2}  }  \delta_{h_1\sim_N h_2 }.
\end{align}
\end{corollary}
Summing over the inputs $(h_2g_2^*,h_1g_1^*)$ of the character, one can define inner products: 
\begin{align}
\langle   \chi_A |\chi_B\rangle :=\sum_{h_1h_2^*,h_1g_1^*}\chi_A^*(h_1h_2^*, h_1g_1^*) \chi_B(h_1h_2^*, h_1g_1^*),
\end{align}
where $\chi^*$ is the complex-conjugated character. 
With the above theorem in place, one can calculate the {\bfseries anyon-tunneling map}, 
\begin{align}
    \varphi_{a\1, a}=\langle \chi_{a\1}\chi_a |\chi_{\mathcal{G}}\rangle /\langle \chi_{a\1}\chi_a|  \chi_{a\1}\chi_a\rangle, \label{eq_anyon_tunneling_map}
\end{align}
 where $\chi_a$ is the character of the irreducible representation of $\mathcal{D}(M/N)$, while $\chi_{a\1}$ is the one of $\mathcal{D}(M\1/N\1)$.
 
%---------
One might consider generalizing the present approach to use Hopf algebras as the input algebra for the theory. Another avenue for exploration is extending this in the direction of the Levin-Wen model \cite{PhysRevB.71.045110,PhysRevB.103.195155} or the Turaev-Viro code \cite{KOENIG20102707,PhysRevX.12.021012}. An open challenge remains in designing the analogs of the restriction operator $T^M$ and the quotient operator $L^N$, or  in crafting the confinement terms for these models. If we can establish a method to characterize the subphases,  it might be feasible to follow similar steps as outlined in this work  using representation theory.
%----------
\subsection{Example: Quantum double of \texorpdfstring{$S_3$}{Lg}}
Let $G=S_3$ be the permutation group of degree 3.  Label the group elements by $\{e,\tau,\tau^2,\sigma,\sigma\tau, \sigma\tau^2\}$ with $\sigma$ the flipping generator and $\tau$ the rotational generator. There are three nontrivial derived phases by partial condensation. We list the anyon tunneling map below.
\begin{itemize}
    \item $M=\mathbb{Z}_2=\{e,\sigma\}$ and $N=\{e\}$. This is the $A+ C$ condensed phase.  The mapping of anyons in $\mathcal{D}(S_3)$ to the (deconfined) anyons in $\mathcal{D}(M/N)=\mathcal{D}(\mathbb{Z}_2)$ is
    \begin{align}
        A\to \mathbf{1}\quad B\to e\quad  C\to \mathbf{1}+e\quad  D\to m\quad E\to f.
    \end{align}  
   \item Another $M/N=\mathbb{Z}_2$ phase is given by $M=S_3$ and $N=\mathbb{Z}_3=\{e,\tau,\tau^2\}$.  This corresponds to the condensation of $A+F$.  The anyon tunneling map is
   \begin{align}
       A\to \mathbf{1}\quad B\to e\quad D\to m\quad E\to f\quad F\to \mathbf{1}+e.
      \end{align}
    \item The last nontrivial phase is $M/N=\{e,\tau,\tau^2\}/\{e\}=\mathbb{Z}_3$.  The anyons of the $\mathcal{D}(\mathbb{Z}_3)$ phase are labeled by two parameters $a_1,a_2\in \mathbb{Z}_3=\{0,1,2\}$ with $a_1$ labeling the conjugacy class and $a_2$ labeling the representation of $\mathbb{Z}_3$.  Label $a_1=1$ as $m$, $a_1=2$ as $\wt{m}$; similarly, label $a_2=1$ as $e$ and $a_2=2$ as $\wt{e}$. The anyon tunneling map is 
    \begin{align}
        A\to \mathbf{1}\quad B\to \mathbf{1}\quad C\to e+\wt{e}\quad F\to m+\wt{m}\quad G\to m\wt{e}+\wt{m} e\quad H\to em+\wt{e}\wt{m}.
   \end{align}
\end{itemize}
For more examples of anyon types and anyon-tunneling maps, see Appendix~\ref{sec_mapping_between_bulk_and_the_hole}.
%------
%------------
%  Floquet code
%------------
\section{Generalized Floquet code}\label{sect_Floquet_code}
Several authors \cite{Hastings2021dynamically,kesselring:2022} have introduced dynamically generated codes, termed "Floquet codes", which are based on either the honeycomb code or the color code. These can be understood as periodic phases brought about by two-qubit check measurements. Intriguingly, the same set of check measurements can also function to infer the syndrome. In this section, we explore the theoretical generalization of the Floquet code to a general finite group $G$, possibly nonabelian. %While the measurements inducing the phase transition remain low-weight (1-qudit), the syndrome can no longer be deduced from these measurements. Instead, the syndrome must be based on four-qudit measurements, akin to the surface code or the quantum double model. As such, we trade the engineering advantages   inherent to the Floquet code for the theoretical geenrality.

The original Floquet based on the honeycomb lattice can be recast in terms of the Hamiltonian system $H_G^{M,N}$ as previously defined, specifically for $G=\mathbb{Z}_2\times \mathbb{Z}_2$. For varied combinations of $(M, N)$ where $N$ is normal in $G$, we can achieve distinct anyon-condensed phases of the parent topological order $\mathcal{D}(\mathbb{Z}_2\times \mathbb{Z}_2)$.

Let $a$ be the representative of conjugacy class $\overline{a}$ in group $G$. For any $b\in \overline{a}$, fix $k_b\in G$ such that $b=k_bak_b^{-1}$. We define $k_a=e$. Cited from \cite{shor2011},  for representation $(\overline{a},\pi)$ of $\mathcal{D}(G)$, the character is given by 
\begin{align}
    \chi_{(\overline{a},\pi)}(hg^*)=\delta_{g\in \overline{a}}\delta_{gh,hg}\tr_\pi(k_g^{-1} hk_g)
\end{align}
The character of boundary $(M,N)$ for abeliangroup G is given by Eq.~\ref{chi_MN_abelian} in Appendix~\ref{app_condensation_bdry}, which is 
\begin{align}
\chi_{\mathcal{A}_{M,N}}(hg^*)=
\frac{|G|}{|M|} \delta_{g\in N} \delta_{h \in M}
\end{align}
for group $N$ normal in $G$, and $N\subset M$.
By decomposing $\chi_{\mathcal{A}_{M,N}}$, we can read off the condensable algebra for subphase $\mathcal{D}(M/N)$.
%-----------
%----------

In the context of QDM with Hamiltonian Eq.~\ref{H_GMN}, we define a Floquet code to be a measurement-induced phase transition through two sets of measurement $\{T^M,1-T^M\}$ and $\{L^N/|N|, 1-L^N/|N|\}$, such that the logical state $\sum_i\alpha_i\ket{i}$ is preserved after the phase transition.

We will also employ the terminology of anyons and condensations, which primarily revolve around the ground state, though it's entirely feasible to consider situations deviating significantly from the ground state. The potential arbitrariness of stabilizer measurements might elevate the state considerably above the ground state(s) $\ket{\Omega(M,N)}$ of $H^{M,N}$. Nonetheless, by deploying local operators, one can realign local excitations of terms $A_v^M$, $B_p^N$, $T^M_e$, and $L_e^N$ to the $+1$ eigenvalue, given their nature as commuting projectors. Therefore, for the sake of discussion, it's both sufficient and justifiable to focus on stabilizer codes, phases, and excitations in the vicinity of ground states.
%-------
%Proposition~\ref{qubit_conditional_unitary_sufficient_condition} can be easily generated to qudits.
%-----
\begin{proposition}\label{proposition_conditional_unitary_iff_condition}
Let $M/N$ and $M\1/N\1$ be two isomorphic quotient groups. Let $d=|M/N|=|M\1/N\1|$. The conditional unitary exists between the two sets of effective qudits $\mathbb{C}[M/N]$ and $\mathbb{C}[M\1/N\1]$ iff  the $d$-dimensional integer matrix $M_{ji}\equiv |g\1_j N\1 \cap g_iN|$ statisfies $M^TM\propto \text{id}$.

The proof is in Remark~\ref{proof_of_proposition_conditional_unitary_iff_condition}.
\end{proposition}
\begin{proposition}\label{proposition_sufficient_condition_for_logical_state_preservation}
    It's sufficient to have logical state preserved if the measurement operators amounts to conditional unitary quditwise.

    The proof is in Remark~\ref{proof_of_proposition_sufficient_condition_for_logical_state_preservation}.
\end{proposition}
%————
To give a concrete example, and to make the proposition above more convenient, we look at the case of qubit. Namely $M/N\cong M\1/N\1\cong\mathbb{Z}_2$.
\begin{proposition}\label{qubit_conditional_unitary_sufficient_condition}
Let the effective qubit be $\ket{0}\equiv \ket{N}=|N|^{-1/2}\sum_{n\in N}\ket{n}$, and $\ket{1}\equiv \ket{\sigma N}=|N|^{-1/2}\sum_{n\in N}\ket{\sigma n}$ for some representative $\sigma$ of coset $M/N$. We also define $\ket{0\1}$ and $\ket{1\1}$ through similar expressions. By construction, $M=N\cup\sigma N$ and $M\1 =N\1 \cup \sigma\1 N\1$.
    It's sufficient to have conditional unitary if 
%    \begin{align}        \langle 1\1 | 0\rangle=\langle 0\1 |1\rangle=0,\quad  \langle 0|0\1\rangle \neq 0 ,\quad  \langle 1|1\1\rangle \neq 0. \end{align}    Or equivalently,
    \begin{align}
& \sigma\1 N\1\cap N=N\1\cap \sigma N=\emptyset, ,\quad  |N\1 \cap N|=|\sigma\1 N\1 \cap \sigma N|\neq 0.
    \end{align}
This is simply a special case of Proposition~\ref{proposition_conditional_unitary_iff_condition} where
\begin{align}
    M=|N\1\cap N| \ \mqty(1& 0 \\ 0& 1).
\end{align}
\end{proposition}
    In fact,  given  $N\1 \cap N\neq \emptyset$ and $\sigma\1 N\1\cap \sigma N\neq \emptyset$,  $|N\1 \cap N|=|\sigma\1 N\1\cap \sigma N|$ is guaranteed. To see this, notice that $N\cap N\1$ is normal in $M\cap M\1$.  Since $\sigma\1 N\cap \sigma\1 N\neq \emptyset$, we can choose $\sigma\1 =\sigma$. Therefore, $|N\cap N\1|=|\sigma (N\cap N\1)|=|\sigma N\cap \sigma N\1|$.
%------------- 
\subsection{Example: The Floquet code in \texorpdfstring{$\mathcal{D}(\mathbb{Z}_2\times\mathbb{Z}_2)$}{Lg} and its implications to general Floquet code}\label{sect_Floquet_code_DZ2Z2}
As detailed in \cite{kesselring:2022}, the anyons in the color code can be equated with two layers of the toric code, as per the subsequent identification:
\begin{align}
\begin{bmatrix}
 {\color{red}rx} & {\color{teal}gx} &  {\color{blue}bx}  \\ 
 {\color{red}ry} & {\color{teal}gy} &  {\color{blue}by}  \\ 
 {\color{red}rz} & {\color{teal}gz} &  {\color{blue}bz} 
\end{bmatrix}
\mapsto 
\begin{bmatrix}
 {\color{red}e1} & {\color{teal}ee} &  {\color{blue}1e}  \\ 
 {\color{red}em} & {\color{teal}ff} &  {\color{blue}me}  \\ 
 {\color{red}1m} & {\color{teal}mm} &  {\color{blue}m1} 
\end{bmatrix}\label{color_code_isomorphism}
\end{align}
All of these entries are bosons, meaning that the corresponding modular $T$-matrix entry is given by $T_{ii}=e^{i2\pi s_i}=1$. Eq.~\ref{color_code_isomorphism} actually serves as an isomorphic functor from the category of the color code—implementable on the hexagonal lattice—to the category $\mathcal{D}(\mathbb{Z}_2\times \mathbb{Z}_2)$, which can be realized on any lattice.

We will now reformulate the toric code without resorting to the honeycomb lattice. 
Take $M=G=\mathbb{Z}_2\times \widetilde{\ZZ}_2$ and $N=\ZZ_2^d=\{(1,1),(-1,-1)\}$, we have 
\begin{align}\chi_{\mathcal{A}_{M,N}}(hg^*)=\delta_{g\in N}=\sum_{a\in N}\chi_{\overline{a},1}=\chi_{\mathbf{1}}+\chi_{m_1m_2}
\end{align}
where $\mathbf{1}\in Irrep(G)$ denotes the trivial representation, and $\mathbf{1}\in \mathcal{TC}$ represents the vacuum excitation in the toric code. As a result, the subphase characterized by $(M=\ZZ_2\times \wt{\ZZ}_2,N=\ZZ_2^d)$ aligns with the condensable algebra $\mathbf{1}+ m_1m_2$. We can delineate other scenarios in a similar manner. Specifically, the Floquet code schedule presented in \cite{kesselring:2022} aligns with the subsequent choices of subphases.
\begin{table}[htb]
\centering
\begin{tabular}{|c|c|c|c|c|c|c|} \hline
$\mathcal{A}_{M,N}$ & $\mathbf{1}+ e$ & $\mathbf{1}+ m\wt{m}$ & $\mathbf{1}+ \wt{e}$ & $\mathbf{1}+ \wt{m}$ & $\mathbf{1}+ e\wt{e}$ & $\mathbf{1}+ m$ \\ \hline
$M$ & $\widetilde{\mathbb{Z}}_2$ & $\mathbb{Z}_2\times \widetilde{\mathbb{Z}}_2$ & $\mathbb{Z}_2$ & $\mathbb{Z}_2\times \widetilde{\mathbb{Z}}_2$ & $\mathbb{Z}_2^d$ & $\mathbb{Z}_2\times \widetilde{\mathbb{Z}}_2$ \\ \hline
$N$ & $\{e\}$ & $\mathbb{Z}_2^d$ & $\{e\}$ & $\widetilde{\mathbb{Z}}_2$ & $\{e\}$ & $\mathbb{Z}_2$ \\ \hline
\end{tabular}
\caption{Table of derived phases $\mathcal{D}(M/N)$ with the corresponding condensable algebra (see Appendix~\ref{app_condensation_bdry}).}
\end{table}
Firstly, we demonstrate that this physical system can emulate the Floquet code based on the honeycomb or hexagonal lattice as presented in \cite{Hastings2021dynamically,kesselring:2022}. The non-condensed original Hamiltonian contains terms, where $G=\ZZ\times \wt{\ZZ}_2$.
\begin{align}
A^{G}_v=&\frac{1}{|G|}\sum_{g\in \ZZ\times \wt{\ZZ}_2}L^g=\frac{1}{4}(1+\prod_{v\in e} X_e+\prod_{v\in e}\wt{X}_e+\prod_{v\in e} X\wt{X}_e)\sim \prod X_e+\prod\wt{X}_e \\
B_p^e=&\sum_{{g_e}}\prod_{v\in e} \ket{g_e}\bra{g_e}\delta(\prod_{v\in e}g_e,e)\sim (1+\prod_{v\in e} Z_e)(1+\prod_{v\in e} \wt{Z}_e)\sim \prod_{v\in e}Z+\prod_{v\in e}\wt{Z},
\end{align}
where we introduce the equivalence relation, denoted as $\sim$, through factors of multiplication, constant differences, or inference of eigenvalues. As an illustration, the eigenvalues of $Z^{\otimes 4}$ and $\wt{Z}^{\otimes 4}$ can be employed to deduce the eigenvalue of $(Z\wt{Z})^{\otimes 4}$. Consequently, this configuration represents two copies of the toric code. Taking the initial phase as a representative example, if we incorporate measurements  $T^M_e=T_e^{\wt{Z}_2}$ on edge $e$, we {\bfseries disturb} the original term $A^G$ in the parent phase (specifically, $T^M$ does not commute with $A^G$). We substitute it with $A^M\sim \prod_{v\in e}\wt{X}e$. From a physical perspective, $T^M$ constrains the degrees of freedom from $\ket{g\in G}$ to $\ket{g\in M}$. Subsequently, the term $\prod_{v\in e}X_e$, which corresponds to an excitation site of anyon $e$, is removed from the Hamiltonian. We thus characterize this as the $\mathbf{1}+ e$-condensed phase. 
%-----
\begin{table}[htb]
\centering
\scalebox{1.0}{
\begin{tabular}{|c|c|c|c|c|c|c|c|c|c|c|} \hline 
color checks &condens. & $M$&  $N $ & meas. & $A^M$  &  $B^N$& $T^M$   & $L^N$  &logical $e$  & logical $m$ \\ \hline
{\color{red}rxx}&$1+e$ & $\widetilde{\mathbb{Z}}_2$   & $\{e\}$ &$Z$   &  $\frac{1}{2}(1+ \widetilde{X}^{\otimes 4}) $ & $\frac{1}{4}(1+Z^{\otimes 4})(1+\widetilde{Z}^{\otimes 4})$ & $Z$ & $id$ & $Z\widetilde{Z}$, $\widetilde{Z}$ & $Z\widetilde{X}$ , $\widetilde{X}$ \\  \hline
%------
 & & & &  $X\widetilde{X}$ & &  disturb& disturb &$X\widetilde{X}$  & &   \\
 &  & & &  $X\widetilde{X}$ & infer   & & &$X\widetilde{X}$   &  & \\
{\color{teal}gzz} &$1+m\wt{m}$ & $G$ & $\mathbb{Z}_2^d$ & $X\widetilde{X}$ & $\frac{1}{4}(1+ X^{\otimes 4})(1+\widetilde{X}^{\otimes 4})$ & $\frac{1}{2}(1+Z^{\otimes 4}\widetilde{Z}^{\otimes 4})$ & $id$ & $X\widetilde{X}$ &$Z\widetilde{Z}$, $Y\widetilde{Y}$& $\widetilde{X}$, $X$ \\ \hline
%-----------
 &  & & &  $\widetilde{Z}$ &disturb  & &  $\widetilde{Z}$  &  disturb & & \\
 & & & &  $\widetilde{Z}$ &  &  infer &  $\widetilde{Z}$  &  & & \\
{\color{blue}bxx} &$1+\wt{e}$ & $\mathbb{Z}_2$ & $\{e\}$& $\widetilde{Z}$ & $\frac{1}{2}(1+X^{\otimes 4})$ &$\frac{1}{4}(1+Z^{\otimes 4})(1+\widetilde{Z}^{\otimes 4})$ & $\widetilde{Z}$& $id$ & $Z$, $Z\widetilde{Z}$ & $X\widetilde{Z}$, $X$\\  \hline
%----------------
& & & &  $\widetilde{X}$ & &  disturb& disturb &  $\widetilde{X}$  & & \\
& & & &  $\widetilde{X}$ & infer & & &  $\widetilde{X}$  & & \\
{\color{red}rzz} &$1+ \wt{m}$ & $G$ & $\widetilde{\mathbb{Z}}_2$ &$\widetilde{X}$ &  $\frac{1}{4}(1+X^{\otimes 4})(1+\widetilde{X}^{\otimes 4})$& $ \frac{1}{2} (1+Z^{\otimes 4})$ &$id$ &  $\widetilde{X}$ & $Z$, $Z\widetilde{X}$ & $X\widetilde{X}$, $X$ \\  \hline
%------------------------
& & & &  $Z\widetilde{Z}$ &disturb  & &  $Z\widetilde{Z}$  & disturb & & \\
 & & & &  $Z\widetilde{Z}$ & &  infer &  $Z\widetilde{Z}$  & & &\\
{\color{teal}gxx} &$1+e\wt{e}$&  $\mathbb{Z}_2^d$ &$\{e\}$ &$Z\widetilde{Z}$ & $\frac{1}{2}(1+X^{\otimes 4}\widetilde{X}^{\otimes 4})$  & $\frac{1}{4}(1+Z^{\otimes 4})(1+\widetilde{Z}^{\otimes 4})$ & $Z\widetilde{Z}$ & $id$  &$Z$, $\widetilde{Z}$ & $Y\widetilde{Y}$, $X\widetilde{X}$ \\  \hline
%----------------------
& & & &  $X$ &  &  disturb& disturb &  $X$ & &  \\
& & & &  $X$ & infer &  &  &  $X$  & & \\
{\color{blue}bzz} &$1+m$ & $G$ & $\mathbb{Z}_2$ &$X$ &$\frac{1}{4}(1+X^{\otimes 4})(1+\widetilde{X}^{\otimes 4})$ &$ \frac{1}{2}(1+\widetilde{Z}^{\otimes 4})$ & $id$ & $X$& $\widetilde{Z}$, $X\widetilde{Z}$ & $\widetilde{X}$, $X\widetilde{X}$ \\  \hline
%-----------------------------
& & & &  $Z$ &disturb  &  &  $Z$  & disturb & & \\
& & & &  $Z$ & &  infer &  $Z$ & & &\\
{\color{red}rxx} &$1+e$ & $\widetilde{\mathbb{Z}}_2$   & $\{e\}$ &$Z$   &  $\frac{1}{2}(1+ \widetilde{X}^{\otimes 4}) $ & $\frac{1}{4}(1+Z^{\otimes 4})(1+\widetilde{Z}^{\otimes 4})$ & $Z$ & $id$  & $Z\widetilde{Z}$, $\widetilde{Z}$ & $Z\widetilde{X}$ , $\widetilde{X}$ \\ \hline
\end{tabular}
}
\caption{Each round the condensation term $T^M$ or  $L^N$ is the $1$-qudit measurement (or $2$ qubit measurement, since the dimension of one qudit is $4$). For reader's convenience, we also listed the corresponding color checks in \cite{kesselring:2022}. "Disturb" indicates that the  measurements of the current round disturbed the corresponding operator of the last round. "Infer" implies the eigenvalue of the operator at the current round can be inferred from other syndromes.}\label{schedule_of_Z2Z2_floquet_code}
\end{table}
%------------------------
Thus, we can reframe the Floquet code schedule presented in \cite{kesselring:2022} as outlined in Table~\ref{schedule_of_Z2Z2_floquet_code}, which operates periodically mod 6. Let's delve deeper into the behavior of the wavefunction as it transitions from one phase to the next.

Moving forward, for the sake of clarity, we'll focus on the square lattice. It's important to note that our analysis remains unaffected by the choice of lattice; any arbitrary lattice would suffice.
First assume without loss of generality that the state is the ground state of $H_{M=\wt{\ZZ_2},N=\{e\}}$, the $(\mathbf{1}+e)$-condensed phase. We then implement $1$-qudit measurements $X\wt{X}$ on all the edges. The measurement procedure disrupts the eigenstate properties of the plaquette operator $B_p^{N=\{e\}}\sim (1+Z^{\otimes 4})(1+\wt{Z}^{\otimes 4})$ within the $\mathbf{1}+e$ phase.   Furthermore, given eigenvalues of $A^M\sim  \wt{X}^{\otimes 4}$ from the $\mathbf{1}+e$ phase, one can infer the eigenvalue of $A^M$ at the phase of $1+m\wt{m}$ through the measurements of $X\wt{X}$.

To get a comprehensive view of the explicit wavefunction associated with these states or phases, let's define a reference state:
$
\ket{\text{ref}}:=\prod_{e\in \Lambda} \ket{+\wt{+}}_e$
which is a product state on the whole lattice $\Lambda$. Each qudit state is the eigenstate of operator $X$ and $\wt{X}$ $\ket{+\wt{+}}=\frac{1}{2}(|0\rangle +|1\rangle)(|\wt{0}\rangle+|\wt{1}\rangle )$.  Each unit square loop represented by $ZZZZ$ effectively "sketches" a loop diagram of length $4$ on the reference state by flipping $+$ to $-$, and a similar operation holds true for the operator $\wt{Z}\wt{Z}\wt{Z}\wt{Z}$.  
Hence, the operator $B^{N=\{e\}}\sim ZZZZ+\wt{Z}\wt{Z}\wt{Z}\wt{Z}$ in the phase $1+ e$ enforces an equal superposition of all diagrams containing two types of unit square loops labeled by $ZZZZ$ and $\wt{Z}\wt{Z}\wt{Z}\wt{Z}$ or any potential mix of the two. Each individual diagram $D$ will correspond to a vector component in the (unnormalized) ground state(s) $\Omega_{M=\wt{Z}_2,N=\{e\}}$
\begin{align}
  \ket{\Omega}_{M=\wt{Z}_2,N=\{e\}}=\sum_{D\in \text{loop diag.}}\ket{D}
\end{align}
Assuming, without loss of generality, that all measurement outcomes yield $X\wt{X}=+1$. We the two qubits $\{|+\wt{+}\rangle,|-\wt{-}\rangle\}$ within each 4-dimensional qudit. The residual diagrams  are made exclusively of $Z\wt{Z}$-loops on the reference state $\ket{\text{ref}}$,  resulting precisely in the ground state of $H_{M=G,N=\ZZ_2^d}$, the $(\mathbf{1}+m\wt{m})$-condensed phase. 

To see allowed Floquet codes in the parent phase $\mathcal{D}(G)$, the simplest way is to calculate the anyon-tunneling map \ref{eq_anyon_tunneling_map}. For example
\begin{align}
   \varphi_{a,b}(M/N=\widetilde{\mathbb{Z}}_2/\{e\},M\1/N\1=G/\mathbb{Z}_2^d)=id_{a,b}
\end{align}
which maps $\mathbf{1}\mapsto \mathbf{1}$, $e\mapsto e$, $m\mapsto m$, and $f\mapsto f$. In fact, each step in the Floquet table corresponds such an identity matrix.  On the other hand, if we try to consider an illegal Floquet code phase transition, we find that
\begin{align}
    \varphi(M/N=\widetilde{\mathbb{Z}}_2/\{e\},M\1/N\1=\mathbb{Z}^d_2/\{e\})=\mqty(1& 1 & 0 & 0\\ 1& 1 & 0 & 0\\ 0& 0 & 0 & 0 \\ 0& 0 & 0 & 0).
\end{align}
Apparently, the mapping $\mathbf{1}+e\to \mathbf{1}$ and $\mathbf{1}+e\to e$ doesn't conform to a good anyon mapping.

It's important to highlight that there are three additional phases in the middle row that the Floquet code schedule does not encompass. However, we have exhausted all permutations of $M/N$. This scenario arises because, quite fortuitously (and well-known), the $e$-$m$ duality exists within the toric code. In the context of operators, this duality is represented between the $A_v$ and $B_p$
terms. Such duality may not be present across all quantum double phases, making it absent in our model. This duality is also evident in $\mathcal{D}(\mathbb{Z}_2\times \mathbb{Z}_2)$, which corresponds to the double-layer toric code. Assuming that after the schedule's second step, instead of transitioning to the $1+\wt{e}$ phase, we apply a Hadamard transformation to the subgroup $\mathbb{Z}_2\subset G$, we would then arrive at the $\mathbf{1}+e\wt{m}$ condensed phase. Subsequently,
\begin{align}
    A^M\mapsto \frac{1}{4}(1+Z^{\otimes 4})(1+\wt{X}^{\otimes 4}),\quad B^N\to \frac{1}{2}(1+X^{\otimes 4} \wt{Z}^{\otimes 4}),\quad T^M\to id,\quad L^N\to Z\wt{X}.
\end{align}
The  form of ``$L^N$" not qualifying as a gauge transformation under some normal group  $N$ corroborates that the three phases in the middle row fall outside the purview of our model characterized by pairs $(T^M,L^N)$. After implementing such a duality transformation, should we proceed with the prior schedule but sandwich it with  Hadamard  $H$, we would traverse the subsequent condensed phases (excluding the trivial anyon  $\mathbf{1}$).
\begin{align}
    \underline{e\wt{1}}\to m\wt{m}\overset{\Delta}{\to} e\wt{m}\to m\wt{m}\to m\wt{e} \to \mathbf{1}\wt{m} \to \underline{ f\wt{f}}\to \underline{m\wt{\mathbf{1}}}
\end{align}
where after the step $\Delta$, the phases are those in the second row and the third row.
The composition of this series amounts to the anyon mapping 
\begin{align}
    \varphi_{ab}=\mqty(1 &  0 & 0 &0 \\ 0&  0 & 1 & 0  \\ 0& 1& 0& 0 \\ 0& 0& 0& 1).
\end{align}
In this series of phase transitions, the step denoted by $\Delta$ assumes is the same to the matrix $\varphi_{ab}$ above, while all other steps are the identity matrix  $id_{4\times 4}$.  The three highlighted phases with underline represent the diagonal part of the colored anyon table in Eq.~\ref{color_code_isomorphism}, aligning with the triangle diagram of the phase transitions explored in \cite{davydova2023quantum}. We wish to reiterate that the model proposed by \cite{bombin:2008} may overlook certain derived phases that, while isomorphic, are distinct from the quantum double 
$\mathcal{D}(M/N)$ for some $M$ and $N$.

However, this model doesn't capture the low-weight syndrome. This limitation stems from the fact that one cannot deduce the eigenvalues of $A^M$ or $B^N$ at round $\ell$, and then, at a subsequent round $\ell+k$, reassess the eigenvalue of the operator without impacting its value during rounds $\ell \leq r\leq r+k$. To acquire a syndrome, it's necessary to measure all $A^M$ and $B^N$  operators after the measurements of $T^M$ and $L^N$ at each round. 
  As a result, while we gain theoretical generalities, we sacrifice the important engineering benefits associated with Floquet code.

Despite that, the model adeptly mirrors various paths of phase transitions and automorphisms that have been scrutinized in the context of $\mathcal{D}(\mathbb{Z}_2\otimes \mathbb{Z}_2)$. As such, this framework stands as a promising tool for predicting potential automorphisms and permissible phase transitions in a dynamic code.
To consider a general Floquet code of schedule $(\mathcal{P}_0, \mathcal{P}_1,\cdots)$, if 
the map $\varphi_{ab}^{(i,i+1)}$ from phase $\mathcal{P}_i$ to phase $\mathcal{P}_{i+1}$  is  an element of the permutation group
$S_n$, where $n$ signifies the count of anyons in the examined phase, and provided the modular data (inclusive of the modular $S$ and $T$ matrices) remain unaltered under the action of $\varphi$, then $\varphi^{(i,i+1)}$ is an automorphism and the Floquet step 
from  $\mathcal{P}_i$ to  $\mathcal{P}_{i+1}$ is legit.

Moreover, it's essential to underscore that extending from the Abelian Floquet code to the non-Abelian scenario presents inherent challenges, especially when qudits are positioned at the vertices of a lattice graph. In the quantum double codes, the qudits residing on the edges align with the commutativity of  left and right group multiplications. When examining the Levin-Wen model \cite{PhysRevB.71.045110}, the objects of the foundational fusion category also situate themselves on the edges of the lattice graph, with the trivalent vertices reflecting the fusions of these particular objects.
Thus, placing qudit degrees of freedom on the vertices for the parent phase may confront some intrinsic difficulties. 
%----
\subsection{Example: The Floquet code in \texorpdfstring{$\mathcal{D}(D_4)$}{Lg}}
%-------
The classification of anyon types in  $\mathcal{D}(D_4)$ can be found in Appendix~\ref{sect_representation_of_DD4}. In this section, we directly employ Proposition~\ref{proposition_conditional_unitary_iff_condition} to uncover all potential measurement-induced phase transitions of phases isomorphic to $\mathcal{D}(\mathbb{Z}_2)$. In this instance, the process is straightforward, given that $M_{ji}\propto I_{ji}$ directly. To identify a phase isomorphic to $\mathcal{D}(\mathbb{Z}_2)$, we have three subgroup choices for $M/N$ by considering their orders: namely, 8/4, 4/2, and 2/1. These are tabulated for clarity in Table~\ref{table_of_D4}. For the sake of brevity in vector notation, we express 
$\sum_{g\in G} c_g g\in \mathbb{C}G$, instead of using the conventional Dirac ket notation $\sum_g c_g\ket{g}$.
\begin{table}[htb]
\centering
\begin{tabular}{|c|c|c|c|c|} \hline
 Label & $M$ & $N$ &  $\ket{0}$ & $\ket{1}$  \\ \hline
 1 &$D_4$ & $\{e,r,r^2,r^3\}$  &  $e+r+r^2+r^3$  & $s+sr+sr^2+sr^3$ \\
2& $D_4$ & $\{e,r^2,s,sr^2\}$  &  $e+r^2+s+sr^2$  & $r+r^3+sr+sr^3$ \\ 
3& $D_4$ & $\{e,r^2,sr,sr^3\}$  &  $e+r^2+sr+sr^3$  & $s+sr^2+r+r^3$ \\  \hline
4& $\{e,r,r^2,r^3\}$ & $\{ e,r^2\}$ &  $e+r^2$ & $r+r^3$  \\ 
5& $\{e,r,s,sr^2\}$ & $\{ e,r^2\}$ &  $e+r^2$ & $s+sr^2$  \\  
6& $\{e,r^2,sr,sr^3\}$ &  $\{ e,r^2\}$ &  $e+r^2$ & $sr+sr^3$  \\ \hline
7&  $\{e,r^2\}$ & $\{e\}$ &  $ e$ & $r^2$ \\ 
8&  $\{e,s\}$ & $\{e\}$ &  $ e$ & $s$ \\ 
9&  $\{e,sr\}$ & $\{e\}$ &  $ e$ & $sr$ \\ 
10&  $\{e,sr^2\}$ & $\{e\}$ &  $ e$ & $sr^2$ \\  \hline
\end{tabular}
\caption{Derived phases isomorphic to the toric code in the parent theory $\mathcal{D}(D_4)$ labeled the pair $M,N\subset G$.}\label{table_of_D4}
\end{table}
By Proposition~\ref{proposition_conditional_unitary_iff_condition}, the allowed phase transitions that preserve logical information are given by pairs
\begin{align}
    &(1,5)\quad (1,6)\quad (2,4)\quad (2,6)\quad (3,4)\quad (3,5)\\ 
    &(1,8)\quad (1,9)\quad (1,10)\quad (2,9)\quad (3,8)\quad (3,10) \\
    &(5,8)\quad (5,10)\quad (6,9)
\end{align}
Note that each phase transition is bidirectional. Any loop or periodic trajectory described by these pairs constitutes a Floquet code. For illustration, we present three distinct Floquet codes of different periodicity lengths.
\begin{align}
    1\to 8\to 1\quad \quad 1\to 8\to 3\to 10\to 1,\quad \quad 1\to 8\to 5\to 10\to 3 \to 8 \to 1.
\end{align}
Certainly, a Floquet code can exhibit no periodicity, or equivalently, have a periodicity length of infinity. Such a category of Floquet codes has been aptly termed as {\bfseries tree codes} in \cite{davydova2023}.
%---
\section{Topological quantum operations assisted by phase transitions}\label{sect_topological_quantum_computation}
In this section we consider topological gates enriched by partially condensed phases. Within the modular tensor category, the algebra under consideration can manifest as either Lagrangian or non-Lagrangian.
\subsection{Phase transition in the spatial directions}
\subsubsection{Ground State}
The ground states depend on the phases. Previously \cite{PhysRevLett.119.170504,koenig2010,fowler2012}, there are only trivial phase embedded in the phase that we are interested. For example, in surface code, we only consider the $e$-condensed puncture or the $m$-condensed puncture (both are trivial phase) embedded in the $\mathcal{D}(\mathbb{Z}_2)$ phase, and the corresponding subspace of the ground states are given by diagrams of fusion tree  with leaves the punctures and branches labeled by anyons.  The $F$-moves of the nodes of branches give rise to different basis of the same ground state subspace. 
In general, we should expect several nontrivial phases to be embedded in a bigger phase, in each of which there might be even more phases.  Hence, given a quantum system of several different levels of phases, a ground state is a tree diagram with nodes representing phases and edges (branches) representing anyons. Since an anyon changes after tunneling through domain walls, we need a pair of anyon labels $(a_+,a_-)$ for each edge of the tree diagram, to represent an anyon before and after the tunneling. 
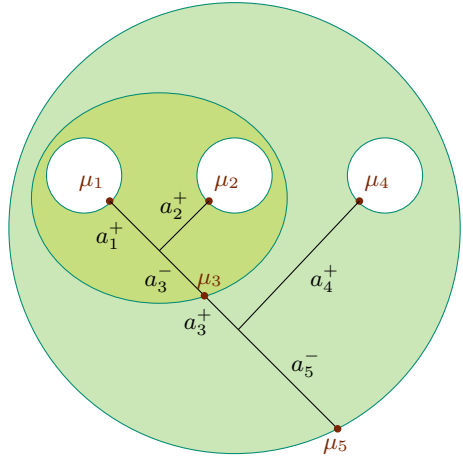
\begin{figure}[htb]
\centering
\begin{tikzpicture}[baseline={(current bounding box.center)}]
  \filldraw[fill=YellowGreen, fill opacity=0.5, draw=PineGreen] (2,-0.7) ellipse (3.cm and 3.cm);
  \filldraw[fill=SpringGreen, fill opacity=0.7, draw=PineGreen, draw opacity=1.0] (1,-0.3) ellipse (1.7cm and 1.4cm);
      % Small oval circles
  \filldraw[fill=white, fill opacity=1.0, draw=PineGreen] (0,0) ellipse (0.5cm and 0.5cm);
  \filldraw[fill=white,fill opacity=1.0, draw=PineGreen] (2,0) ellipse (0.5cm and 0.5cm);
  \filldraw[fill=white, fill opacity=1.0, draw=PineGreen] (4,0) ellipse (0.5cm and 0.5cm);
    \node at (3.9,-0.1) [text=Brown]{$\mu_4$};
    \fill[Brown] (3.66,-0.34) circle (0.05cm);
    \draw[black] (2.05,-2.05) -- (3.66,-0.34);
      %---------
    \node at (2-0.1,-0.1) [text=Brown]{$\mu_2$};
    \fill[Brown] (2-0.34,-0.34) circle (0.05cm);
    \draw[black] (2-0.34,-0.34) -- (1,-1.) ;
      %---------
   \node at (0.1,-0.1) [text=Brown]{$\mu_1$};
    \fill[Brown] (0.34,-0.34) circle (0.05cm);
    \draw[black] (0.34,-0.34) -- (3.35,-3.35);
    %------------------
    \node at (0.35,-0.8) {$a_{1}^+$};
    \node at (1.2,-0.4) {$a_{2}^+$};
    \node at (1.,-1.37) {$a_{3}^-$};
    \node at (1.67,-1.4) [text=Brown]{$\mu_3$};
    \fill[Brown] (1.6,-1.6) circle (0.05cm);
    \node at (1.52,-1.95) {$a_{3}^+$};
    \node at (3.2,-1.37) {$a_{4}^+$};
    \node at (2.95,-2.5) {$a_{5}^-$};
    \node at (3.35,-3.6) [text=Brown]{$\mu_5$};
    \fill[Brown] (3.37,-3.37) circle (0.05cm);
  % Bigger oval circle around the first two small circles
  % Fifth oval circle around all small circles
  %SkyBlue  CornflowerBlue  TealBlue YellowGreen  SpringGreen
\end{tikzpicture}
  \caption{A subtree of the phase and the corresponding diagram.}\label{fig_phase_tree}
\end{figure}
%-------
\subsubsection{Fusing and Braiding}  
Two algebras in $\mathcal{D}(G)$ can undergo fusion. Let's consider two punctures labeled by condensable algebras $\mathcal{A}_1$ and $\mathcal{A}_2$. They are termed as {\bfseries fusible} relative to $\mathcal{A}_3$ if a condensable algebra $\mathcal{A}_3$ exists such that $\mathcal{A}_1\cup \mathcal{A}_2\subset \mathcal{A}_3$. For instance, in $\mathcal{D}(\mathbb{Z}_2)$, the algebras $\mathbf{1}+ e$ and $\mathbf{1}+ m$ are deemed non-fusable.
In a parallel vein, an algebra $\mathcal{A}$ is termed {\bfseries splitable} if there are algebras $\mathcal{A}_1$ and $\mathcal{A}_2$ such that they are fusible with respect to $\mathcal{A}$.

Generally, the phase structure is depicted as a tree diagram, with its edges representing the   relation of containing. The diagram's vertices signify the derived phases contained in a branch, and the root is represented by $\mathcal{A}_0$. 

The fusion of two tree algebras, namely $\mathcal{T}$ and $\mathcal{T}\1$, progresses level by level. The fusion process begins by merging the roots, $\mathcal{T}_0$ and $\mathcal{T}\1_0$. It's imperative to recognize that our model of TQC, enhanced by phase transitions, won't encompass this kind of operation. Typically, such operations alter the topological logical qudits under consideration. However, these operations are intimately tied to the code deformation protocol, as expounded in \cite{kesselring:2022}.
%------
\subsubsection{Wrapping} 
Before delving into the intricacies of braiding, it is pertinent to establish some foundational definitions for the sake of clarity. An anyon, denoted as $a$, is characterized as being {\bfseries eventually condensed} to a puncture associated with a (sub)tree of algebras $\mathcal{T}$, provided there's a tunneling series $a_0\overset{\mu_0}{\to} a_1 \overset{\mu_1}{\to} \cdots \overset{\mu_n}{\to} \mathbf{1}$ that ends as the trivial particle at some node/subphase/puncture $\mathcal{P}_k$ within the tree hierarchy (and continues as $\mathbf{1}$ subsequently). Given that, at our focal level, the anyon manifests as $a$, we say that an $a$-ribbon is {\bfseries attached} to the puncture $\mathcal{P}_k$.

When examining the $i$-th subphase $\mathcal{P}_i$ at the focal level of  phase tree, consider the scenario where an anyon $a_i$ encircles another anyon $b_i$ attached to $\mathcal{P}_i$. Drawing from \cite{PhysRevLett.119.170504,bakalov2001}, this configuration generates a holonomy operation.
\begin{align}
   \begin{tikzpicture}[baseline={(current bounding box.center)}]
    % Draw and fill a smaller circle with green color at (0,0)
    \filldraw[fill=gray, draw=black] (0,0) circle (0.3);  
    % Draw a bigger circle at (0,0)
    \draw (0,0) circle (0.6);
    % Draw a line from (0,0) to (3,0)
    \draw (0.3,0) -- (1.3,0);
     \node at (-0.9,0) [text=black]{$a_i$};
     \node at (1,0.3) [text=black]{$b_i$};
\end{tikzpicture}= W_{a_i}\ket{\cdots, b_i,\cdots }\ =\frac{S_{a_ib_i}}{d_{b_i}}\ket{\cdots, b_i,\cdots}.
\end{align} 
By wrapping around different levels of the phase tree, this type of operation can give us rich phase gates for various (sub)phases. 
\subsubsection{Braiding}
Within the context of the fusion of punctures as previously delineated, our primary focus narrows to the braiding of two punctures of subphases, both residing within the same branch of the tree. The braiding of only punctures would be trivial, and hence we turn our attention to the braiding of two punctures attached with ribbons in the standard basis \cite{PhysRevLett.119.170504}:
\begin{equation}
\sigma^2\qty(
\begin{tikzpicture}[baseline={(current bounding box.center)}]
    % Drawing the top circles
    \filldraw[fill=gray, draw=black] (-1,1) circle (0.3); 
    \filldraw[fill=gray, draw=black] (1,1) circle (0.3);
    % Drawing the lines with labels
    \draw (-1,-0.7) -- (-1,0) node[midway, left] {$a_2$};
    \draw (-1,0) -- (-1,0.7) node[midway, left] {$a_1$} ;
    \draw (1,-0.7) -- (1,0) node[midway, right] {$b_2$};
    \draw (1,0) -- (1,0.7) node[midway, right] {$b_1$};
    \draw (-1,0) -- (1,0) node[midway,above]{$c$};
\end{tikzpicture}) 
=\sum_{c\1} 
\sum_{c\2} (F_{b_2}^{a_2 a_1 b_1})^{-1}_{c\2 c\1} R_{c\1}^{b_1 a_1} R^{a_1 b_1}_{c\1}  (F^{a_2a_1 b_1}_{b_2})_{c\1 c}
%----
\quad 
\begin{tikzpicture}[baseline={(current bounding box.center)}]
    % Drawing the top circles
    \filldraw[fill=gray, draw=black] (-1,1) circle (0.3); 
    \filldraw[fill=gray, draw=black] (1,1) circle (0.3);
    % Drawing the lines with labels
    \draw (-1,-0.7) -- (-1,0) node[midway, left] {$a_2$};
    \draw (-1,0) -- (-1,0.7) node[midway, left] {$a_1$} ;
    \draw (1,-0.7) -- (1,0) node[midway, right] {$b_2$};
    \draw (1,0) -- (1,0.7) node[midway, right] {$b_1$};
    \draw (-1,0) -- (1,0) node[midway,above]{$c^{\prime\prime}$};
\end{tikzpicture}
\end{equation}
%------
\subsubsection{Ribbon-path operator}
The primary objective of this section is to employ the 6j-symbol to map the action ribbon $\gamma$ (shown as the red line in Eq.~\ref{eq_path_operator_first_line}) onto the standard basis. 
This is a generalization of acting a string of Pauli-$X$  between two punctures in surface code \cite{fowler2012}, which flips the logical $\ket{\overline{0}}$ and $\ket{\overline{1}}$. 

Back to the ribbon operator we consider here, the path $\gamma$  connects $\mathcal{T}_0$ and $\mathcal{T}_1$ at $\alpha$ and $\beta$, respectively, circumventing any other punctures from below. 
Even though a general path connecting $\mathcal{T}_0$ and $\mathcal{T}_1$ might be homotopically distinct from $\gamma$ due to its potential to wrap around different punctures multiple times, it's crucial to understand that such paths can be simplified to a basic "straight line" accompanied by some independent loops encircling the punctures.  This is because tadpole diagrams don't exist in modular tensor categories.
Because of that, we can, for the sake of simplicity and without any loss of generality, limit our consideration to  paths  like  $\gamma$ in Eq.~\ref{eq_path_operator_first_line}.
\begin{align}
   & \begin{tikzpicture}[baseline={(current bounding box.center)}]
\def\d{1.2}
\def\r{0.3}
\def\delta{0.07}
\def\endpt{3.0}
\def\ratio{1.6}
    %\draw (0,0)--(2.5*\d,0);
    \draw[midarrow](\d,0)-- (\r,0) node[midway,above]{$b$};
    \draw[midarrow](2*\d,0)-- (\d,0) node[midway,above]{$c$};
    \draw (2*\d,0)--(2*\d+\ratio*\r-\delta,0);
    \draw[midarrow] (\endpt*\d,0)--(2*\d+\ratio*\r+\delta,0) node[midway,above]{$d$};
    \draw[midarrow] (\d,0)--(\d,\d-\r) node[midway,left]{$e$};
    \draw[midarrow] (2*\d,0)--(2*\d,\d-\r) node[midway,left]{$f$};
    \draw[red](0,0)--(0,-\ratio*\r ) ;
    \draw[red](0,-\ratio*\r )-- ( 2*\d+\ratio*\r,-\ratio*\r)node[midway,below]{$a$};
    \draw[red,midarrow](\d+\ratio*\r/2+\ratio*\r*0.15-0.001,-\ratio*\r)--(\d++\ratio*\r*0.15+\ratio*\r/2+0.001,-\ratio*\r);
    \draw[red]( 2*\d+\ratio*\r,-\ratio*\r)--( 2*\d+\ratio*\r,\d)--(2*\d,\d);
    \filldraw[fill=gray, draw=black] (0,0) circle (\r);
    \filldraw[fill=gray, draw=black] (\d,\d) circle (\r);
    \filldraw[fill=gray, draw=black] (2*\d,\d) circle (\r);
    %----
    \def\dotr{0.06}
    \fill[black] (\r,0) circle (\dotr);
    \node at (1.15*\r,0.75*\r) {$\mu$};
    %-------
    \fill[red] (0,-\r) circle (\dotr);
    \node[red] at (-\r,-1.15*\r) {$\alpha$};
    %-------
    \fill[red] (2*\d+\r,\d) circle (\dotr);
    \node[red] at (2.1*\d+\r,\d+1.05*\r) {$\beta$};
    \fill[black] (\d,\d-\r) circle (\dotr);
    \node at (1.14*\d,0.91*\d-\r) {$\nu$};
    \fill[black] (2*\d,\d-\r) circle (\dotr);
    \node at (2.14*\d,0.91*\d-\r) {$\lambda$};
\end{tikzpicture}
%--------
=\sum_g C_g
\begin{tikzpicture}[baseline={(current bounding box.center)}]
\def\d{1.2}
\def\r{0.3}
\def\delta{0.07}
\def\endpt{3.0}
\def\ratio{1.6}
    %\draw (0,0)--(2.5*\d,0);
    \draw[midarrow](\d,0)-- (\r,0) node[midway,above]{$b$};
    \draw[midarrow](2*\d,0)-- (\d,0) node[midway,above]{$c$};
    \draw (2*\d+\ratio*\r,0)--(2*\d,0) node[midway,above]{$d$};
    \draw[midarrow] (2*\d+\ratio*\r/3+0.001,0)--(2*\d+\ratio*\r/3-0.001,0);%draw arrow for d
    \draw[midarrow](\endpt*\d,-\ratio*\r)--(2*\d+\ratio*\r,-\ratio*\r) node[midway,below]{$d$};
    \draw[midarrow] (\d,0)--(\d,\d-\r) node[midway,left]{$e$};
    \draw[midarrow] (2*\d,0)--(2*\d,\d-\r) node[midway,left]{$f$};
    \draw[black](0,0)--(0,-\ratio*\r ) ;
    \draw[black](0,-\ratio*\r )-- ( 2*\d+\ratio*\r,-\ratio*\r)node[midway,below]{$a$};
    \draw[midarrow](\d+\ratio*\r/2+\ratio*\r*0.15-0.001,-\ratio*\r)--(\d++\ratio*\r*0.15+\ratio*\r/2+0.001,-\ratio*\r);
    \draw[teal]( 2*\d+\ratio*\r,-\ratio*\r)--( 2*\d+\ratio*\r,0)node[midway,right]{$g$};
    \draw[teal,midarrow]( 2*\d+\ratio*\r,-\ratio*\r/3-0.001)--( 2*\d+\ratio*\r,-\ratio*\r/3+0.001);%draw arrow for g
    \draw[midarrow] ( 2*\d+\ratio*\r,0)--( 2*\d+\ratio*\r,\d)node[midway,right]{$a$};
    \draw( 2*\d+\ratio*\r,\d)--(2*\d,\d);
    \filldraw[fill=gray, draw=black] (0,0) circle (\r);
    \filldraw[fill=gray, draw=black] (\d,\d) circle (\r);
    \filldraw[fill=gray, draw=black] (2*\d,\d) circle (\r);
    %---------
    \def\dotr{0.06}
    \fill[black] (\r,0) circle (\dotr);
    \node at (1.15*\r,0.75*\r) {$\mu$};
    %-------
    \fill[black] (0,-\r) circle (\dotr);
    \node[black] at (-\r,-1.15*\r) {$\alpha$};
    %-------
    \fill[black] (2*\d+\r,\d) circle (\dotr);
    \node[black] at (2.1*\d+\r,\d+1.05*\r) {$\beta$};
    \fill[black] (\d,\d-\r) circle (\dotr);
    \node at (1.14*\d,0.91*\d-\r) {$\nu$};
    \fill[black] (2*\d,\d-\r) circle (\dotr);
    \node at (2.14*\d,0.91*\d-\r) {$\lambda$};
\end{tikzpicture} \label{eq_path_operator_first_line} 
%------------
=\sum_{g,h,i} C^{\delta R}_g C^{MM}_{h,i}
\begin{tikzpicture}[baseline={(current bounding box.center)}]
\def\d{1.2}
\def\r{0.3}
\def\delta{0.07}
\def\endpt{3.0}
\def\ratio{1.6}
    %\draw (0,0)--(2.5*\d,0);
    \draw[midarrow](\d/2+\r/2,0)--(\r,0) node[midway,above]{$h$};
    \draw[midarrow](\d,0)-- (\d/2+\r/2,0) node[midway,above]{$b$};
    \draw[midarrow](2*\d,0)-- (\d,0) node[midway,above]{$c$};
    \draw (2*\d+\ratio*\r,0)--(2*\d,0) node[midway,below]{$d$};
    \draw[midarrow] (2*\d+\ratio*\r/3+0.001,0)--(2*\d+\ratio*\r/3-0.001,0);%draw arrow for d
    \draw[midarrow](\endpt*\d,-\ratio*\r)--(2*\d+\ratio*\r,-\ratio*\r) node[midway,below]{$d$};
    \draw[midarrow] (\d,0)--(\d,\d-\r) node[midway,right]{$e$};
    \draw[midarrow] (2*\d,0)--(2*\d,\d/2-\r/2) node[midway,left]{$f$};
    \draw[midarrow] ((2*\d,\d/2-\r/2) --(2*\d,\d-\r) node[midway, left]{$i$};
    \draw[black](\d/2+\r/2,0)--(\d/2+\r/2,-\ratio*\r ) ;
    \draw[black](\d/2+\r/2,-\ratio*\r )-- ( 2*\d+\ratio*\r,-\ratio*\r)node[midway,below]{$a$};
    \draw[midarrow](1.25*\d+\ratio*\r/2+\ratio*\r*0.15-0.001,-\ratio*\r)--(1.25*\d+\ratio*\r/2+\ratio*\r*0.15+\ratio*\r/2+0.001,-\ratio*\r);
    \draw[black]( 2*\d+\ratio*\r,-\ratio*\r)--( 2*\d+\ratio*\r,0)node[midway,right]{$g$};
    \draw[black,midarrow]( 2*\d+\ratio*\r,-\ratio*\r/3-0.001)--( 2*\d+\ratio*\r,-\ratio*\r/3+0.001);%draw arrow for g
    \draw[midarrow] ( 2*\d+\ratio*\r,0)--( 2*\d+\ratio*\r,\d/2-\r/2)node[midway,right]{$a$};
    \draw( 2*\d,\d/2-\r/2)--(2*\d+\ratio*\r,\d/2-\r/2);
    \filldraw[fill=gray, draw=black] (0,0) circle (\r);
    \filldraw[fill=gray, draw=black] (\d,\d) circle (\r);
    \filldraw[fill=gray, draw=black] (2*\d,\d) circle (\r);
    %-------
    \def\dotr{0.06}
    \fill[black] (\r,0) circle (\dotr);
    \node at (1.15*\r,-0.75*\r) {$\gamma$};
    %-------
    %\fill[black] (0,-\r) circle (\dotr);
    %\node[black] at (-\r,-1.15*\r) {$\alpha$};
    %-------
   % \fill[black] (2*\d+\r,\d) circle (\dotr);
    %\node[black] at (2.1*\d+\r,\d+1.05*\r) {$\beta$};
    \fill[black] (\d,\d-\r) circle (\dotr);
    \node at (1.14*\d,0.91*\d-\r) {$\nu$};
    \fill[black] (2*\d,\d-\r) circle (\dotr);
    \node at (2.14*\d,0.91*\d-\r) {$\epsilon$};
    %-----add vacuum line----
    \draw[dashed](1.3*\d,0)--(1.3*\d,-\ratio*\r);
\end{tikzpicture}\nnb\\
=&\sum_{g,h,i,k,j} C_g^{\delta R} C^{MM}_{h,i}  C^{FF}_{k,j}
\begin{tikzpicture}[baseline={(current bounding box.center)}]
\def\d{1.2};
\def\r{0.3};
\def\delta{0.07};
\def\endpt{3.0};
\def\ratio{1.6};
    %\draw (0,0)--(2.5*\d,0);
    \draw[midarrow](\d-\r,0)--(\r,0) node[midway,above]{$h$};
    %\draw[midarrow](\d,0)-- (\d/2+\r/2,0) node[midway,above]{$b$};
    %\draw[midarrow](2*\d,0)-- (\d,0) node[midway,above]{$c$};
   % \draw (2*\d+\ratio*\r,0)--(2*\d,0) node[midway,below]{$d$};
    %\draw[midarrow] (2*\d+\ratio*\r/3+0.001,0)--(2*\d+\ratio*\r/3-0.001,0);%draw arrow for d
    \draw[midarrow](\endpt*\d,-\ratio*\r)--(2*\d+\ratio*\r,-\ratio*\r) node[midway,below]{$d$};
    \draw[midarrow] (\d,+\r)--(\d,\d-\r) node[midway,right]{$e$};
    \draw[midarrow] (2*\d,0)--(2*\d,\d/2-\r/2) node[midway,left]{$f$};
    \draw[midarrow] ((2*\d,\d/2-\r/2) --(2*\d,\d-\r) node[midway, left]{$i$};
    %\draw[black](\d/2+\r/2,0)--(\d/2+\r/2,-\ratio*\r ) ;
    %\draw[black](\d/2+\r/2,-\ratio*\r )-- ( 2*\d+\ratio*\r,-\ratio*\r)node[midway,below]{$a$};
    %\draw[midarrow](1.25*\d+\ratio*\r/2+\ratio*\r*0.15-0.001,-\ratio*\r)--(1.25*\d+\ratio*\r/2+\ratio*\r*0.15+\ratio*\r/2+0.001,-\ratio*\r);
    %\draw[teal]( 2*\d+\ratio*\r,-\ratio*\r)--( 2*\d+\ratio*\r,0)node[midway,right]{$g$};
    %\draw[teal,midarrow]( 2*\d+\ratio*\r,-\ratio*\r/3-0.001)--( 2*\d+\ratio*\r,-\ratio*\r/3+0.001);%draw arrow for g
    \draw[midarrow] ( 2*\d+\ratio*\r,0)--( 2*\d+\ratio*\r,\d/2-\r/2)node[midway,right]{$a$};
    \draw(2*\d,0) --( 2*\d+\ratio*\r,0);
    %---------
    \draw( 2*\d,\d/2-\r/2)--(2*\d+\ratio*\r,\d/2-\r/2);
    \filldraw[fill=gray, draw=black] (0,0) circle (\r);
    \filldraw[fill=gray, draw=black] (\d,\d) circle (\r);
    \filldraw[fill=gray, draw=black] (2*\d,\d) circle (\r);
    %-------
    \def\dotr{0.06}
    \fill[black] (\r,0) circle (\dotr);
    \node at (1.15*\r,-0.75*\r) {$\gamma$};
    %-------
    %\fill[black] (0,-\r) circle (\dotr);
    %\node[black] at (-\r,-1.15*\r) {$\alpha$};
    %-------
   % \fill[black] (2*\d+\r,\d) circle (\dotr);
    %\node[black] at (2.1*\d+\r,\d+1.05*\r) {$\beta$};
    \fill[black] (\d,\d-\r) circle (\dotr);
    \node at (0.86*\d, 0.91*\d-\r) {$\nu$};
    \fill[black] (2*\d,\d-\r) circle (\dotr);
    \node at (2.14*\d,0.91*\d-\r) {$\epsilon$};
    %-----add vacuum line----
    %\draw[dashed](1.3*\d,0)--(1.3*\d,-\ratio*\r);
    %----add bubbles----
    \filldraw[fill=white,draw=black] (\d,0) circle(\r);
    %-------
    \node at (\d-1.05*\r,1.2*\r) {$b$};
    \draw[midarrow] (\d,\r) arc [start angle=90, end angle=180, radius=\r] ;%node[midway, above]{$b$};
    \draw[midarrow] (\d,-\r) arc [start angle=-90, end angle=90, radius=\r] node[midway,right]{$c$};
%    \node at (\d+\r,1.1*\r) {$c$};
    \draw[midarrow] (\d-\r,0) arc [start angle=180, end angle=270, radius=\r];
    \node at (\d-1.05*\r,-1.2*\r) {$a$};
    %----------
    \draw (\d,-\r)--(\d,-\ratio*\r);
    \draw[midarrow](\d,-\ratio*\r)--(1.5*\d,-\ratio*\r)node[midway, below]{$k$};
    %-------
    \draw (1.5*\d, -\ratio*\r)-- (1.5*\d, -\ratio*\r-1.4*\r);
    \draw[midarrow](1.5*\d, -\ratio*\r-1.4*\r) -- (2.5*\d, -\ratio*\r-1.4*\r) ;
    \node at (2*\d,-\ratio*\r-0.8*\r) {$a$};
    \draw(2.5*\d, -\ratio*\r-1.4*\r)--(2.5*\d,-\ratio*\r) ;
    %\draw[midarrow] (2*\d-\r,-\ratio*\r) arc [start angle=180, end angle=360, radius=\r] node[midway,below]{$a$};
    \draw[midarrow](2*\d,-\ratio*\r) -- (1.5*\d-\r,-\ratio*\r);
    \node at (1.75*\d,-\ratio*\r-0.5*\r) {$c$};
    %node[midway,above]{$c$};
    \draw[midarrow](2.5*\d+\r,-\ratio*\r)--(2*\d,-\ratio*\r);
    \node at (2.25*\d,-\ratio*\r-0.5*\r) {$g$};
    % node[midway,below]{$g$};
    %-----
    \draw[midarrow](2*\d,-\ratio*\r)--(2*\d,0)node[midway,left]{$j$};
\end{tikzpicture}
%---------
=\sum_{g,h,i,k,j} C_g^{\delta R} C^{MM}_{h,i} C^{FF}_{k,j}C^{\delta FF}
\begin{tikzpicture}[baseline={(current bounding box.center)}]
\def\d{1.2}
\def\r{0.3}
\def\delta{0.07}
\def\endpt{3.0}
\def\ratio{1.6}
    \filldraw[fill=gray, draw=black] (0,0) circle (\r);
    \filldraw[fill=gray, draw=black] (\d,\d) circle (\r);
    \filldraw[fill=gray, draw=black] (2*\d,\d) circle (\r);
    \def\dotr{0.06};
    \fill[black] (\r,0) circle (\dotr);
    \node at (1.15*\r,-0.75*\r) {$\gamma$};
    \fill[black] (\d,\d-\r) circle (\dotr);
    \node at (0.86*\d, 0.91*\d-\r) {$\nu$};
    \fill[black] (2*\d,\d-\r) circle (\dotr);
    \node at (2.14*\d,0.91*\d-\r) {$\epsilon$};
    \draw[midarrow](\d,0)--(\r,0) node[midway,above]{$h$};
    \draw[midarrow](\d,0)--(\d,\d-\r) node[midway,right]{$e$};
    \draw[midarrow](\d,0)--(2*\d,0) node[midway,above]{$k$};
    \draw[midarrow](2*\d,0)--(2*\d,\d-\r) node[midway,right]{$i$};
    \draw[midarrow](2.5*\d,0)--(2*\d,0) node[midway,below]{$d$};
\end{tikzpicture}
\end{align}
where
\begin{align}
    C_g^{\delta R}=&\sqrt{\frac{d_g}{d_ad_d}}\delta_{adg^*}R^{ad}_g
    \quad \quad 
    C_{h,i}^{MM}=(M^{\bullet\bullet,h}_{\bullet,a^*b})_\gamma^{\alpha\mu}
(M^{\bullet\bullet,i}_{\bullet,fa})_\epsilon^{\lambda\beta} 
\quad \quad 
C_{k,j}^{FF}= (F^{cc^*a}_a)_{k\mathbf{1}} (F^{cfa}_g)_{jd} \\
C^{\delta FF}=&\qty(\delta_{ij}\sqrt{\frac{d_ad_f}{d_i}})\qty(F^{hb^*c}_{k^*;ea^*}\sqrt{\frac{d_{b^*}d_c}{d_e}})\qty(F^{k^*c^*g}_{d; ja^*}\sqrt{\frac{d_{c^*}d_g}{d_j}}),
\end{align}
%------------
and we have used the following identities
\begin{align}
    \begin{tikzpicture}[baseline={(current bounding box.center)}]
    \draw[midarrow](0,0)--(.6,0)node[midway,above]{$x$};
    \draw[midarrow] (.6,0) arc [start angle=180, end angle=0, radius=0.3] node[midway,above]{$i$};
    \draw[midarrow] (.6,0) arc [start angle=180, end angle=360, radius=0.3] node[midway,below]{$j$};
    \draw[midarrow](.6+.6,0)--(1.2+.6,0)node[midway,above]{$y$};
    \end{tikzpicture}
    =&\delta_{xy}\sqrt{\frac{d_i d_j}{d_x}} \ 
    \begin{tikzpicture}%[baseline={(current bounding box.center)}]
     \draw[midarrow](0,0)--(1,0)node[midway,above]{$x$};
    \end{tikzpicture} \quad \quad \quad 
    %------
    \begin{tikzpicture}[baseline={(current bounding box.center)}]
    \def\r{0.3}
    \draw[midarrow](2*\r,0)--(0,0)node[midway,below]{$x$};
    \draw[midarrow] (2*\r,0) arc [start angle=180, end angle=90, radius=\r];
    \node at (2.15*\r,1.2*\r) {$y$};
    \draw[midarrow] (3*\r,\r)--(3*\r,3*\r) node [midway,right]{$u$};
    \draw[midarrow] (4*\r,0) arc [start angle=0, end angle =90, radius=\r];
    \node at (4*\r,1.05*\r) {$z$};
    \draw[midarrow] (4*\r,0) arc [start angle=0, end angle =-180, radius=\r] node [midway,below]{$v$};
    \draw[midarrow](6*\r,0)--(4*\r,0)node[midway,below]{$w$};
    \end{tikzpicture}
    =F^{xyz}_{w;u v}\sqrt{\frac{d_y d_z}{d_u}} \ 
    \begin{tikzpicture}[baseline={(current bounding box.center)}]
    \def\r{0.3}
    \draw[midarrow](3*\r,0)--(0,0)node[midway,below]{$x$};
    \draw[midarrow] (3*\r,0)--(3*\r,3*\r) node [midway,right]{$u$};
    \draw[midarrow](6*\r,0)--(3*\r,0)node[midway,below]{$w$};
    \end{tikzpicture}.
\end{align}
%----
%\subsection{Puncture crossing a domain wall}
%---------
\subsection{Phase transition in the temporal direction}\label{temporal_phase_transition_I}
Up to this point, our discussion has revolved around phase transitions in spatial directions. However, the same theory can be extended to investigate temporal phase transitions, which is intimately related to the Floquet code.

Consider the scenario where a ribbon connects two phases. This can be visualized in a different light. Imagine the parent phase, represented by $\mathcal{D}(G)$, sandwiched between the two derived phases, $\mathcal{D}(M\1/N\1)$ and $\mathcal{D}(M/N)$. This parent phase is so thin that our attention is primarily on the derived phases.

Now, at time $t=0$, let's set $M=M\1=M_0$ and $N=N\1=N_0$. In this case, the two phases are identical, making the domain wall separating them trivial. Picture an anyon $a$ traversing this negligible domain wall, its ends situated on both sides of the wall.

Now, envision that by the time $t=1$, we have transitioned to a new phase, with $M=M\1=M_1$ and $N=N\1=N_1$. This transition is brought about by adjusting the local terms $T^M$ and $L^N$. In a practical scenario, these terms wouldn't shift simultaneously. This leads us to an interesting possibility: within a certain time frame $\Delta T$ around time $t=0.5$, we first change the right half phase using the parameters $T^{M=M_1}$ and $L^{N=N_1}$. The left half, meanwhile, persists in its original state, $\mathcal{D}(M_0/N_0)$. It follows that the anyon-tunneling map $\varphi_{ab}$ that we considered in Eq.~\ref{eq_anyon_tunneling_map}  can also be treated temporally as a map of anyons from phase $\mathcal{P}_0$ at $t=0$ to $\mathcal{P}_1$ at $t=1$. 
In other words, one can view a ribbon graph in both spatial and temporal directions.  Specifically, a ribbon, when treated as a morphism of a modular tensor category $\mathcal{C}$, can be perceived either as a temporal trajectory within the phase or as a component of a fusion diagram in space at a specific time slice.

\begin{figure}[htb]
\centering
\begin{tikzpicture}
\def\d{1}
\def\linew{1.3}
\def\r{0.06}
\def\phigh{1.3}
\def\delta{0.04}
%----------
\filldraw[fill=gray!50!white] (\d,0) ellipse (2. and 0.6);
\draw (0,0) ellipse (0.1 and 0.05);
\draw (\d,0) ellipse (0.1 and 0.05);
\draw (2*\d,0) ellipse (0.1 and 0.05);
%-----
\filldraw[fill=pink!50!white] (\d,\phigh) ellipse (2. and 0.5);
\draw (0.27*\d,\phigh) ellipse (0.1 and 0.05);
\draw (0.63*\d,\phigh) ellipse (0.1 and 0.05);
\draw (1.3*\d,0.875*\phigh) ellipse (0.1 and 0.05);
%-------------
%\draw[->, thick] (-1,0) -- (4,0) node[right] {$x$};
%\draw[->, thick] (0,-1) -- (0,3) node[above] {$y$};
\draw[line width=\linew,color=red] (\d,0) .. controls (1*\d,1*\d) and (0*\d,0.5*\d) .. (0,3*\d) ;
\filldraw[fill=pink!50!white,draw=none](0.42*\d,1.015*\d) circle (\r);
\draw[line width=\linew,color=blue] (0,0) .. controls (-.2*\d,1*\d) and (1*\d,0.7*\d) .. (\d,3*\d) ;
%-------
\filldraw[fill=pink!50!white,draw=none](0.8*\d,1.65*\d) circle (\r);
%\draw[line width=\linew,color=teal](2*\d,0) .. controls(2*\d,0.6*\d) and (2*\d,0.9*\d) .. (1.1*\d, 1.65*\d);
\draw[line width=\linew,color=teal](2*\d,0) .. controls(2*\d,1*\d) and (0.*\d,1.8*\d) .. (0.93*\d-1.5*\delta, 2.2*\d);
\draw[line width=\linew,color=teal](0.93*\d+2*\delta, 2.2*\d+\delta) .. controls(1.5*\d,2.4*\d) and (2*\d,2.7*\d) .. (2*\d, 3*\d);
%\draw (-\d,\phigh)--(3*\d,\phigh);
%-----
%----------
%----------
\filldraw[fill=gray!50!white] (\d,3*\d) ellipse (2. and 0.45);
\draw (0,3*\d) ellipse (0.1 and 0.05);
\draw (\d,3*\d) ellipse (0.1 and 0.05);
\draw (2*\d,3*\d) ellipse (0.1 and 0.05);
%-------
\draw[->,thick] (-1.75*\d,0.5\d)--(-1.75*\d, 2.5*\d) node[midway,left]{time};
\node[anchor=west] at (3.7*\d, 0) {Initialization};
\node[anchor=west] at (3.7*\d, \phigh) {Phase transition};
\node[anchor=west] at (3.7*\d, 3*\d) {Readout};
\end{tikzpicture}
\caption{An illustration of a simple TQC process  assisted by phase transition. The time direction is upward. The logical qudits are encoded by fusion channels and logical gates are implemented through braiding of anyons. Here the pink time slice in the middle represents the temporal domain wall of a phase transition; the bottom and the top gray times slices represent the initialization and readout, respectively.}
\label{General_quantum_computation_with_phases}
\end{figure}
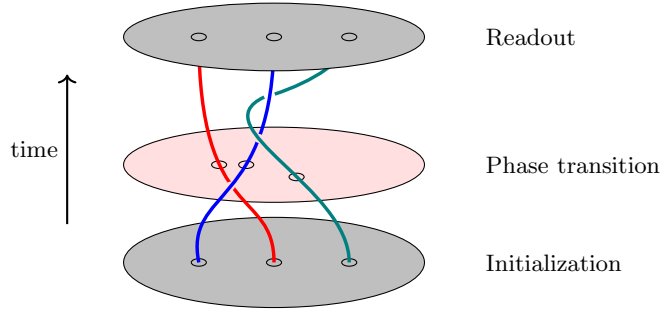

 By leveraging the utility of introducing diverse phases, a more comprehensive framework for quantum computation is depicted in Figure \ref{General_quantum_computation_with_phases}.
%-----

Let's consider a concrete example to explore the conditions  for preserving logical information during phase transitions. It's noteworthy that the following representation can be interpreted as either a spatial or a temporal ribbon diagram of anyons transitioning through domain walls.
For fixed $a,b,c$ and $d$, we use the fusion spaces labeled by the intermediate anyon $x$ as  qudits for quantum computation.
\begin{align}
\ket{x}:=
&\begin{tikzpicture}[baseline={(current bounding box.center)}]
\def\d{.7}
   \draw (0,0)  -- (\d/2,-\d/2);
    \node at (0,0.15*\d) {$a$};
    %-----------
    \draw ( +\d, 0) -- (\d/2,-\d/2);
    \node at (\d,0.15*\d) {$b$};
    %-------
    \draw (2*\d,0) -- (\d,-\d);
    \node at (2*\d,0.15*\d) {$c$};
    %----
    \draw (\d/2,-\d/2)--(\d,-\d);
    \node at (0.62*\d,-0.88*\d) {$x$};
    %-----------
    \draw(\d,-\d)--(1.4*\d ,-1.4*\d)node[midway,below]{$d$};
\end{tikzpicture}\label{ketx_fusion_tree}
\end{align}
%-----------
The domain wall in the diagram is drawn spatial, but this process can also be understood as temporal (a time scan), see Sect.~\ref{temporal_phase_transition_I}.
After the phase transition,  these diagrams (vectors) are mapped to:
%-----
\begin{align}
   & \begin{tikzpicture}[baseline={(current bounding box.center)}]
\def\d{.7}
\def\h{1.0}
\def\ratio{0.55}
\def\dratio{0.17}
    %-----horizental line-----
    \draw(-0.7*\d,\ratio*\h)--(2.7*\d,\h*\ratio);
%------
    \draw (0,0)  -- (\d/2,-\d/2);
    \draw(0,0) -- (0,\h);
    \node at (-\dratio*\d,\dratio*\d) {$a$};
    %-----------
    \draw ( +\d, 0) -- (\d/2,-\d/2);
    \draw(\d,0) -- (\d,\h);
    \node at (-\dratio*\d+\d,\dratio*\d) {$b$};
    %-------
    \draw (2*\d,0) -- (\d,-\d);
    \draw(2*\d,0) -- (2*\d,\h);
    \node at (2*\d-\dratio*\d,\dratio*\d) {$c$};
    %----
    \draw (\d/2,-\d/2)--(\d,-\d);
    \node at (0.62*\d,-0.88*\d) {$x$};
    %-----------
    \draw(\d,-\d)--(1.4*\d ,-1.4*\d)node[midway,below]{$d$};
    %------
    \def\dotr{0.06}
    \fill[black] (0,\h*\ratio) circle (\dotr);
    \fill[black] (\d,\h*\ratio) circle (\dotr);
    \fill[black] (2*\d,\h*\ratio) circle (\dotr);
    \node at (-\dratio*\d, \dratio*\d+\h*\ratio) {$\mu$};
    \node at (-\dratio*\d+\d, \dratio*\d+\h*\ratio) {$\nu$};
    \node at (2*\d-\dratio*\d , \dratio*\d+\h*\ratio ) {$\lambda$};
    %----
    \node at (-0*\d, 2*\dratio*\d+\h) {$e$};
    \node at (-0*\d+\d, 2*\dratio*\d+\h) {$f$};
    \node at (-0*\d+2*\d , 2*\dratio*\d+\h) {$g$};
\end{tikzpicture}
=\sum_{x\1,\rho}[M^{ef;x}_{x\1;ab}]_\rho^{\mu\nu} 
\begin{tikzpicture}[baseline={(current bounding box.center)}]
\def\d{.7}
\def\h{1.0}
\def\r{0.3}
\def\ratio{0.55}
\def\dratio{0.17}
    %-----horizental line-----
    \draw(-0.3*\d,\ratio*\h)--(2.7*\d,\h*\ratio);
%------
    %\draw (0,0)  -- (\d/2,-\d/2);
    %\draw(0,0) -- (0,\h);
    %\node at (-\dratio*\d,\dratio*\d) {$a$};
    %-----------
    %\draw ( +\d, 0) -- (\d/2,-\d/2);
    %\draw(\d,0) -- (\d,\h);
    %\node at (-\dratio*\d+\d,\dratio*\d) {$b$};
    %-------
    %\draw (2*\d,0) -- (\d,-\d);
    \draw(2*\d,0)--(\r/2+\d,\r/2-\d);
    \draw(2*\d,0) -- (2*\d,\h*\ratio);
    \node at (2*\d-\dratio*\d,\dratio*\d) {$c$};
    %-----------
    %\draw(\d,-\d)--(1.4*\d ,-1.4*\d)node[midway,below]{$d$};
    %------
    \def\dotr{0.06}
    %\fill[black] (0,\h*\ratio) circle (\dotr);
    %\fill[black] (\d,\h*\ratio) circle (\dotr);
    \fill[black] (2*\d,\h*\ratio) circle (\dotr);
    %\node at (-\dratio*\d, \dratio*\d+\h*\ratio) {$\mu$};
    %\node at (-\dratio*\d+\d, \dratio*\d+\h*\ratio) {$\nu$};
    \node at (2*\d-\dratio*\d , \dratio*\d+\h*\ratio ) {$\lambda$};
    %----
    %\node at (-0*\d, 2*\dratio*\d+\h) {$e$};
    %\node at (-0*\d+\d, 2*\dratio*\d+\h) {$f$};
    %\node at (+0.1*\d+2*\d , 2*\dratio*\d+\h) {$g$};
    %------
    \def\l{.7/4};
    %----
    \draw (\d/2,-\d/2+\r)--(\r/2+\d,\r/2-\d);
    \draw (\r/2+\d,\r/2-\d)-- (\r/2+\d+\d/2,\r/2-\d-\d/2)node[midway, below]{$d$};
    %\draw (\d/2,-\d/2+\r)--(\d/2+\d,-\d/2+\r-\d);
    \node at (0.8*\d,-0.75*\d) {$x$};
    \filldraw[fill=white,draw=black](\d/2,-\d/2+\r) circle (\r);
    %-----
    \def\hh{.9}
    \draw[color=teal] (\d/2,-\d/2+2*\r)--(\d/2,\h*\ratio);%node[midway,left]{$x$};
    \node[text=teal] at (\d/2+0.2*\d,\h*\ratio-0.2*\d ) {$x$};
    \draw[color=red] (\d/2,\h*\ratio)--(\d/2,\hh);
    \node[text=red] at (\d/2+0.3*\d,\h*\ratio+0.3*\d ) {$x\1$};
    %-----
    \draw (\d/2-\r,\hh+\r) arc [start angle=180, end angle=360, radius=\r];
    %\draw(2*\d,\h*\ratio) -- (2*\d,\hh+\r);
    \draw(2*\d,\h*\ratio) -- (2*\d,\h);
    %---
    \fill[blue] (\d/2,\h*\ratio) circle (\dotr);
    \node[text=blue] at (\d/2-0.2*\d,\h*\ratio-0.2*\d ) {$\rho$};
    %----
    \node at (-0.1*\d, 2*\dratio*\d+\h) {$e$};
    \node at (+0.1*\d+\d, 2*\dratio*\d+\h) {$f$};
    \node at (0.*\d+2*\d , 2*\dratio*\d+\h) {$g$};
    %-----
    \node at (\d/2-1.4*\r,-\d/2+\r) {$a$};
    \node at (\d/2+1.4*\r,-\d/2+\r) {$b$};
\end{tikzpicture}
=\sum_{x\1,\rho}[M^{ef;x}_{x\1;ab}]_\rho^{\mu\nu} \sqrt{\frac{d_ad_b}{d_x}}
\begin{tikzpicture}[baseline={(current bounding box.center)}]
\def\d{.7}
\def\h{1.0}
\def\r{0.3}
\def\ratio{0.55}
\def\dratio{0.17}
    %-----horizental line-----
    \draw(-0.3*\d,\ratio*\h)--(2.7*\d,\h*\ratio);
%------
    %\draw (0,0)  -- (\d/2,-\d/2);
    %\draw(0,0) -- (0,\h);
    %\node at (-\dratio*\d,\dratio*\d) {$a$};
    %-----------
    %\draw ( +\d, 0) -- (\d/2,-\d/2);
    %\draw(\d,0) -- (\d,\h);
    %\node at (-\dratio*\d+\d,\dratio*\d) {$b$};
    %-------
    %\draw (2*\d,0) -- (\d,-\d);
    \draw(2*\d,0)--(\r/2+\d,\r/2-\d);
    \draw(2*\d,0) -- (2*\d,\h*\ratio);
    \node at (2*\d-\dratio*\d,\dratio*\d) {$c$};
    %-----------
    %\draw(\d,-\d)--(1.4*\d ,-1.4*\d)node[midway,below]{$d$};
    %------
    \def\dotr{0.06}
    \fill[black] (2*\d,\h*\ratio) circle (\dotr);
    \node at (2*\d-\dratio*\d , \dratio*\d+\h*\ratio ) {$\lambda$};
    %------
    \def\l{.7/4};
    %----
    \draw (\d/2,-\d/2+\r)--(\r/2+\d,\r/2-\d);
    \draw (\r/2+\d,\r/2-\d)-- (\r/2+\d+\d/2,\r/2-\d-\d/2)node[midway, below]{$d$};
    %\draw (\d/2,-\d/2+\r)--(\d/2+\d,-\d/2+\r-\d);
    %\node at (0.8*\d,-0.75*\d) {$x$};
    %\filldraw[fill=white,draw=black](\d/2,-\d/2+\r) circle (\r);
    %-----
    \def\hh{.9}
    %\draw[color=teal] (\d/2,-\d/2+2*\r)--(\d/2,\h*\ratio);%node[midway,left]{$x$};
    \draw[color=black] (\d/2,-\d/2+\r)--(\d/2,\h*\ratio);
    %\node[text=teal] at (\d/2+0.2*\d,\h*\ratio-0.2*\d ) {$x$};
    \draw[color=red] (\d/2,\h*\ratio)--(\d/2,\hh);
    \node[text=red] at (\d/2+0.3*\d,\h*\ratio+0.3*\d ) {$x\1$};
    %-----
    \draw (\d/2-\r,\hh+\r) arc [start angle=180, end angle=360, radius=\r];
    %\draw(2*\d,\h*\ratio) -- (2*\d,\hh+\r);
    \draw(2*\d,\h*\ratio) -- (2*\d,\h);
    %---
    \fill[blue] (\d/2,\h*\ratio) circle (\dotr);
    \node[text=blue] at (\d/2-0.2*\d,\h*\ratio-0.2*\d ) {$\rho$};
    %----
    \node at (-0.1*\d, 2*\dratio*\d+\h) {$e$};
    \node at (+0.1*\d+\d, 2*\dratio*\d+\h) {$f$};
    \node at (0.*\d+2*\d , 2*\dratio*\d+\h) {$g$};
    \node at (0.5*\d+\dratio*\d,\dratio*\d) {$x$};
    %-----
   % \node at (\d/2-1.4*\r,-\d/2+\r) {$a$};
    %\node at (\d/2+1.4*\r,-\d/2+\r) {$b$};
\end{tikzpicture}
\nnb\\
=&\sum_{x\1,\rho}[M^{ef;x}_{x\1;ab}]_\rho^{\mu\nu} \sqrt{\frac{d_ad_b}{d_x}} 
\sum_{h,\alpha}[M^{x\1 g; d}_{h;xc}]^{\rho\lambda}_\alpha
\begin{tikzpicture}[baseline={(current bounding box.center)}]
\def\d{.7}
\def\r{.3}
\def\ratio{1.5}
\def\dotr{0.06}
   \draw (0,0)  -- (\d/2,-\d/2);
    \node at (0,0.2*\d) {$e$};
    %-----------
    \draw ( +\d, 0) -- (\d/2,-\d/2);
    \node at (\d,0.2*\d) {$f$};
    %-------
    \draw (2*\d,0) -- (\d,-\d);
    \node at (2*\d,0.2*\d) {$g$};
    %----
    \draw (\d/2,-\d/2)--(\d,-\d);
    \node at (0.5*\d,-1*\d) {$x\1$};
    %-----------
    \draw[color=red](\d,-\d)--(\d ,-\ratio*\d)node[midway,right]{$h$};
    %-----horizental line-----
    \draw(-0.5*\d,-\ratio*\d)--(2.5*\d,-\ratio*\d);
    %---------
    \draw[color=teal](\d ,-\ratio*\d)--(\d,-\ratio*\d-0.5*\d)node[midway,left]{$d$};
    \filldraw[fill=white,draw=black](\d,-\ratio*\d-0.5*\d-\r) circle (\r);
    \draw(\d+\r,-\ratio*\d-0.5*\d-\r)--(1.7*\d+\r,-\ratio*\d-0.5*\d-\r)node[midway,below]{$d$};
    \node at (\d -1.2*\r,  -\ratio*\d-0.5*\d-\r -1.2*\r)  {$x$};
    \node at (\d +1.1*\r,  -\ratio*\d-0.5*\d-\r +1.1*\r)  {$c$};
    %-------
    \fill[blue] (\d ,-\ratio*\d) circle (\dotr);
    \node[text=blue] at (\d-0.2*\d ,-\ratio*\d+0.2*\d) {$\alpha$};
\end{tikzpicture}
=
\sum_{x\1,\rho}[M^{ef;x}_{x\1;ab}]_\rho^{\mu\nu}  
\sum_{h,\alpha}[M^{x\1 g; d}_{h;xc}]^{\rho\lambda}_\alpha
\sqrt{\frac{d_ad_bd_c}{d_d}}
\begin{tikzpicture}[baseline={(current bounding box.center)}]
\def\d{.7}
\def\r{.3}
\def\ratio{1.5}
\def\dotr{0.06}
   \draw (0,0)  -- (\d/2,-\d/2);
    \node at (0,0.2*\d) {$e$};
    %-----------
    \draw ( +\d, 0) -- (\d/2,-\d/2);
    \node at (\d,0.2*\d) {$f$};
    %-------
    \draw (2*\d,0) -- (\d,-\d);
    \node at (2*\d,0.2*\d) {$g$};
    %----
    \draw (\d/2,-\d/2)--(\d,-\d);
    \node at (0.5*\d,-1*\d) {$x\1$};
    %-----------
    \draw[color=black](\d,-\d)--(\d ,-\ratio*\d)node[midway,right]{$h$};
    %-----horizental line-----
    \draw(-0.5*\d,-\ratio*\d)--(2.5*\d,-\ratio*\d);
    %---------
    \draw[color=black](\d ,-\ratio*\d)--(\d,-\ratio*\d-0.7*\d)node[midway,left]{$d$};
    \fill[black] (\d ,-\ratio*\d) circle (\dotr);
    \node[text=black] at (\d+0.2*\d ,-\ratio*\d-0.2*\d) {$\alpha$};
\end{tikzpicture}. \label{eq_fusion_tree_tunnelign_process}
\end{align}
The root and the leaves of a fusion tree are fixed for a TQC through braiding.
Hence, we arrive at the following theorem.
\begin{theorem}\label{theorem_Uxpx}
Let $\{a,b,c,d,e,f,g,h\}$ be fixed anyons.
A TQC of logical qudits encoded by a three-anyon-fusion diagram
 as shown in Eq.~\ref{ketx_fusion_tree} preserves logical qudit under phase transition iff 
\begin{align}
    U_{x\1 x}=\sum_{\rho,\alpha}[M^{ef;x}_{x\1;ab}]_\rho^{\mu\nu}  
[M^{x\1 g; d}_{h;xc}]^{\rho\lambda}_\alpha \label{eq_Uxpx}
\end{align}
is unitary up to a normalization constant.  In general, for qudits encoded by fusion diagrams of $n$ anyons, there will be $n$ 6j-symbols appearing in the matrix. The derivation is similar. 
\end{theorem} 
It's noteworthy that the phase transition itself is a quantum gate  in the sense of Eq.~\ref{eq_Uxpx}. 
%----
\subsection{Topological quantum computation with both temporal and spatial phase transitions}
We examine the mixed scenario depicted in Fig.\ref{fig_phase_tree} and Fig.\ref{General_quantum_computation_with_phases}. In this scenario, logical qudits are represented by fusion trees situated on different phases (an extension of gapped boundaries). As time $t$ advances, these phases might vary. For the sake of clarity, we currently exclude phase fusion, splitting, creation and annihilation from our consideration. This means that each subphase $\mathcal{P}_i^t$ at time $t$ corresponds to a subphase $\mathcal{P}_i^{t+1}$ at the subsequent time step.

The first step is to consider quantum computation within each Floquet phase. 
This confines the anyon mappings within each phase $\mathcal{P}_i$ exclusively to automorphisms, simplifying the scenario.
 Every anyon $a\in \mathcal{P}_i^t$ is mapped to $\varphi^{t,t+1}_i(a)$. This retains the modular data $S^{ab}$, $T^{ab}$, and the fusion rules $N_{a,b}^c$. As this is an automorphism of the fusion basis, the logical information remains conserved.

However, only considering transitions of isomorphic phases will not harness the full power of phase-transition-assisted TQC.
We first focus on level one trees, where the subphases represent the trivial topological order. An elementary example involves two punctures linked by a ribbon. Assume the background or parent phase is denoted by $\mathcal{P}^0$. Let $\varphi$ symbolize the anyon-tunneling map transitioning from $\mathcal{P}^{t=0}$ to $\mathcal{P}^{t=1}$. Define $S^0=\{x\}$ as the set of anyons constituting the qudit of interest. Intrinsically, the anyon set $\{x\}$ is required to meet the condensation criteria, being contained within both condensable algebras $\mathcal{A}^0_l$ and $\mathcal{A}_r^0$ at time $t=0$. For the logical qudit to remain intact, it's both necessary and sufficient for $\varphi(x)$ to be a unique anyon in $\mathcal{P}^1$, with ${\varphi(x)}$ being a subset of $\mathcal{A}_l^1\cap \mathcal{A}_r^1$.
\begin{align}
\varphi
\bigg|
\begin{tikzpicture}[baseline=-0.5ex]
\def\d{1.3}
\def\r{.3}
   \draw (0,0)  -- (\d,0);
      \filldraw[draw=black, fill=white] (0,0) circle (\r);
      \filldraw[draw=black, fill=white] (\d,0) circle (\r);
      \node[text=black] at (\d*0.5,0.1*\d) {$x$};
      \node[text=black] at (0,0) {$\mathcal{A}_l^0$};
      \node[text=black] at (\d,0) {$\mathcal{A}_r^0$};
\end{tikzpicture}
\bigg\rangle
=
\bigg|
\begin{tikzpicture}[baseline=-0.5ex]
\def\d{1.5}
\def\r{.3}
   \draw (0,0)  -- (\d,0);
      \filldraw[draw=black, fill=white] (0,0) circle (\r);
      \filldraw[draw=black, fill=white] (\d,0) circle (\r);
      \node[text=black] at (\d*0.5,0.15*\d) {$\varphi(x)$};
      \node[text=black] at (0,0) {$\mathcal{A}_l^1$};
      \node[text=black] at (\d,0) {$\mathcal{A}_r^1$};
\end{tikzpicture}
\bigg\rangle
\end{align}
%---------------
The simple structure of the formula stems from the simplicity of the fusion tree. 
In a general context, one would expect intricate coefficients and linear combinations.
Next, let's explore a level one phase tree constructed from a parent phase $\mathcal{P}^0$ accompanied by four gapped punctures. 
Let the qudit of TQC be 
\begin{align}
\ket{x}:=
&\begin{tikzpicture}[baseline={(current bounding box.center)}]
\def\d{.7}
\def\r{0.2}
    \draw (0,0)  -- (\d/2,-\d/2);
    \node at (0.1*\d,-0.33*\d) {$a$};
    %-----------
    \draw ( +\d, 0) -- (\d/2,-\d/2);
    \node at (\d-0.1*\d,-0.35*\d) {$b$};
    %-------
    \draw (2*\d,0) -- (\d,-\d);
    \node at (1.6*\d,-0.65*\d) {$c$};
    %----
    \draw (\d/2,-\d/2)--(\d,-\d);
   \node at (0.62*\d,-0.88*\d) {$x$};
    %-----------
    \draw(\d,-\d)--(1.4*\d ,-1.4*\d);
    \node at (1.1*\d,-1.37*\d) {$d$};
    %------
    \filldraw[draw=black, fill=white] (-\r*0.707,\r*0.707) circle (\r);
      \filldraw[draw=black, fill=white] (\d+0.707*\r,\r*0.707) circle (\r);
    \filldraw[draw=black, fill=white] (2*\d+0.707*\r,\r*0.707) circle (\r);
    \filldraw[draw=black, fill=white] (1.4*\d+0.707*\r,-1.4*\d-\r*0.707) circle (\r);
\end{tikzpicture} 
\label{eq_x_with_boundary}
\end{align}
for fixed $a,b,c$ and $d$ in the condensable algebra at each corresponding puncture. 
Consider the phase transition $\varphi$ that takes $\mathcal{P}^0$ to $\mathcal{P}^1$. Let $\varphi(a)=e$, $\varphi(b)=f$, $\varphi(c)=g$, and $\varphi(d)=h$ be fixed in the subsequent phase, with each contained their respective condensable algebra. Therefore, the qudit we select in the subsequent phase is represented by $Span\{\ket{\varphi(x)}:x\in S_0\}$, with every branch label substituted by its corresponding mapped counterpart. 
Throughout the interval 
$0\leq t\leq 1$, a variety of TQC operations (such as braiding, wrapping, and tunneling) can be applied to this qudit, which can be resolved back to linear combinations of the standard basis of the tree diagrams through 
 basic ribbon moves.
Due to the  linearity of  ribbon calculus, 
the superposition coefficients of the diagrams are carried over from $t=0$ to $t=1$. 
Consequently, it suffices to focus solely on the phase transition of the basis tree diagrams given by Eq.~\ref{eq_x_with_boundary}, irrespective of the specific operations executed during the interval $0\leq t\leq 1$.
Theorem~\ref{theorem_Uxpx} suggests that the phase transition $\varphi:\mathcal{P}^{0}\to \mathcal{P}^{1}$ retains logical information if and only if $U_{\varphi(x)x}$, as defined in Eq.\ref{eq_Uxpx}, is unitary up to  a normalization factor. Though this example gets more complexity compared to the previous one, it still only mirrors our discourse in Sect.\ref{temporal_phase_transition_I}, since  we're focusing solely on a level-one tree.

To evaluate the phase transition for a tree of a higher level, one might adopt an inductive approach, progressing from one level to the next. At the first level, the primary deviation from the depiction in Eq.\ref{eq_x_with_boundary} is that the anyon $d\equiv d_-$ isn't attached to a gapped puncture. Instead, it tunnels from the current phase $\mathcal{P}^t_0$ to the external phase $\mathcal{P}^t_1$. As a result, the ensuing state will encompass a summation over $h=\varphi(d)$, as outlined in Eq.\ref{eq_fusion_tree_tunnelign_process}. When moving beyond $\mathcal{P}_0^t$, we transition to the tree's higher level, where the initial phase $\mathcal{P}_0^t$ becomes a subphase within $\mathcal{P}_1^t$.
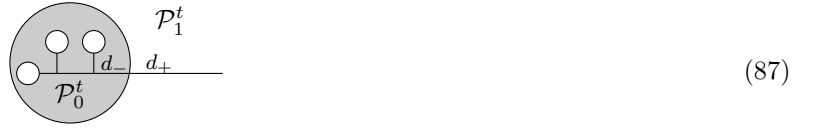
\begin{align}
\begin{tikzpicture}[baseline=-0.5ex]
\def\d{.7}\def\r{0.4}  
\def\delta{0.15} 
\filldraw[draw=black, fill=gray, fill opacity=0.4] (0.1*\d,0.2*\d) circle (2*\r); 
\node at (0.1*\d,-0.4*\d) {$\mathcal{P}_0^t$};
\node at (2*\d,1*\d) {$\mathcal{P}_1^t$};
\node[font=\footnotesize] at (0.95*\d,0.15*\d) {$d_-$};
\node[font=\footnotesize] at (1.8*\d,0.17*\d) {$d_+$};
%----
\draw (-0.7*\d ,0)  -- (3*\d,0);   
\filldraw[draw=black, fill=white] (-0.7*\d ,0) circle (\delta);
\draw (-0.15*\d ,0)  -- (-0.15*\d,0.6*\d);
\filldraw[draw=black, fill=white] (-0.15*\d,0.6*\d) circle (\delta);
\draw (0.55*\d ,0)  -- (0.55*\d,0.6*\d);
\filldraw[draw=black, fill=white] (0.55*\d,0.6*\d) circle (\delta);
\end{tikzpicture} 
\end{align}
Let the anyon-tunneling map that corresponds to the spatial domain wall between $\mathcal{P}^t_0$ and $\mathcal{P}^t_1$ be denoted by $\theta^t$. Hence, $d_+\in \theta^{t=0}(d_-)$. It's crucial to recognize that there are two temporal anyon mappings and two spatial anyon mappings involved. The compatibility condition is given by:
\begin{align}
   \theta^{1} (\varphi_{\mathcal{P}_0}(d_-)) \cap  \varphi_{\mathcal{P}_1}(\theta^{0}(d_-))\neq \emptyset.
\end{align}
In $\mathcal{P}_1$, the evaluation of the phase transition of the tree diagrams in the qudit's standard basis follows the same procedure as in $\mathcal{P}_0$. We persist in advancing to higher tree levels in this manner until the mapping of the entire tree diagram is finalized.

The final intricacy to address involves 
operations on subphases themvelves
during a phase transition (creation, removal, fusing, splitting). This means that tree diagrams at time $t=1$ can differ starkly from those at time $t=0$.
Let $\mathcal{T}^t={(T, a_{j,-}^T, a^T_{j,+}, \mu^T_j)_t}$ represent the set of standard basis of tree diagrams with anyon labels and channel labels at time $t$. These labeled trees contain the qudits under consideration.
For the preservation of logical information, it's both necessary and sufficient that the phase transition from $\mathcal{T}^0$ to $\mathcal{T}^1$, when confined to the logical subphase give rise to a unitary (refer to Eq.~\ref{eq_fusion_tree_tunnelign_process} for a straightforward example).
However, due to constraints on our time, we won't delve into comprehensive examples and instead defer such exploration to future endeavors.

Thus, as is evident, by orchestrating diverse phases, we can facilitate a range of topological gates. This expands the repertoire of available gates beyond the typical scope of traditional topological quantum computations, which predominantly rely on a singular phase.
%---
%--------
\section{Conclusions}
In this work, we have considered the two-point excitations of deconfined anyons in subphases of quantum double models, employing character theory. We have presented an explicit formula for computing anyon-tunneling maps across domain walls.

Interestingly, the model can replicate the Floquet codes that have been widely studied by recent literature. 
While this approach sacrifices the low-weight engineering advantage, it generalizes Floquet codes to nonabelian quantum doubles and correctly predicts potential automorphisms between subphases of a parent quantum double. 

In the last chapter, we examined phases of general modular tensor categories and detailed the conditions for preserving logical qudits in fusion bases during phase transitions.
Our discussion also illuminates how introducing various phases, both temporally and spatially, could offer a more diverse choice of quantum gates. This has potential implications for devising more efficient circuits than previously considered. 
%--------
\section*{Acknowledgements}
We would like to express my profound gratitude to Xiao-Gang Wen, Liyuan Chen, Margarita Davydova, Zhenghan Wang, and Markus Kesselring for their invaluable discussions and insights. Their expertise and contributions have significantly enriched this work. Additionally, I wish to extend my appreciation to the Physics Department of MIT for their unwavering support throughout this endeavor.
%------
%------------
\appendix
%--------------
%---------------
%------
\section{Mapping between bulk and the hole}\label{sec_mapping_between_bulk_and_the_hole}
%------------------
%-----
\begin{remark}\label{proof_of_proposition_list_of_AgN_BgN_properties}
(Proof of Proposition~\ref{proposition_list_of_AgN_BgN_properties}).
We list some  properties of $A^{gN}$ and $B^{gN}$ located at the same site.
\begin{itemize}
\item[1.] 
 $A^{gN}A^{g\1 N}=\frac{1}{|N|^2}\sum_{n,n\1\in N} A^{gn}A^{g\1n\1}=\frac{1}{|N|^2}\sum_{n,n\1\in N} A^{gng\1n\1}=\frac{1}{|N|^2}\sum_{n,n\1\in N} A^{gg\1 n_{g\1}n\1}=\frac{1}{|N|}\sum_{n}A^{gg\1 n}=A^{gg\1 N}$, where we have defined $n_\ell \in N$ s.t. $n \ell =\ell n_\ell$.
\item[2.] 
$B^{gN}B^{g\1 N}=\sum_{n,n\1\in N} \delta_{gn,g\1 n\1}B^{gn}=\sum_{n \in N}\delta_{(g\1)^{-1}gn\in N}B^{gn}=\delta_{gN,g\1 N}B^{gN}$.
\item[3.] 
\begin{align}
B^{gN}A^{hN}=&\frac{1}{|N|}\sum_{n,n\1\in N} B^{gn}A^{hn\1}=\frac{1}{|N|}\sum_{n,n\1\in N}A^{hn\1}B^{(hn\1)^{-1} gn (hn\1)}=\frac{1}{|N|}\sum_{n,n\1\in N}A^{hn\1}B^{(n\1)^{-1}h^{-1} gn hn\1}\nnb\\
=&\frac{1}{|N|}\sum_{n,n\1\in N}A^{hn\1}B^{(n\1)^{-1}h^{-1} gh  n_g n\1} 
=\frac{1}{|N|}\sum_{n\1\in N}\sum_{n_g\in N}A^{gn\1}B^{h^{-1} gh \ [(n\1)^{-1}]_{h^{-1} gh} n_h n\1}\\
=&\frac{1}{|N|}\sum_{n\1\in N}\sum_{n_h\in N}A^{hn\1}B^{h^{-1} gh n_h}=A^{hN}B^{h^{-1}ghN}
\end{align}
where we have again applied the definition "$n_\ell$" defined above for $n\mapsto [(n\1)^{-1}] $ in $N$ and $\ell\mapsto h^{-1} gh$, and we have used the fact that $n_h\in N$ for fixed $h$, $g$, as well as $n\1$, is a dummy index.
\item[4.] $
(A^{h_1 N}B^{g_1N})(A^{h_2 N}B^{g_2N})= A^{h_1 N}A^{h_2 N}B^{h_2^{-1}g_1h_2N} B^{g_2N}=A^{h_1 h_2 N} B^{g_2 N}\delta_{h_2^{-1} g_1 h_2N, g_2 N}$.
%See the footnote\footnote{
%\begin{align}(A^{h_1}_{s_0}B^{g_1}_{s_0}) (A^{h_2}_{s_0}B^{g_2}_{s_0})=A^{h_1}_{s_0} A^{h_2}_{s_0}B^{h_2^{-1}g_1h_2}_{s_0} B^{g_2}_{s_0}=\delta_{h_2^{-1}g_1h_2, g_2}A^{h_1h_2}_{s_0} B^{g_2}_{s_0} \end{align}}.
\end{itemize}
\end{remark}
%-----
%----
%------
\begin{remark}\label{proof_of_proposition_properties_of_G_gamma}
(Proof of Proposition~\ref{proposition_properties_of_G_gamma}).
Using Eq.~(B15) of \cite{bombin:2008} 
\begin{align}
    F^{h_1,g_1}_\gamma F^{h_2,g_2}_\gamma=F^{h_1h_2,g_1}\delta_{g_1,g_2},
\end{align}
We have
\begin{align}
G^{h_1,g_1}_\gamma G^{h_2,g_2}_\gamma
 =&\sum_{n\1_1,n\1_2,n_1,n_2} F_\gamma^{\overline{n\1}_1 h_1n_1\1,n\1_1 g_1 n_1 }
 F_\gamma^{\overline{n\1}_2 h_2n_2\1,n\1_2 g_2 n_2 } 
 \delta_{n_1,n_2}\delta_{g_1,g_2} \delta_{n\1_1,n\1_2}\nnb\\
 =&\sum_{n_1,n_1\1 } F_\gamma^{\overline{n\1}_1 h_1h_2 n_1\1,n\1_1 g_1 n_1 } \delta_{g_1,g_2}
 = G^{h_1h_2,g_1}\delta_{N\1 g_1N,N\1 g_2N}
\end{align}
%Since the equivalence class by $\sim_{N,N\1}$ partitions $G\times G$, the result of $G^{h_1,g_1}_\gamma G^{h_2,g_2}_\gamma$ 
The proof for the second part is straightforward.  
\begin{align}
(G_\gamma^{h,g})^\dagger= (\sum_{n,n\1} F_\gamma^{\overline{n\1} h n\1,\overline{n\1}gn})^\dagger 
=  (\sum_{n,n\1} F_\gamma^{\overline{n\1} \overline{h} n\1,\overline{n\1}gn})=G_\gamma^{h^{-1},g}
\end{align}
\end{remark}
%-----
%-------------
\begin{remark}\label{proof_of_lemma_expval_of_G}
(Proof of Lemma~\ref{lemma_expval_of_G}).
First, we insert $B^N_{s_0}$ located at the starting point of $\xi$.
\begin{align}
\bra{\Omega}     G^{h,g}_\gamma \ket{\Omega}
=&\bra{\Omega}  B_{\partial_1\gamma}^{N}  G^{h,g}_\gamma \ket{\Omega}
=\bra{\Omega}  G^{h,g}_\gamma  B_{\partial_1\gamma}^{ \overline{g}h gN} \ket{\Omega}
=\delta_{ h N, N}\bra{\Omega} G^{h,g}_\gamma\ket{\Omega}\\
=&\bra{\Omega}  B_{\partial_0\gamma}^{N\1}  G^{h,g}_\gamma \ket{\Omega}
=\bra{\Omega}  G^{h,g}_\gamma  B_{\partial_0\gamma}^{ \overline{h} N\1} \ket{\Omega}
=\delta_{ h N\1, N\1}\bra{\Omega} G^{h,g}_\gamma\ket{\Omega}
\end{align}
Hence $\bra{\Omega}G^{h,g}_\gamma\ket{\Omega}=\delta_{h,e}\bra{\Omega}G^{e,g}\ket{\Omega}$, where we have assumed $N\cap N\1=\{e\}$ (see the comments below Lemma~\ref{lemma_quotient_by_Ncap}). 
Next we insert $A^{\ell N}_{s_0}$, which acts as an identity on $\ket{\Omega}$ (to see this, simply notice the fact that $A^{\ell N}_{s_0}A^M_{s_0}=A^M_{s_0}$ act on $\ket{\Omega}$ as an identity). 
\begin{align}
    \langle \Omega | G^{e, g} |\Omega\rangle=&
    \langle \Omega |A^{\ell\1 N\1}_{\partial_0 \gamma} G^{e, g} |\Omega\rangle= 
    \langle \Omega | G^{\ell\1 e \overline{\ell\1}, \ell\1 g}A^{\ell\1 N\1 }_{\partial_0\gamma} |\Omega\rangle
     = \langle \Omega | G^{e, \ell\1 g} |\Omega\rangle \quad \forall \ell\1 \in M\1\\
    %------------------
     \langle \Omega | G^{e, g} |\Omega\rangle=&
     \langle \Omega |A^{\ell N}_{\partial_1\gamma} G^{e, g} |\Omega\rangle=
      \langle \Omega | G^{e, g\ell^{-1}}A^{\ell N}_{\partial_1 \gamma} |\Omega\rangle   
      =\langle \Omega | G^{e, g\overline{\ell}}|\Omega\rangle, \quad \forall \ell \in M
\end{align}
The implies that $C_{M\1gM}:=\bra{\Omega}G^{h,g}\ket{\Omega}$ is a constant that depends only on the double closet $M\1 gM$. 
\end{remark}
%----
\begin{remark}\label{calculation_of_chi_of_phase_tunneling_map}
(Proof of Theorem~\ref{theorem_chi_of_phase_tunneling_map}). Let $h,g\in Q_S(\sim_{N\1, N})$ be the representatives of the equivalence classes governed by  the relation $\sim_{N\1, N}$ specifically restricted to $S=\{(h,g): h\in M\1\cap gM\overline{g},g\in G\}$.
If both $h_1N=h_2N$ and  $h_1N\1=h_2N\1$ are satisfied, we express the equivalence as $h_1\sim_{N\&N\1} h_2$ or alternatively as  $h_1\sim_{N_\cap}h_2$.  If one has executed the quotient mapping  with$G\to G/N_\cap, N\to N/N_\cap$, and $N\1\to N\1/N_\cap$,  the relation $h_1\sim_{N\&N\1}h_2$  reduces to the straightforward $h_1=h_2$ in $G$, rendering the  $\sim_{N\&N\1}$ relation irrelevant.  Nonetheless, for ease of reference, we initiate our discussion without applying the quotient operations.
\begin{align}
    \chi_{\mathcal{G}}(h_2g_2^*,h_1g_1^*)=& \sum_{h,g \in Q_S}\langle \psi^{h, g}| h_2g_2^*\otimes h_1g_1^* |\psi^{h, g}\rangle\nnb\\
    =& \sum_{h,g}\langle \psi^{ h, g}| (A^{h_2N\1 
 }_{\partial_0\gamma}B^{g_2N\1}_{\partial_0\gamma}) (A^{h_1N}_{\partial_1\gamma}B^{g_1N}_{\partial_1\gamma}) |\psi^{h, g}\rangle  \nnb \\
    =&\sum_{h,g } \sum_{gN} \delta_{g^{-1} \overline{h} g g_1N,N} \delta_{N\1,g_2hN\1}\langle \psi^{h,gN}|\psi^{ h_2hh_2^{-1},h_2gh_1^{-1}N}\rangle  \nnb \\
    =&\sum_{h,g} \delta_{  g_1N,\overline{g}hgN} \delta_{g_2hN\1,N\1}
    \ \delta_{\overline{g}hgN, h_1\overline{g} hg\overline{h_1}N }  \delta_{hN\1,h_2hh_2^{-1}N\1 }   \delta_{gNN\1 , h_2gh_1^{-1} NN\1}\nnb\\
      =&\delta_{h_2g_2N\1,g_2h_2N\1 } \delta_{g_1h_1N,h_1g_1N}
\sum_{h,g} \delta_{  g_1N,\overline{g}hgN}\delta_{g_2hN\1,N\1}
    \ \delta_{\overline{g}hgN, h_1\overline{g} hg\overline{h_1}N }
 \delta_{hN\1,h_2hh_2^{-1}N\1 }   \delta_{gNN\1 , h_2gh_1^{-1} NN\1}\nnb\\
      =&\delta_{h_2g_2N\1,g_2h_2N\1 } \delta_{g_1h_1N,h_1g_1N}
\sum_{h,g\in Q_S} \delta_{  g_1N,\overline{g}hgN}\delta_{g_2hN\1,N\1}
       \delta_{NgN\1 , N h_2gh_1^{-1} N\1}\\
    =&(|N\1|\ |N|)^{-1} \delta_{h_2g_2N\1,g_2h_2N\1 } \delta_{g_1h_1N,h_1g_1N}
\sum_{(h,g)\in S} \delta_{  g_1N,\overline{g}hgN}\delta_{g_2hN\1,N\1}
       \delta_{NgN\1 , N h_2gh_1^{-1} N\1},
\end{align}
where the factor $|N|\ |N\1|$ count the redundancy resulted from the relation $\sim_{N,N\1}$. 
In the above calculation of the inner product $\langle \psi^{h,gN}|\psi^{ h_2hh_2^{-1},h_2gh_1^{-1}N}\rangle $, the second parameter $h_2gh_1^{-1}$ may not be the representative of double coset $N\1 h_2gh_1^{-1} N$ in $Q(\sim_{N,N\1})$.  Suppose the representative instead is 
$n\1 h_2gh_1^{-1} n$, where $n\1$ and $n$ are the gauge factors. The first component of the representative in $Q(\sim_{N,N\1})$ is $n\1 h_2 h h_2^{-1} \overline{n\1}$. In the equivalence classes of relation $\sim_{N\&N\1}$, we need the two coset relations $\sim_N$ and $\sim_{N\1}$ to characterize such an element, and $n\1 h_2 h h_2^{-1} \overline{n\1} N\1= h_2 h h_2^{-1} N\1$. However, to describe the class by $\sim_{N}$, the term depends on $n\1\in N\1$. We use $\overline{g}hg$ as a representative to avoid such trouble, which gives $(h_2 g\overline{h_1})^{-1}h_2 hh_2^{-1}(h_2 g\overline{h_1})= h_1 \overline{g} h g \overline{h_1}$, and indeed the class 
$h_1 \overline{g} h g \overline{h_1}N=\overline{n}h_1 \overline{g} h g \overline{h_1}nN$ does not depend on the gauge factors $n$ and $n\1$. 
\end{remark}
%---------
\subsection{Example: Quantum double of \texorpdfstring{$\mathbb{Z}_2\times \widetilde{\mathbb{Z}}_2$}{Lg}}
In this section we consider anyon tunneling between $\mathcal{P}_1=\mathcal{D}(M/N)$ and $\mathcal{P}_2=\mathcal{P}_0=\mathcal{D}(G)$ for $G=\mathbb{Z}_2\times \mathbb{Z}_2$. 
We label the 2nd copy of $\ZZ_2$ group with a tilde above to distinguish them. In the case of $\mathbb{Z}_2\times \widetilde{\mathbb{Z}}_2$, 
\begin{align}
\chi_{\mathcal{G}}(h_1g_1^*,h_2g_2^*)=\frac{|G|}{|N|} \delta_{h_1,g_1\in M} \delta_{g_1\sim g_2}\delta_{h_1\sim h_2}
\end{align}
On the other hand, since the abelian $G$ is a product group,
\begin{align}
    \chi_{(a, R)}(hg^*)=\delta_{g_1,a_1}\Tr_{\pi_1}(h_1)\ \delta_{g_2,a_2}\tr_{\pi_2}(h_2),\quad  \pi_1,\pi_2\in Irr(\mathbb{Z}_2)
\end{align}
\begin{itemize}
    \item
Take $M=\ZZ_2\times \wt{\ZZ}_2$, $N=\wt{\mathbb{Z}}_2$ for example, 
\begin{align}
\chi_{\mathcal{G}}((h,\widetilde{h}) (g,\widetilde{g})^*,ij^*)
    =2 \delta_{ g  , j} 
    \delta_{ h, i },\quad \forall (h,\widetilde{h}), (g,\widetilde{g})\in G,\quad i,j\in Q\equiv M/N= \mathbb{Z}_2
\end{align}
For the quantum double of $\ZZ_2$, we list the excitations of here.
\begin{table}[htb]
\centering
\begin{tabular}{|c|c|c|c|c|} \hline
 & $\mathfrak{1}$ &e  &m   & f\\ \hline
$(a,R)$ & $(1,1)$  & $(1 ,Sgn)$  & $( -1,1)$ & $(-1,Sgn)$\\ \hline 
$\chi(hg^*)$& $\delta_{g,1}$ & $\delta_{g,1}Sgn(h)$ & $\delta_{g,-1}$ & $\delta_{g,-1}Sgn(h) $\\ \hline
\end{tabular}
\caption{We represent the elements of $\mathbb{Z}_2$ as $\{1,-1\}$. The anyon classifications within $\mathcal{D}(\ZZ_2)$ are provided, with each anyon uniquely identified by an element $a\in G$ and an irreducible representation $R$ of $G$. 
Here $\mathbf{1}$ denotes vacuum excitation.
The notation '1' can refer to either the unit element in 
$\ZZ_2$  or the trivial representation, while '$-1$' serves as the generator of $\mathbb{Z}_2$.  Additionally, '$Sgn$' signifies the sign representation of $\mathbb{Z}_2$}
\end{table}
Note that
\begin{align}\sum_{(a,\pi)\in \{\mathfrak{1},m\}} \chi(hg^*)=\delta_{g,1}+\delta_{g,-1}=1.
\end{align}
Compute
\begin{align}
&\chi(\mathfrak{1})\chi(\mathfrak{1}_{\mathcal{P}_1})+\chi(e)\chi(e_{\mathcal{P}_1})+\chi(m)\chi(m_{\mathcal{P}_1})+\chi(f)\chi(f_{\mathcal{P}_1}) \nnb\\
=&\delta_{g,1}\delta_{j,1}+\delta_{g,1}Sgn(h)\delta_{j,1} Sgn(i)+\delta_{g,-1}\delta_{j,-1}+\delta_{g,-1}Sgn(h)\delta_{j,-1}Sgn(i)\nnb\\
=&(1+Sgn(h)Sgn(i))\delta_{g,1}\delta_{j,1}+(1+Sgn(h)Sgn(i))\delta_{g,-1}\delta_{j,-1}\nnb\\
=&2\delta_{h,i}\delta_{g,1}\delta_{j,1}+2\delta_{h,i}\delta_{g,-1}\delta_{j,-1}\nnb\\
=&2\delta_{g,j}\delta_{h,i}=\chi_{\mathcal{G}}
\end{align}
The following anyon tunneling can be read off:
\begin{align}
       \mathbf{1}( \wt{\mathbf{1}}+ \wt{m})_{\mathcal{P}_0}\to  \mathbf{1}_{\mathcal{P}_1},  \quad 
    e( \wt{\mathbf{1}}+ \wt{m})_{\mathcal{P}_0}\to e_{\mathcal{P}_1},\quad 
        m( \wt{\mathbf{1}}+ \wt{m})_{\mathcal{P}_0}\to  m_{\mathcal{P}_1},\quad     
        f( \wt{\mathbf{1}}+ \wt{m})_{\mathcal{P}_0}\to f_{\mathcal{P}_1}
\end{align}
%-------
\item Next we consider the case where $M=\mathbb{Z}_2$ and $N=\{e\}$ is the trivial group. As we will see, this corresponds to the condensation of $\mathbf{1}+ \wt{e}$.
\begin{align}  
\chi_{\mathcal{G}}((h ,\wt{h})(g,\wt{g})^* ,ij^*)= &4\delta_{g,j}\delta_{h,i}\delta_{\wt{g},\wt{1}}\delta_{\wt{h},\wt{1}},\quad i,j\in \mathbb{Z}_2.
\end{align}
To start with, note that
\begin{align}
    \chi_{\wt{\mathbf{1}}}+\chi_{\wt{e}}=\delta_{\wt{1},\wt{g}}(1+Sgn(\wt{h}))=2\delta_{\wt{g},\wt{1}}\delta_{\wt{h},\wt{1}}.
\end{align}
This constraint will restrict the group from $\ZZ_2\times \wt{\ZZ_2}$ to only $\ZZ_2$. 
Therefore 
\begin{align}
\qty(\chi(\mathfrak{1})\chi(\mathfrak{1}_{\mathcal{P}_1})+\chi(e)\chi(e_{\mathcal{P}_1})+\chi(m)\chi(m_{\mathcal{P}_1})+\chi(f)\chi(f_{\mathcal{P}_1})(hg^*,ij^*)) \ (\chi_{\mathbf{1}}+\chi_{\wt{e}})(\wt{h}\wt{g}^*)=(2\delta_{g,j}\delta_{h,i})(2\delta_{\wt{g},\wt{1}}\delta_{\wt{h},\wt{1}})=\chi_{\mathcal{G}}.\nnb
\end{align}
%-----
\item  Take $M=\mathbb{Z}_d$ and $N=\{e\}$.  This corresponds to the condensation of $\mathbf{1}+e\wt{e}$. In such a case the multiplication/fusion of anyon won't commute with taking character (unlike the previous cases). 
The character is 
\begin{align}
&(\chi(\mathbf{1})+\chi(e\wt{e}))_{\mathcal{P}_0}\chi(\mathbf{1})_{\mathcal{P}_1}
+(\chi(e)+\chi(\wt{e}))_{\mathcal{P}_0}\chi(e)_{\mathcal{P}_1}
+(\chi(mm)+\chi(f\wt{f}))_{\mathcal{P}_0}\chi(m)_{\mathcal{P}_1}
+(\chi(f\wt{m})+\chi(m\wt{f}))_{\mathcal{P}_0}\chi(f)_{\mathcal{P}_1}\nnb\\
=&\delta_{g,1}\delta_{\wt{g},\wt{1}}(1+Sgn(h)Sgn(\wt{h}))\delta_{j,1} 
+ (\delta_{g,1}Sgn(h)\delta_{\wt{g},\wt{1}} +\delta_{g,1}\delta_{\wt{g},\wt{1}}Sgn(\wt{h})) \delta_{j,1}Sgn(i)\nnb\\
&+\delta_{g,-1}\delta_{\wt{g},\wt{-1}}(1+Sgn(h)Sgn(\wt{h}))  \ \delta_{j,-1}
+ \delta_{g,-1}\delta_{\wt{g},\wt{-1}}(Sgn(h)+Sgn(\wt{h}))\ \delta_{j,-1} Sgn(i)\nnb\\
=&\delta_{g,1}\delta_{\wt{g},g}\delta_{j,g} \qty(2\delta_{h,\wt{h}}+[Sgn(h)+Sgn(\wt{h})]Sgn(i))
+ \delta_{g,-1}\delta_{g,\wt{g}}\delta_{j,g} \qty(2\delta_{h,\wt{h}} +[Sgn(h)+Sgn(\wt{h})]Sgn(i) )\nnb\\
=&\delta_{g,\wt{g}}\delta_{j,g} \qty(2\delta_{h,\wt{h}}+2\delta_{h,\wt{h}}Sgn(h) Sgn(i))\nnb\\
=&4\delta_{h,\wt{h}}\delta_{g,\wt{g}}\delta_{j,g} \delta_{h,i}
\end{align}
%-----
The character of algebra $\mathcal{G}$ is (for $i,j\in \mathbb{Z}_2^d$)
\begin{align}
    \chi_{\mathcal{G}}(h\wt{h}(g\wt{g})^*,ij^*)= &4(\delta_{h\wt{h},g\wt{g}\in \mathbb{Z}_d}) 
    \delta_{j,g\wt{g}}\delta_{i,h\wt{h}}=4 \delta_{h,\wt{h}}\delta_{g,\wt{g}}\delta_{j,g}\delta_{i,h},
\end{align}
which is the same as the decomposition above.
\end{itemize}
Other cases are similar and we will omit the calculations. 
%----------------------
\subsection{Example: Quantum double of \texorpdfstring{$S_3$}{Lg}} 
We consider $G=S_3$ for illustration. Since this is the simplest nonabelian instance, we will detail explicit computations herein to provide a concrete example.
\begin{table}[htb]
\centering
\begin{tabular}{|c|c|c|c|c|c|c|c|c|c|} \hline
Notations & A & B & C & D & E & F & G & H  \\ \hline  
$(C,R)$ & e,1 & e,Sgn & e, Std & $\overline{\sigma},1$ & $\overline{\sigma}$, Sgn & $\overline{\tau}, 1$ & $\overline{\tau},[\omega]$ & $\overline{\tau},[\omega^*]$ \\ \hline
$Z(r_C)$ & $S_3$ & $S_3$ & $S_3$ & $\mathbb{Z}_2$ & $\mathbb{Z}_2$ & $\mathbb{Z}_3$ & $\mathbb{Z}_3$ & $\mathbb{Z}_3$ \\ \hline
dim & 1 & 1 & 2& $3$ & $3$ & $2$ & $2$ & $2$ \\ \hline
\end{tabular}
\caption{Table of anyons of $\mathcal{D}(S_3)$. The symbol $C$ denotes a conjugacy class  within $G$. The element $r_C\in C$ is a representative of $C$. $R$ is an irreducible representative of the centralizer $Z(r_C)$. The term "Sgn" refers to the sign representation in $S_3$ or $\mathbb{Z}_2$. For the group $\mathbb{Z}_3$ we use the generators to signify  the associated representations. The letter $\omega$ denotes the complex number $e^{2\pi i/3}$.}\label{table_of_S3_anyons} 
\end{table}
The group elements are ordered by $\{e,\tau,\tau^2,\sigma,\sigma\tau ,\sigma\tau^2\}$ for matrix presentations.
\begin{itemize}   
\item 
We first take $M=\mathbb{Z}_2=\{e,\sigma\}$ and $N=\{e\}$ as an example. The symbols $e\equiv +1\in \mathbb{Z}_2$ $\sigma\equiv -1\in\mathbb{Z}_2$ are identified. This scenario corresponds to the condensation of $A+ C$.
We list here the characters, and group them by conjugacy classes:
\begin{align}
    \chi_A(hg^*)=&\chi_{\overline{e},1}(hg^*)= \delta_{g,e} \nnb \\
     \chi_B(hg^*)=& \chi_{\overline{e},Sgn}(hg^*)=\delta_{g,e}\Tr(h)=\delta_{g,e}Sgn(h)\nnb \\
     \chi_C(hg^*)= &\chi_{\overline{e},Std}(hg^*)=\delta_{g,e}\Tr(h)=\delta_{g,e}(2,-1,-1,0,0,0)_h =\delta_{g,e}(2\delta_{h,e}-\delta_{h,\tau}-\delta_{h,\tau^2})
\end{align}
%----
\begin{align}
        \chi_D(hg^*)=& \chi_{\overline{\sigma},1}(hg^*)=\delta_{g\in \overline{\sigma}}  \delta_{h\in \{g,e\}} \Tr_\pi(k_g^{-1} hk_g)
    =\delta_{g\in \{\sigma,\sigma\tau,\sigma\tau^2\}}  (\delta_{h,e} \Tr_\pi(e)+\delta_{h,g}\Tr_\pi(\sigma))\nnb\\
    =&\delta_{g\in \{\sigma,\sigma\tau,\sigma\tau^2\}}  \delta_{h\in\{e,g\}}  \nnb \\
    \chi_E(hg^*)=& \chi_{\overline{\sigma},Sgn}(hg^*)=\delta_{g\in \{\sigma,\sigma\tau,\sigma\tau^2\}}  (\delta_{h,e} -\delta_{h,g})
\end{align}
%------
\begin{align}
    \chi_F(hg^*)=&\chi_{\overline{\tau},1}(hg^*)=\delta_{g\in \{\tau,\tau^2\}}\delta_{h\in \{e,\tau,\tau^2\}} \Tr_\pi(k_g^{-1} hk_g)\nnb\\ 
    =&\delta_{h\in \{e,\tau,\tau^2\}} (\delta_{g,\tau} \Tr_\pi( h)+\delta_{g,\tau^2}\Tr_\pi(\sigma h\sigma^{-1}))
    =\delta_{h\in \{e,\tau,\tau^2\}} (\delta_{g,\tau} \Tr_\pi( h)+\delta_{g,\tau^2}\Tr_\pi(h^{-1})) \nnb \\
    =&\delta_{h\in \{e,\tau,\tau^2\}} \delta_{g\in \{\tau,\tau^2\}} \nnb \\
    \chi_G(hg^*)=&\chi_{\overline{\tau},[\omega]}(hg^*)
    =\delta_{h,e}(\delta_{g,\tau}+\delta_{g,\tau^2})
    +\delta_{h,\tau}(\delta_{g,\tau}\omega+\delta_{g,\tau^2}\omega^2)
    +\delta_{h,\tau^2}(\delta_{g,\tau}\omega^2+\delta_{g,\tau^2}\omega) \nnb\\
    \chi_H(hg^*)=&\chi_{\overline{\tau},[\omega]}(hg^*)
    =\delta_{h,e}(\delta_{g,\tau}+\delta_{g,\tau^2})
    +\delta_{h,\tau}(\delta_{g,\tau}\omega^2+\delta_{g,\tau^2}\omega)
    +\delta_{h,\tau^2}(\delta_{g,\tau}\omega+\delta_{g,\tau^2}\omega^2) 
\end{align}
Compute
\begin{align}
&\chi_A(h_2g_2^*)\chi_{\mathfrak{1}}(h_1g_1^*)
+\chi_B(h_2g_2^*)\chi_e(h_1g_1^*)
+\chi_D(h_2g_2^*) \chi_m(h_1g_1^*) 
+\chi_E(h_2g_2^*) \chi_f(h_1g_1^*)\nnb\\
&+\chi_C(h_2g_2^*)(\chi_{\mathfrak{1}}+\chi_e)(h_1g_1^*)
\nnb\\
=&\delta_{g_2,e}\delta_{g_1,1} 
+\delta_{g_2,e}Sgn(h_2) \ \delta_{g_1,1} Sgn(h_1)
+\delta_{g_2\in \{\sigma,\sigma\tau,\sigma\tau^2\}} \delta_{h_2\in \{e,g_2\}} \ \delta_{g_1,-1}
+\delta_{g_2\in \{\sigma,\sigma\tau,\sigma\tau^2\} } (\delta_{h_2,e}-\delta_{h_2,g_2})\ \delta_{g_1,-1}Sgn(h_1) \nnb\\
&+\delta_{g_2,e}(2\delta_{h_2,e}-\delta_{h_2,\tau}-\delta_{h_2,\tau^2}) \delta_{g_1,1}(1+Sgn(h_1))\nnb\\
=&2\delta_{g_2,e}\delta_{g_1,1}\delta_{h_2\sim_{\ZZ_3}h_1}+\delta_{g_2\in \overline{\sigma}}\delta_{g_1,-1}\qty{\delta_{h_2,e}(1+Sgn(h_1)+\delta_{h_2,g_2}(1-Sgn(h_1))} \nnb\\
&+2\delta_{g_2,e}(2\delta_{h_2,e}-\delta_{h_2,\tau}-\delta_{h_2,\tau^2}) \delta_{g_1,1}\delta_{h_1,1}  \nnb\\
=&2\delta_{g_2,e}\delta_{g_1,1}\delta_{h_2\sim_{\ZZ_3}h_1}
+2\delta_{g_2\in \overline{\sigma}}\delta_{g_1,-1}\qty{\delta_{h_2,e}\delta_{h_1,1}+\delta_{h_2,g_2}\delta_{h_1,-1}} \nnb\\
&+2\delta_{g_2,e}(2\delta_{h_2,e}-\delta_{h_2,\tau}-\delta_{h_2,\tau^2}) \delta_{g_1,1}\delta_{h_1,1}  \nnb\\
=&\delta_{g_2,e}\delta_{g_1,1}\qty{( (2+4)\delta_{h_2,e}+(2-2)\delta_{h_2,\tau}+(2-2)\delta_{h_2,\tau^2})\delta_{h_1,1}+2\delta_{h_2\in\overline{\sigma}} \delta_{h_1,-1}}\nnb\\
&+2\delta_{g_2\in \overline{\sigma}}\delta_{g_1,-1}\qty{\delta_{h_2,e}\delta_{h_1,1}+\delta_{h_2,g_2}\delta_{h_1,-1}}\nnb\\
=&\delta_{g_2,e}\delta_{g_1,1}\qty{ 6\delta_{h_2,e}\delta_{h_1,1}+2\delta_{h_2\in\overline{\sigma}} \delta_{h_1,-1}}
+2\delta_{g_2\in \overline{\sigma}}\delta_{g_1,-1}\qty{\delta_{h_2,e}\delta_{h_1,1}+\delta_{h_2,g_2}\delta_{h_1,-1}} \nnb\\ 
=&\delta_{g_2,e} \delta_{h_2,e}\sum_{g\in G} \delta_{g_1,g^{-1} \overline{g_2}g}\delta_{h_1,g^{-1} h_2 g} + \delta_{g_2,e}\delta_{h_2\in \overline{\sigma}}\sum_{g\in G}\delta_{g_1,g^{-1} \overline{g_2}g}\delta_{h_1\equiv \sigma,g^{-1} h_2 g}  \nnb\\
&{\color{gray}+
\delta_{g_2\in \overline{\tau}}
\sum_{g\in G} 
(\delta_{h_2, e}
+\delta_{h_2\in \overline{\tau}} )\delta_{g_1, g^{-1}\overline{g_2} g} \delta_{h_1, g^{-1} h_2g} } \nnb\\
&+\delta_{g_2,\overline{\sigma}}\delta_{g_1,-1\equiv \sigma} (\delta_{h_2,e}\sum_{g\in G}\delta_{g_1\equiv \sigma , g^{-1} \overline{g_2} g}\delta_{h_1,g^{-1}e g}+\delta_{h_2,g_2}\sum_{g\in G}\delta_{g_1\equiv \sigma, g^{-1} \overline{g_2} g}\delta_{h_1\equiv \sigma,g^{-1} g_2 g})\\
=&\delta_{h_2g_2,g_2h_2}\sum_{g\in G}\delta_{g_1\sim_N \overline{g} g_2^{-1} g}\delta_{h_1\sim_N \overline{g} h_2 g}= \chi_{\mathcal{G}},
\end{align}
where the gray term is identically zero.
This tells us
\begin{align}
    A\to \mathbf{1}\quad B\to e\quad C\to \mathbf{1}+e  \quad D\to m\quad  E\to f.
\end{align}
Consequently,  $A$ and $C$ undergo condensation.  Anyons $B,C, D$, and $E$ remain deconfined, whereas $F$, $G$, and $H$ are confined. It's noteworthy that a part of $C$ is consended while another part remains  deconfined.
%-----------
\item
Take  $N=\mathbb{Z}_3=\{e,\tau,\tau^2\}$ as an example. Such a case corresponds to the condensation of $1+b=A+F$. The character of the representation is
\begin{align}
 \chi_{\mathcal{G}}(h_2g_2^*,h_1g_1^*)=&
 2\delta_{h_2g_2,g_2h_2}\delta_{g_1h_1\sim h_1 g_1}\delta_{h_1\sim h_2}\delta_{g_1\sim g_2}
\end{align}
On the other hand, compute
\begin{align}
&\chi_A(h_2g_2^*)\chi_{\mathfrak{1}}(h_1g_1^*)
+\chi_B(h_2g_2^*)\chi_e(h_1g_1^*)
+\chi_D(h_2g_2^*) \chi_m(h_1g_1^*) 
+\chi_E(h_2g_2^*) \chi_f(h_1g_1^*)\nnb\\
&+\chi_F(h_2g_2^*)(\chi_{\mathfrak{1}}+\chi_e)(h_1g_1^*)
\nnb\\
=&\delta_{g_2,e}\delta_{g_1,1} 
+\delta_{g_2,e}Sgn(h_2) \ \delta_{g_1,1} Sgn(h_1)
+\delta_{g_2\in \{\sigma,\sigma\tau,\sigma\tau^2\}} \delta_{h_2\in \{e,g_2\}} \ \delta_{g_1,-1}
+\delta_{g_2\in \{\sigma,\sigma\tau,\sigma\tau^2\} } (\delta_{h_2,e}-\delta_{h_2,g_2})\ \delta_{g_1,-1}Sgn(h_1) \nnb\\
&+\delta_{h_2\in \{e,\tau,\tau^2\}}\delta_{g_2\in \{\tau,\tau^2\}} \delta_{g_1,1}(1+Sgn(h_1))\nnb\\
=&\delta_{g_2,e}\delta_{g_1,1} (1+Sgn(h_2)Sgn(h_1))
%---------
+\delta_{g_2\in \{\sigma,\sigma\tau,\sigma\tau^2\}} \delta_{g_1,-1}  \qty{ 
\delta_{h_2,e}(1+Sgn(h_1))+\delta_{h_2,g_2}(1-Sgn(h_1)) } \nnb\\
%-------------
&+ \delta_{h_2\in \{e,\tau,\tau^2\}}\delta_{g_2\in \{\tau,\tau^2\}} \delta_{g_1,1}2\delta_{h_1,1}\nnb\\
%---------------
=&\delta_{g_2,e}\delta_{g_1,1} 2\delta_{h_2\sim h_1}
%---------
+2\delta_{g_2\in \{\sigma,\sigma\tau,\sigma\tau^2\}} \delta_{g_1,-1}  \qty{ 
\delta_{h_2,e}\delta_{h_1 ,1}+\delta_{h_2,g_2}\delta_{h_1,-1} } \nnb\\
%-------------
&+ 2\delta_{g_2\in \{\tau,\tau^2\}}\delta_{h_2\in \{e,\tau,\tau^2\}}\delta_{g_1,1}\delta_{h_1,1}\nnb \\
%------
=&2\delta_{g_2,e}\delta_{g_2h_2,h_2g_2}\delta_{g_1\sim g_2} \delta_{h_2\sim h_1}
%---------
+2\delta_{g_2\in \{\sigma,\sigma\tau,\sigma\tau^2\}} \delta_{g_1\sim g_2}  \delta_{g_2h_2,h_2g_2}\delta_{h_2 \sim h_1} \nnb\\
%-------------
&+ 2\delta_{g_2\in \{\tau,\tau^2\}}\delta_{g_2h_2,h_2g_2}\delta_{g_1\sim 
 g_2}\delta_{h_1\sim h_2}\nnb \\
=&2\delta_{g_2h_2,h_2g_2}\delta_{g_1\sim g_2} \delta_{h_2\sim h_1}
=2\delta_{g_2h_2,h_2g_2}\delta_{g_1h_1\sim h_1g_1}\delta_{g_1\sim g_2} \delta_{h_2\sim h_1}=\chi_{\mathcal{G}}(h_2g_2^*,h_1g_1^*).
%----
\end{align}
We hence conclude that the ribbon operators realize the following morphisms
\begin{align}
  A  \to \mathbf{1},\quad 
   B  \to e,\quad 
D  \to m ,\quad 
   E  \to f,\quad 
      F  \to \mathbf{1}+e.
\end{align}
This is similar to the previous case except that $C$ and $F$ are interchanged. The result is  expected, as the $C$-$F$ duality regarding the two subphases has been observed in \cite{PhysRevB.108.075105}.
\item $M=\mathbb{Z}_3=\{e,\tau,\tau^2\}$ and $N=\{e\}$.  The anyon type of $\mathcal{D}(\mathbb{Z}_3)$ is given by the following formula
\begin{align}
    \chi_{\overline{a_1},\rho(a_2)}(hg^*)\equiv \chi_{a_1,a_2}(hg^*)
    =\delta_{g=a_1}\omega^{a_2h},\quad a_1,a_2,h,g\in \mathbb{Z}_3= \{0,1,2\}.
\end{align}
where the representation $\rho_{a_2}(h)=\omega^{a_2 h}$,  for $\omega=e^{2\pi i/3}$.  For $h_1,g_1\in G$, 
$h_2,g_2\in M$ the character of algebra $\mathcal{G}$ is 
\begin{align}
    \chi_{\mathcal{G}}
    =&\chi_A  \chi_{0,0}
    +\chi_B\chi_{0,0}
    +\chi_C(\chi_{0,1}+\chi_{0,2})
    +\chi_{F}(\chi_{1,0}+\chi_{2,0})
    \nnb\\
    &+\chi_G(\chi_{1,2}+\chi_{2,1}) +\chi_H(\chi_{1,1}+\chi_{2,2})
\end{align}
We label $a_1=1$ as $m$, $a_1=2$ as $\wt{m}$; similarly, label $a_2=1$ as $e$,  $a_2=2$ as $\wt{e}$. We get 
\begin{align}
    A\mapsto \mathbf{1},\quad B\mapsto \mathbf{1},\quad C\to e+\wt{e},\quad F=m+\wt{m},\quad G\to m\wt{e}+ \wt{m}e,\quad H\to em+\wt{e}\wt{m}.
\end{align}
\end{itemize}
%-----
\subsection{Example: \texorpdfstring{$\mathcal{D}(D_4)$}{Lg}}\label{sect_representation_of_DD4}
The group is 
\begin{align}
D_4=&\langle r,s: r^4=s^2=e, rs=sr^{-1},sr=r^{-1}s\rangle\\
=&\{s^mr^k:s^2=r^4=1, r^ks =sr^{-k},\ sr^k=r^{-k } s,\  m=0,1;\ k=0,\cdots, 3\}
\end{align}
There are five conjugacy classes:
\begin{align}
[e]=&\{e\}\quad 
[r^2]=\{r^2\}\quad 
[r]=\{r,r^3\}\quad 
[s]=\{s,sr^2\}\quad 
[sr]=\{sr,sr^3\}
\end{align}
The centralizers are given by the following,
\begin{align}
Z(e)=&D_4\quad\quad  Z(r^2)=D_4\\
Z(r)=&\{r^j:j=0,\cdots n-1\}\cong \mathbb{Z}_4\\
Z(s)=&\{e,r^2,s,sr^2\}\cong D_2\\
Z(sr)=&\{e,r^2,(sr),(sr)r^2\}\cong D_2
\end{align}
For the irreps of $D_4$, there are four 1-dim irreps
\begin{align}
\rho_0(s^mr^k)=&1=\text{trivial}\quad
\rho_1(s^mr^k)=(-1)^m\quad 
\rho_2(s^mr^k)=(-1)^k\quad 
\rho_3(s^mr^k)=(-1)^{k+m}.
\end{align}
We document the anyon types in Table~\ref{table_D4_anyons}.
\begin{table}[htb]
\centering
\begin{tabular}{|c|c|c|c|}\hline
excitations& conj.  & $Z(r_C)$ &  representation \\ \hline
 $A_i$& $[e]$  & $D_4$ &  $\rho_i$ for $i=0,1,2,3,4$\\
$B_i$ &  $[r^2]$ & $D_4$ &   $\rho_i$ for $i=0,1,2,3,4$\\
$C_k$ & $[r]$ & $\mathbb{Z}_4=\langle r\rangle$ &generator being $\omega^k$ for $k=0,1,2,3$\\
$D_{(i,j)}$ & $[s]$ & $D_2\cong \mathbb{Z}_2\times \mathbb{Z}_2=\langle s, r^2\rangle$ & $i,j\in\{tri,Sgn\}$\\
$E_{(i,j)}$ & $[sr]$ &$D_2\cong \mathbb{Z}_2\times \mathbb{Z}_2=\langle (sr), r^2\rangle$          & $i,j\in\{tri,Sgn\}$ \\ \hline
\end{tabular}
\caption{We have defined $\omega=e^{i\pi /2}=i$ for $\mathbb{Z}_4$.  There are in total $5+5+4+4+4=22$ irreps for $\mathcal{D}(D_4)$.}\label{table_D4_anyons}
\end{table}
Take the case $M/N=\{e,r^2,s,sr^2\}/\{e,r^2\}$ as an example. By expanding the formula over characters of $\mathcal{D}(D_4)$ anyons tensoring the characters of $\mathcal{D}(M/N)\cong \mathcal{D}(\mathbb{Z}_2)$, one arrives at the following decomposition
\begin{align}
    \chi_{\mathcal{G}}=&\chi_{A_0}\chi_{\mathbf{1}}+\chi_{A_1}\chi_e+\chi_{A_2}\chi_{\mathbf{1}} +\chi_{A_3}\chi_{e} +\chi_{B_0}\chi_{\mathbf{1}}+\chi_{B_1}\chi_e+\chi_{B_2}\chi_{\mathbf{1}} +\chi_{B_3}\chi_{e} +2\chi_{D_{00}} \chi_{m}+2\chi_{D_{01}}\chi_{f}
\end{align}
%------------ 
Hence, the anyon-tunneling map is
\begin{align}
&A_0\to \mathbf{1},\quad A_1\to  e  ,\quad A_2\to \mathbf{1},\quad A_3\to e\\
&B_0\to \mathbf{1},\quad B_1\to e  ,\quad B_2\to \mathbf{1},\quad B_3\to e\\
&D_{00}\to 2m ,\quad D_{01}\to 2f.
\end{align}
%-----------
\section{Anyon Condensation}\label{app_condensation_bdry}
Consider a quantum double model, denoted as $\mathcal{D}(G)$, implemented on lattice $\Lambda$. Let's define a subphase as $\mathcal{P}_1=\mathcal{D}(M/N)$. To compute the condensable algebra transitioning from phase $\mathcal{D}$ to subphase $\mathcal{D}(M/N)$, we'll employ character theory.
Since the determination of the condensable algebra can be read off from the anyon-tunneling computations we previously presented in Sect.~\ref{subsect_anyon_tunneling}, we will streamline our discussion by outlining only the pivotal steps, foregoing exhaustive proof details. For those seeking deeper insights into the proofs, one might turn to \cite{shor2011}, where similar algebraic structures are discussed.
Introduce $\xi$ as a ribbon that bridges an interior site $s_1$ to a boundary site $s_0$.
Define algebra
\begin{align}
\mathcal{C}_\xi:=\{T\in \mathcal{F}_\xi:[T,A_{s_0}^M]=[T,B_{s_0}^N]=0\}
\end{align}
Let 
$
T=\sum_{h,g}c_{h,g}F_\xi^{h,g} 
$
be in $\mathcal{C}_\xi$.   Recall from \cite{bombin:2008}
\begin{align}
A_{s_0}^kF_\xi^{h,g}=&F_\xi^{khk^{-1},kg} A_{s_0}^k\\
B_{s_0}^k F_\xi^{h,g}=&F_\xi^{h,g}B_{s_0}^{kh}
\end{align}
Solving the equation $[T,A_{s_0}^M]=[T,B^N_{s_0}=0$, 
we arrive at 
\begin{align}
c_{m^{-1} hm,m^{-1}g} =c_{h,g},\quad c_{h,g}=c_{h,g}\delta_{h\in N}
\end{align} 
Hence the operators
\begin{align}
T^{n,g}:=\sum_{\ell \in M}F_\xi^{\ell n\ell^{-1}, \ell g^{-1}},\quad n\in N\subseteq M,\quad  g\in G
\end{align}
span $\mathcal{C}_\xi$. Furthermore, they exhibit the following properties
\begin{align}
T^{n,gk}=& T^{knk^{-1},g}  \label{T_op_properties1} \\
T^{n,g}T^{n^\prime, g^\prime}=&0,\quad \text{if }gM\neq g\1 M \label{T_op_properties2}\\
T^{n,g}T^{n\1, g}=&T^{nn\1,g} \label{T_op_properties3}\\
(T^{n,g})^\dagger=&T^{n^{-1},g}. \label{T_op_properties4}
\end{align}
Define $
\ket{\psi^{n,g }}:=T^{n, g}\ket{\psi}$ and the spanned vector space
\begin{align}
\mathcal{A}_{M,N}:=Span\qty{\ket{\psi^{n,g }}: n\in N,g\in G}
=span\qty{T^{n, g}\ket{\psi}: n\in N,g\in G}
\end{align}
Recall  Eq.~(B41) and (B42) in \cite{bombin:2008}
\begin{align}
A_{s_1}^k F_\xi^{h,g}=&F_\xi^{h,gk^{-1} }A_{s_1}^k\\
B_{s_1}^k F_\xi^{h,g}= &F_\xi^{h,g} B_{s_1}^{g^{-1} h^{-1} gk}.
\end{align}
One can compute
\begin{align}
A_{s_1}^h \ket{\psi^{n,g}}
%=\sum_{\ell \in M}A_{s_1}^hF_\xi^{\ell n\ell^{-1}, \ell g^{-1}}\ket{\psi}
% \nnb \\ =&\sum_{\ell \in M}F_\xi^{\ell n\ell^{-1}, \ell g^{-1} h^{-1}}A_{s_1}^h\ket{\psi}
=\ket{\psi^{n,hg}} 
\quad \quad 
B_{s_1}^h \ket{\psi^{n,g}}
%=\sum_{\ell \in M}F_\xi^{\ell n\ell^{-1}, \ell g^{-1}}  B_{s_1}^{(\ell g^{-1})^{-1}  (\ell n\ell^{-1})^{-1} ( \ell g^{-1}) h}\ket{\psi}
% \nnb\\ =&\sum_{\ell \in M}F_\xi^{\ell n\ell^{-1}, \ell g^{-1}}  B_{s_1}^{(g\ell^{-1})   \ell n^{-1}\ell^{-1} ( \ell g^{-1}) h}\ket{\psi}\nnb\\
%=&\sum_{\ell \in M}F_\xi^{\ell n\ell^{-1}, \ell g^{-1}}  B_{s_1}^{gn^{-1} g^{-1} h}\ket{\psi}\nnb\\
%=&\delta_{gn^{-1}g^{-1}h,e}\ket{\psi^{n,g}}
=\delta_{h,gng^{-1}}\ket{\psi^{n,g}}
\end{align}
They form a representation of $D(G)$ on $\mathcal{A}_{M,N}$.
Now suppose  $G=g_1 M\cup \cdots \cup g_r M$. Then,
% in the following calculate we use comment % to skip some calculations
\begin{align}
\langle\psi^{n,g_i} |\psi^{n\1, g_j}\rangle
%=&\langle \psi| (T^{n,g_i})^\dagger T^{n\1,g_j}\ket{\psi}\\
%=&\delta_{i,j}\langle \psi | T^{n^{-1} n\1, g_i}|\psi\rangle \\
%=&\delta_{i,j}\langle\psi | \sum_{\ell \in M}F_\xi^{\ell n^{-1} n\1\ell^{-1}, \ell g_i^{-1}}  |\psi\rangle\nnb\\
%=&\delta_{i,j}\sum_{\ell \in M}\frac{\delta_{\ell n^{-1} n\1 \ell^{-1},e}}{|G|}\nnb\\
=&\frac{|M|}{|G|}\delta_{i,j}\delta_{n,n\1}=\frac{1}{r}\delta_{i,j}\delta_{n,n\1}
\end{align}
Therefore
\begin{align}
\{\sqrt{r}|\psi^{n,g_i}\rangle : n\in N, i=1,\cdots , r\equiv |G|/|M|  \}
\end{align}
is a set of  orthonormal basis of $\mathcal{A}_{M,N}$.
The character is calculated through taking the trace of teh algebra
\begin{align}
\chi_{\mathcal{A}_{M,N}}(hg^*)=&r\sum_{n\in N}\sum_{i=1}^r \bra{\psi^{n,g_i}} A_{s_1}^h B_{s_1}^g\ket{\psi^{n,g_i}}\\
=&r\sum_{n\in N}\sum_{i=1}^r\delta_{g,g_i ng_i^{-1}} \langle \psi^{n,g_i} | \psi^{n,hg_i}\rangle
\end{align}
We know that by partitioning $G$ into cosets of $M$, for $h\in G$, there exists some $m_i\in M$ s.t. $hg_i=g_{\epsilon(i)}m_i$  for some integer label  $\epsilon(i)\leq r$. Then using Eq.~\ref{T_op_properties1} above
%--------
% We skip some calculation by putting comment % ahead of the lines
\begin{align}
\chi_{\mathcal{A}_{M,N}}(hg^*)=&
r\sum_{n\in N}\sum_{i=1}^r\delta_{g,g_i ng_i^{-1}} \langle \psi^{n,g_i} | \psi^{n,g_{\epsilon(i)}m_i}\rangle\\
=&r\sum_{n\in N}\sum_{i=1}^r\delta_{g,g_i ng_i^{-1}} \langle \psi^{n,g_i} | \psi^{m_inm_i^{-1},g_{\epsilon(i)}} \rangle\\
%=&r\sum_{n\in N}\sum_{i=1}^r\delta_{g,g_i ng_i^{-1}}\delta_{n,m_inm_i^{-1} } \delta_{i,\epsilon(i)}\frac{|M|}{|G|}\\
%=&\sum_{n\in N}\sum_{i=1}^r \delta_{g_i^{-1}gg_i,n}  \delta_{nm_i,m_in } \delta_{g_i,hg_im_i^{-1}}\\
%=&\sum_{n\in N}\sum_{i=1}^r \delta_{g_i^{-1}gg_i,n}  \delta_{nm_i,m_in } \delta_{m_i,g_i^{-1}hg_i}\\
%=&\sum_{n\in N}\sum_{i=1}^r \delta_{g_i^{-1}gg_i,n}  \delta_{gh,hg} \delta_{m_i,g_i^{-1}hg_i}\\
=&\delta_{gh,hg}\sum_{i=1}^r  \delta_{g_i^{-1}gg_i\in N} \delta_{g_i^{-1} h g_i\in M}\nnb\\
=&\frac{1}{|M|}\delta_{gh,hg}\sum_{x\in G}  \delta_{x^{-1}gx\in N} \delta_{x^{-1} h x\in M}
\end{align}
Decomposing the character into irreducible representations of $\mathcal{D}(G)$,  labeled as $a$
\begin{align}
    \chi_{\mathcal{A}_{M,N}}=\oplus_a m_a \chi_a
\end{align}
The corresponding condensable algebra is $\oplus_a a$ for $a\in Rep(\mathcal{D}(G))$. 
For abelian group, the character is
\begin{align}
\chi_{\mathcal{A}_{M,N}}(hg^*)=
\frac{|G|}{|M|} \delta_{g\in N} \delta_{h \in M},\quad N\subseteq M\subseteq G\label{chi_MN_abelian}
\end{align}
%---------
\subsection*{Example: the quantum double of \texorpdfstring{$\mathcal{D}(S_3)$}{Lg}}
\begin{itemize}
    \item
We first consider the case $N=\mathbb{Z}_3=\{e,\tau,\tau^2\}$, which corresponds to the condensation of $1+b=A+F$ and $M=G$. Then
\begin{align}
\frac{|M|}{|G|}\chi_{\mathcal{A}_{M,N}}(hg^*)=& \delta_{g\in \mathbb{Z}_3} \delta_{gh,hg} 
=\delta_{g,e}+\delta_{g\in \{\tau,\tau^2\}}\delta_{h\in \mathbb{Z}^3}=(\chi_A+\chi_F)(hg^*).
\end{align}
\item Now consider the case of $N=\{e\}$ and $M=\{e,\sigma\}$, which corresponds to the condensation of $A+C$.
\begin{align}
   |M| \chi_{\mathcal{A}_{M,N}}=& \delta_{g,e}\sum_{x\in G} \delta_{x^{-1} h x\in \{e,\sigma\}}+\delta_{g,\sigma}0
   = \delta_{g,e} (\delta_{h,e}|G|+2\delta_{h\in \overline{\sigma}}+ \delta_{h\in \{\tau,\tau^2\}}0)\nnb\\
   =&2\delta_{g,e} (3\delta_{h,e} +\delta_{h\in \overline{\sigma}})
   =2(\chi_A+\chi_C)(hg^*)
\end{align}
where we used the fact that for each element $\in \overline{\sigma}$, there are two elements $x\in G$ s.t. $x^{-1} hx=\sigma$ (one in $\mathbb{Z}_3$ and the other in $\sigma \mathbb{Z}_3\equiv \overline{\sigma}$), and no $x\in G$ can send either $\tau$ or $\tau^2$ to $e$.
\end{itemize}
%-----
%--------
\section{Proofs of claims about Floquet codes}
\begin{remark}\label{proof_of_proposition_conditional_unitary_iff_condition}
(Proof of Proposition~\ref{proposition_conditional_unitary_iff_condition}). 
Let $\{g_iN\}$ be the cosets of $M/N$.
We begin by measuring $T^{M\1}$, followed by  $L^{N\1}$. The proof is consequently divided into two parts corresponding to the two steps of measurement.
Within this proof, define $\ket{S}:=\sum_{s\in S}\ket{s}$. The overline symbol will represent normalized  vectors: $\ket{\overline{v}}:=\ket{v}/\sqrt{\norm{\ket{v}}_2}$.
Throughout this analysis, it's worth noting that we will occasionally disregard the normalization factors of vectors. Additionally, the edge label of the lattice $\Lambda$ will be omitted since our proof operates on a quditwise basis.
\begin{itemize}
    \item For $T^{M\1}=1$, the state is  mapped to 
    \begin{align}\sum_i\alpha_i\ket{g_iN}\mapsto  \sum_i \alpha_i\ket{g_iN\cap M\1 }=\sum_i \qty(\alpha_i \sqrt{|g_iN\cap M\1|})\ket{\overline{g_iN\cap M\1}}
    \end{align}
 For $T^{M\1}=0$, the state is mapped to 
    \begin{align}\label{TM1_qudit_map}
        \sum_i\alpha_i\ket{g_iN}\mapsto  \sum_i \alpha_i\ket{g_iN\setminus M\1 }=\sum_i\qty(\alpha_i\sqrt{|g_iN\setminus M\1|})\ket{\overline{g_iN\setminus M\1}}.
    \end{align}
Let $V_i^N$ be the subspace $Span\{\ket{g_i n}:n\in N\}$. 
Within each $V_i^N$, there exists a unitary $u_i$ that turns $\ket{\overline{g_iN\setminus M\1} }$ to $\ket{\overline{g_iN\cap M\1}}$. Then the conditional unitary is $U_{M\1}=\oplus_i u_i$.  Yet, the resulting state will undergo a "stretching" by a factor $\sqrt{|g_iN\cap M\1|}$ or $\sqrt{|g_i N\setminus M\1|}$ depending on the measurement. The qudit is only preserved at the first measurement step iff the overlapping number $|(g_iN)\cap M\1|$ is independent of the coset label $i$.  However, if $(g_iN)\cap M\1 \neq \emptyset $ holds true for all $i$, then the condition $|(g_iN)\cap M\1|=const$ will be inherently satisfied. This is because $(g_iN)\cap M=g_i(N\cap M\1)$ form a set of cosets of $M\1/(N\cap M\1)$.
%\footnote{Without loss of generality, one can choose the representatives $g_i$ of cosets $M/N$ such that $g_i\in M$. (1). Suppose $x\in (g_i N)\cap M$. Then $g_i^{-1} x\in N\cap M$. So $x=g_i(g_i^{-1}x)\in g_i (N\cap M)$. (2) Suppose $x=g_in\in g_i(N\cap M)$, where $n\in N\cap M$. By construction, $g_i n\in g_i N$ and $g_i n\in M$. Hence $x\in (g_iN)\cap M$. We hence conclude that $g_i (N\cap M)=(g_iN)\cap M$. }.
    \item 
    Let $\{g\1_j N\1\}$ be the collection of cosets in $M\1/N\1$. 
In the case of measurement result being $L^{N\1}=+1$, we get
\begin{align}
    &\sum_i \alpha_i\ket{g_iN\cap M\1 }=
    \sum_i \alpha_i(\sum_j \ket{g_i N\cap g_j\1N\1})
    = \sum_i \alpha_i \bigg (\sum_j |g_iN\cap g_j\1 N\1|^{1/2} \ket{\overline{g_iN\cap g_j\1 N\1}}\bigg)  
\end{align} 
Define  $S_{j,i}:= g_iN\cap g\1_j N\1$, and $M_{ji}:=|S_{j,i}|$.  %We denote its cardinality by $\#(S_{ji})\equiv |S_{ji}|$
A general vector in the case of $L^{N\1}=1$ is mapped to
\begin{align}
    &\sum_i\alpha_i\ket{g_i N\cap M\1}\overset{L^N=1}{\mapsto}   \sum_{j} \bigg(\sum_i|g_j\1 N\1 \cap  g_iN| \alpha_i \bigg)\ket{g_j\1 N\1}\nnb\\
    \sim& \frac{1}{ \sqrt{\sum_q |\sum_p \#(S_{q,p})\alpha_p|^2}}\sum_j |S_{j,i}| \alpha_i \ket{\overline{g\1_j N\1}}
    = \frac{1}{\sqrt{\alpha^*M^T M\alpha }} \sum_j (M\alpha)_j\ket{\overline{g_j\1 N\1}}.
\end{align}
where $\sim$ refers to the equivalence relation in $\mathbb{CP}^{d}$. 
The action 
\begin{align}
    \alpha_j \mapsto  \frac{1}{\sqrt{\alpha^*M^T M\alpha }}  (M\alpha)_j
\end{align}
is unitary on a general vector $\alpha \in \mathbb{C}^d$ iff $M^T M$ is a multiple of identity, or equivalently, $M$ is proportional to an orthogonal matrix. 
In such a case we can simply apply the unitary $M^T/\sqrt{M^TM}$ to correct it to $\sum_j\alpha_j\ket{\overline{g\1_j N\1}}$. Note that, if $M$ is proportional to an orthogonal matrix, the condition $g_iN\cap M\1\neq \emptyset$ for all $i$ is guaranteed.

Consider the case of $L^N=0$. For each $j$, using the identity
$\ket{g\1_jN\1}=\ket{S_{j,i}}+\ket{g\1_jN\1 \backslash S_{j,i} }$,
one arrives at
\begin{align}
    \ket{S_{j,i}}\equiv \ket{g_j\1 N\1 \cap g_iN}
     = \frac{|S_{j,i}|}{|N\1|}(\ket{g\1_j N\1}+\ket{v_{i,j}}) \label{eqc6}
\end{align}
where 
\begin{align}
 \ket{v_{i,j}}:= \frac{|N\1|-|S_{j,i}|}{|S_{j,i}|} \ket{S_{j,i}} -\ket{g_{j}\1 N\1\setminus S_{j,i}}, %\quad \norm{\ket{v_{i,j}}}_2=|N\1|\qty(\frac{|N\1|-|S_{ji}|}{|S_{ji}|})
\end{align}
It has zero eigenvalue under the action of $L^{N\1}$:
\begin{align}
    L^{N\1} \ket{v_{i,j}}=\frac{1}{|N\1|}\qty{(|N\1|-|S_{j,i}|)\ket{g\1_{j}N\1}- |g\1_{j} N\1\setminus S_{j,i}|\ket{g\1_{j}N\1} }=0.
\end{align}
and hence $\ket{v_{i,j}}$ is orthogonal to $\ket{g\1_{j} N\1}$. 
Therefore in the case of $L^{N\1}=0$, the state is mapped to (ignore normalization)
\begin{align}
    \sum_{i}\alpha_i\ket{g_iN\cap M\1}=\sum_{i,j}\alpha_i\ket{g_iN \cap g_{j}\1N\1}\equiv \sum_{i,j}\alpha_i\ket{S_{i,j}}\mapsto 
    \sum_{i,j}\alpha_i\frac{|S_{j,i}|}{|N\1|}\ket{v_{i,j}}
    %\sum_{i,j}  \alpha_i \sqrt{|S_{j,i}| \frac{|N\1|-|S_{ji}|}{|N\1|} }\ket{\overline{v_{i,j}}}
\end{align}
%For each fixed $j$ labeling $g_jN\1$, we can find linear map $\ell_j$ that turns all $v_{ij}$ to $s_{ij}$, since vectors $\{s_{ij}:i\}\subset \mathbb{C}g_jN\1$ are linearly independent.  Then the desired linear map is $L=+_j \ell_j$. We effectively have the map\begin{align}    \sum_i \alpha_i \ket{g_iN\cap M\1}\mapsto \sum_{i,j}\alpha_i |S_{i,j}|\ket{s_{i,j}}\end{align}
To see the existence of unitary that does the desired map, we first take a closer look at vector 
\begin{align}
    \ket{b_i}:=&\sum_j |S_{ji}|\ket{v_{i,j}} =\sum_j(|N\1| -|S_{j,i}|)\ket{S_{j,i}} -\sum_j |S_{j,i}| \ket{g\1_j N\1 \setminus S_{j,i}}\nnb\\
    =&|N\1|\sum_j \ket{S_{ji}} -\sum_j |S_{ji}| (\ket{S_{ji}} + \ket{g\1_j N\1 \setminus S_{ji}})\nnb\\
    =&|N\1 | \ket{g_{i} N} -\sum_j |S_{ji}| \ket{g\1_j N\1}
\end{align}
Take the inner product
\begin{align} 
\langle b_m | b_i\rangle =& \qty(\bra{g_{m}N} |N\1| -\sum_n |S_{n,m}| \bra{g\1_n N\1})  \qty(|N\1 | \ket{g_{i }N} -\sum_j |S_{ji}| \ket{g\1_j N\1}) \nnb\\ 
=&|N\1|^2|N|\delta_{mi}-|N\1| \sum_n |S_{nm}|  \langle g\1_n N\1 | g_i N\rangle  -|N\1| \sum_j |S_{ji}| \langle g_{mN} |g\1_j N\1\rangle 
+\sum_{j,n} |S_{ji}|\ |S_{nm}| \delta_{nj}|N\1|\nnb\\
=&|N\1|^2 |N|\delta_{mi} -|N\1| \sum_n M_{nm} M_{ni}  -|N\1| \sum_j M_{ji} M_{jm}
+|N\1|\sum_{j} M_{ji}\ M_{j m}  \nnb\\
=&(|N\1|^2|N|-|N\1| (M^TM)_{11})\delta_{mi}
\end{align}
where in the last step we have used the "iff" condition derived before that $MTM$ is a multiple of identity.  Consequently, after normalization, for $L^N=0$ 
\begin{align}
        \sum_{i}\alpha_i\ket{g_iN\cap M\1}\mapsto \sum_i \alpha_i\ket{\overline{b_i}},
\end{align}
and there exists a unitary in $\mathbb{C}[G]$ that maps the orthonormal set $\{\ket{\overline{b_i}}\}$ to the orthonormal set $\{\ket{\overline{g\1_i N\1}}\}$ by basic linear algebra.
\end{itemize}
\end{remark}
%--------
\begin{remark}\label{proof_of_proposition_sufficient_condition_for_logical_state_preservation}
(Proof of Proposition~\ref{proposition_sufficient_condition_for_logical_state_preservation}).
The proof is straightforward. Namely we just re-examinate the steps in Proposition~\ref{proposition_conditional_unitary_iff_condition}.
Let $\ket{g_jN}_A$ label the single qudit (cosets) we are measuring, and let $\ket{i}$ label the rest of the qubits $\overline{A}$ on the lattice (the complement of $A$).  
    Consider the state $\ket{\psi}=\sum_{i,j}\alpha_{ij}\ket{i}_{\overline{A}}\ket{g_jN}_A$. After measuring $T^M$ and making corrections when $T^M=0$, we eventually obtain
    \begin{align}
    \sum_{ij}\alpha_{ij}\ket{i}_{\overline{A}}\ket{g_jN}_A \mapsto      \sum_{ij}\alpha_{ij}\ket{i}_{\overline{A}}\ket{g_jN\cap M\1}_A
    \end{align}
    Then after the measurement of $L^N$ and making corrections, we arrive at
      \begin{align}
    \sum_{ij}\alpha_{ij}\ket{i}_{\overline{A}}\ket{g_jN\cap M\1}_A \mapsto      \sum_{ij}\alpha_{ij}\ket{i}_{\overline{A}}\ket{g_j\1 N\1}_A 
    \end{align}
 Since the entanglement contained in coefficients $\alpha_{ij}$ is intact, the logical qubit is preserved.
\end{remark}
%----------------------

%-----

%-----------
\bibliography{main.bib}% Produces the bibliography via BibTeX.
\end{document}